pdfoutput=1
\documentclass[final,colophon]{phduio}
\DeclareRobustCommand{\ion}[2]{\textup{#1\,\textsc{\lowercase{#2}}}}
\newcommand*\farcs{\ensuremath{\overset{\prime\prime}{.}}}
\providetoggle{blx@lang@captions@english}
\def\arcs{$^{\prime\prime}$}
\usepackage{phdstyle}   % Custom style
\usepackage{kantlipsum} % Dummy text
\usepackage{amsmath}
\usepackage{epstopdf}
\usepackage{txfonts}
\usepackage{subcaption}
\DeclareLanguageMapping{english}{english-apa}
\DefineBibliographyStrings{english}{%
  andothers = {et\addabbrvspace al\adddot},
}
\DeclareUnicodeCharacter{0301}{\'{e}}
\def\kms{\hbox{km$\;$s$^{-1}$}}
\def\halpha{H$\mathrm{\alpha}$}
\def\cak{\hbox{\ion{Ca}{ii}~K}}
\def\Mgk{\hbox{\ion{Mg}{ii}~k}}
\def\radsun{R$_{\odot}$}

\def\Si{\hbox{\ion{Si}{iv}}}
\def\alfvenic{Alfv\'enic}
\let\citet\cite 
\let\citep\parencite

\usepackage{graphicx}
\author{Souvik~Bose}
\title{On the dynamics of spicules and mass flows in the solar atmosphere}
\department{Institute of Theoretical Astrophysics}
\faculty{Faculty of Mathematics and Natural Sciences}
\affiliation
{
    Rosseland Center for Solar Physics
}
\ISSN{1501-7710}           % Request correct number from repro@uio.no
\dissertationseries{2443}  % Request correct number from repro@uio.no

\usepackage{aas_macros}
\begin{document}

    \frontmatter        % Folios in Roman numerals, unnumbered chapters.

    \uiotitle
    
    \thispagestyle{empty}
\vspace*{\stretch{1}}
\begin{flushright}
    \emph{To Maa and Baba}
\end{flushright}
\vspace*{\stretch{3}}
    \thispagestyle{empty}
\vspace*{\stretch{1}}
\begin{flushright}
    \emph{"I start early, and I stay late, day after day after day, year after year. It took me 17 years and 114 days to become an overnight success."}\\
    --Lionel Andr\'es Messi
\end{flushright}
\vspace*{\stretch{3}}
    \chapter{Preface}

This thesis is submitted in partial fulfillment of the requirements
for the degree of \emph{Philosophiae Doctor} at the University of Oslo.
The research presented here was conducted at the Rosseland Center for Solar Physics that is a part of the Institute of Theoretical Astrophysics, University of Oslo, and under the primary supervision of Luc Rouppe van der Voort. The research of this thesis was partially funded by the Norwegian Research Council and it represents an effort towards understanding the role of spicules in mass balance and heating of the solar atmosphere. 

The results and analysis presented in this thesis are based on the datasets that have been obtained from multiple instruments, namely, the Swedish 1-m Solar Telescope, Interface Region Imaging Spectrograph, and Solar Dynamics Observatory. In addition, strong theoretical support has been provided with the help of a state-of-the-art magnetohydrodynamic (MHD) simulation of spicules.

This thesis is a collection of three papers, presented in chronological order of writing, and is spread over four chapters. \cref{chap:intro} provides a basic introduction to the physical processes taking place in the solar interior and an overview of the atmosphere of the Sun. Special emphasis has been given to highlight the coupling of the solar atmosphere. In \cref{chap:Instruments_simulations}, I have attempted to provide an overview of the different instruments along with their datasets that have been used in this thesis. In addition, I have also attempted to discuss the various challenges and advantages of using multiple instruments from the ground and space co-temporaneously. Moreover, this chapter also provides a brief description of the MHD simulation of spicules that has been used to compare with the observations. \cref{chap:spicules} focuses on spicules, their associated mass flows, and their role in heating the solar atmosphere starting from a historical perspective. Moreover, this chapter also serves as a platform on which the research carried out as a part of this thesis lays. Finally, \cref{chap:paper_summary} provides a summary of the papers (referred to as Papers I, II, and III) that form a major part of this thesis along with concluding remarks and future prospects.

In addition to the studies on spicules, I have also had the opportunity to work (or participate) in several other projects in the field of Solar physics during the course of my doctoral research, which has led to various publications. They have been listed separately.

    \chapter{Acknowledgments}

The journey over the past three and a half years of my life in Norway has been a very rewarding (and at times challenging) adventure. Challenging because, firstly, this is the first time I had been away from India, and adjusting to the Norwegian culture and climate (with long and dark winters) took some time. However, from the perspective of my work, it has been an extremely rewarding journey that allowed me to grow and broaden my professional outlook. For this, I would like to express my gratitude to those without whom this thesis would have been incomplete.

First and foremost, I am extremely grateful to my primary supervisor, Luc Rouppe van der Voort, for his invaluable advice, stimulating discussions, and for pushing me in the right direction. Thank you, Luc, for introducing me to the fascinating world of high-resolution Solar physics, latex tips and tricks, for your friendship, and your constant support even before I formally began my Ph.D. It was really fun to work with you, and I hope we can continue our wonderful collaboration in the next phase of my journey.

This thesis would be nowhere close to where it is today without the continuous support and enthusiasm of my co-supervisors, Vasco Henriques and Jayant Joshi. Thank you, Vasco, for helping me to start my Ph.D., teaching me about the basics of SST observations, and for those (immensely) long hours of CRISPEXing and having radiative transfer discussions together, while trying to look for fascinating new stuff with our datasets. Your enthusiasm and guidance was instrumental in maneuvering the steep learning curve during the early days of my Ph.D. 

Jayant, thank you for introducing the concept of $k$-means clustering to me, which has played an important role in the studies leading up to this thesis. Thanks also for those (unlimited) discussions on various aspects in Solar physics (also on Indian politics) that we had during our highly “productive” coffee breaks. I have enjoyed collaborating with both of you throughout my Ph.D. and I would like to continue doing so. 

Thanks, are also due to Mats Carlsson, for teaching me the basics of radiative transfer, for his help in interpreting the Bifrost simulation snapshots, and for organizing the annual \textit{barbeque-boule party}! 

Thank you, Tiago Pereira, for introducing and teaching me the basics of the RH code and for being there when it crashed! It was also wonderful to collaborate with you. 

Mikolaj Szydlarski, thank you so much for helping me to get grips on the art of high-performance computing and for teaching me all those computational tips and tricks that have made my life easier! I am sure there are more new kinds of stuff to learn from you! 

I would like to acknowledge my other collaborators and co-authors: Daniel N\'obrega-Siverio, Juan Mart\'inez-Sykora, and Bart De Pontieu. Daniel, it was wonderful to finally collaborate with you on paper after so many years of discussing and exchanging ideas. Let’s do it again! Juan, thanks a lot for sharing your impressive MHD simulation of spicules with me. Thank you, Bart, for all the stimulating spicule discussions we have had while you hosted me at LMSAL and for motivating me to apply for the position of an IRIS planner. You each had your indispensable contribution to the final paper of this thesis.

Thanks, are also due to current -- Viggo Hansteen, Sven Wedemeyer, Boris Gudiksen, Shahin Jafarzadeh, Atul Mohan, Nancy Narang, K. Chandrasekhar, Ana Bel\'en, Helle Bakke, Frederik Clemmensen, Sneha Pandit, Kilian Krikova, former – Clara Froment, Ada Ortiz-Carbonell, and Carlos Quintero Noda, and other scientific members of the wonderful Rosseland Center for Solar Physics. Without their support, it would have been difficult to conduct this research. 

Special thanks go to my office mates, Henrik Eklund and Juan Camilo Guevara Gomez, without whom, daily life at the office would be super boring! Thank you, guys, for the “secret” closed-door conversations and for keeping things alive and light until the very end. 

Also, I would like to thank all the IT and administrative staff at the institute. Special thanks to Torben Leifsen and Kjell Andresen for providing quick solutions to my complaints about the workstation! Thanks, are also due to Kristine Aall Knudsen, Benedikte Fagerli Karlsen, Brenda J.A. Atubo, and Sara Ansgari Nettum for always being there for all sorts of bureaucratic technicalities. 

I want to thank my family – my parents and my sister, for their constant support and encouragement without which none of this would have been possible. Thank you, Maa, Baba, and didi, for being there from halfway across the world.

Last but in no way the least, I would like to express my heartfelt gratitude to Swastika for tagging along with me over the past two years now. Thanks for being a patient listener (whenever I needed you) and for being super curious about spicules! I look forward to the wonderful times that lay ahead of us. Much love! 

\vskip\onelineskip
\begin{flushleft}
    \sffamily
    \uiocolon\textbf{\theauthor}
    \\
    Oslo,\MONTH\the\year
\end{flushleft}
    \chapter{List of publications}

\section*{Paper I}
\textbf{Characterization and formation of on-disk spicules in the \cak{}
and \Mgk{} spectral lines}\\
\textbf{S.~Bose}, V.~M.~J.~Henriques, J.~Joshi\ and L.~Rouppe~van~der~Voort\\
% \enquote{\nameref{pap:first}}.
In: \emph{Astronomy \& Astrophysics}.
Vol.\ 631,
no.\ L5
(2019)\\
%pp.~123--456.
\doi{10.1051/0004-6361/201936617}\\
arXiv link: \url{https://arxiv.org/abs/1910.05533}

\section*{Paper II}
\textbf{Spicules and downflows in the solar chromosphere}\\
\textbf{S.~Bose}, J.~Joshi, V.~M.~J.~Henriques\ and 
L.~Rouppe~van~der~Voort\\
%\enquote{\nameref{pap:A4}}.
In: \emph{Astronomy \& Astrophysics}
Vol.\ 647,
no.\ 147
(2021)\\
\doi{10.1051/0004-6361/202040014}\\
arXiv link: \url{https://arxiv.org/abs/2101.07829}

\section*{Paper III}
\textbf{S.~Bose}, L.~Rouppe~van~der~Voort, J.~Joshi, V.~M.~J.~Henriques, D.~N\'obrega-Siverio, J.~Mart\'inez-Sykora\ and B.~De~Pontieu\\
\textbf{Evidence of multithermal nature of spicular downflows. Impact on solar atmospheric heating}\\
In: \emph{Astronomy \& Astrophysics}
Vol.\ 654,
no.\ 51
(2021)\\
\doi{10.1051/0004-6361/202141404}\\
arXiv link:
\url{https://arxiv.org/abs/2108.02153}

% \section*{List of papers not included in the thesis}

% Thanks for all the fish!
    \chapter{Publications not included in this thesis}
\label{paper_not_in_thesis}
\begin{itemize}
    \item \textbf{Multichannel autocalibration for the Atmospheric Imaging Assembly using machine learning}\\
    L.~F.~G.~ Dos Santos, \textbf{S.~Bose}, V.~Salvatelli, B.~Neuberg, M.~C.~M.~Cheung et al.\\
    In: \emph{Astronomy \& Astrophysics}.
    Vol.\ 648,
    no.\ A53
    (2021)\\
    \doi{10.1051/0004-6361/202040051}.
    
    \item \textbf{Signatures of ubiquitous magnetic reconnection in the deep atmosphere of sunspot penumbrae}\\
    L.~H.~M.~Rouppe van der Voort, J.~Joshi, V.~M.~J.~Henriques\ and \textbf{S.~Bose}\\
% \enquote{\nameref{pap:first}}.
    In: \emph{Astronomy \& Astrophysics}.
    Vol.\ 648,
    no.\ A54
    (2021)\\
%pp.~123--456.
    \doi{10.1051/0004-6361/202040171}.
    
    \item \textbf{High-resolution observations of the solar photosphere, chromosphere, and transition region. A database of coordinated IRIS and SST observations}\\
    L.~H.~M.~Rouppe van der Voort, B.~De~Pontieu, M.~Carlsson, J.~de~la~Cruz~Rodriguez, \textbf{S.~Bose} et al.\\
% \enquote{\nameref{pap:first}}.
    In: \emph{Astronomy \& Astrophysics}.
    Vol.\ 641,
    no.\ A146
    (2020)\\
%pp.~123--456.
    \doi{10.1051/0004-6361/202038732}.
    
    \item \textbf{Semi-empirical model atmospheres for the chromosphere of the sunspot penumbra and umbral flashes}\\
    \textbf{S.~Bose}, V.~M.~J.~Henriques, L.~Rouppe~van~der~Voort\ and T.~M.~D.~Pereira \\
% \enquote{\nameref{pap:first}}.
    In: \emph{Astronomy \& Astrophysics}.
    Vol.\ 627,
    no.\ A46
    (2019)\\
    %pp.~123--456.
    \doi{10.1051/0004-6361/201935289}.
    
    \item \textbf{On the Variability of the Solar Mean Magnetic Field: Contributions from Various Magnetic Features on the Surface of the Sun}\\
    \textbf{S.~Bose}\ and K.~Nagaraju\\
% \enquote{\nameref{pap:first}}.
    In: \emph{The Astrophysical Journal}.
    Vol.\ 862,
    no.\ 35
    (2018)\\
%pp.~123--456.
    \doi{10.3847/1538-4357/aaccf1}.
    
    \item \textbf{Role of the background regimes towards the Solar Mean Magnetic Field (SMMF)}\\
    \textbf{S.~Bose}\ and K.~Nagaraju\\
% \enquote{\nameref{pap:first}}.
    In: \emph{Proceedings of the International Astronomical Union}.
    Vol.\ 340,
    no.\ 85
    (2018)\\
%pp.~123--456.
    \doi{10.1017/S1743921318001758}.

\end{itemize}
%\section*{Paper IV}

    \cleartorecto
    \microtypesetup{protrusion = false}
    \tableofcontents    % Or \tableofcontents*
    \cleartorecto
    \listoffigures      % Or \listoffigures*
    \cleartorecto
%    \listoftables       % Or \listoftables*
    \microtypesetup{protrusion = true}

    \mainmatter         % Folios in Arabic numerals, numbered chapters.
   
% \begin{refsegment}
    \chapter{Introduction}
\label{chap:intro}

% \begin{refsection}
% {
% \end{refsection}
The Sun is a hot spherical ball of plasma that lies at the center of our solar system. It is comprised of about 99.86\% mass of the solar system and is roughly 4.6~billion years old. Though it is a very common star of our galaxy, it is quintessential to us since it is the most important source of energy that helps sustain life on Earth. It is roughly 109 times bigger than the Earth (with a diameter of 1.39~million kilometers), and has a mass of $\approx$1.98~$\times$~10$^{30}$~kg that is roughly 330,000 times that of Earth. It is classified as a G-type main sequence star with an effective surface temperature of 5780~K. Being relatively closer to us ($\approx$1.49~$\times$~10$^{8}$~km) in comparison with the other stars of our galaxy, it is therefore possible to study this star and its atmosphere in an unprecedented detail both from the ground as well as from space. 

The Sun also has a profound impact on the space-weather around the Earth. The variation in the local space environment is driven largely by the radiation and energetic particles that emanate from the Sun during a period of enhanced solar activity (such as, flares and coronal mass ejections). Unfortunately, if these energetic particles are bound towards the Earth they may have a significant impact on our climate and communication systems. Some of their effects include electronic failures in satellites, radiation hazards to astronauts in space or on-board the International Space Station (ISS), navigation problems in airplanes, or loss of satellites due to atmospheric drag. Atmospheric drag can be referred to as the friction between the satellites (or space stations) and their immediate surroundings that can increase due to sufficient heating and expansion of the Earth's atmosphere due to an increase in the number of energetic particles released during flares or coronal mass ejections. In addition, the electric grids supplying electricity  to our homes can also be impacted by the so-called solar magnetic storms which carry vast amounts of magnetized plasma into space \citep[such as the March 1989 solar storm in Qu\'ebec. See,][]{Boteler_2019}. All the above potential impacts highlight the importance of studying and continuously monitoring the Sun, its atmosphere and the energy transfer from the solar interior to its outer atmospheric layers. The sections that follow henceforth attempt to provide a brief introduction to the different layers of the solar atmosphere starting from its interior to the corona, with an intention to illustrate the coupling between the different layers. 

\section{The solar structure}
\label{sec:solar_atmosphere}
As a main sequence star, the Sun converts hydrogen \citep[that accounts for nearly 74.6\% of its mass][]{2003ApJ...591.1220L} to helium in its core via the process of nuclear fusion. This conversion process also generates a tremendous amount of energy which is continuously leaked into its immediate surrounding layers, namely, the radiative zone and the convection zone. The names of these zones are given based on the most dominant process by which energy is transported across them. Photons are absorbed and emitted many times, via the process of radiative diffusion, as they propagate through the radiative zone that lies between 0.25~\radsun\ and 0.7~\radsun\ (\radsun\ being the radius of the Sun). The continuous absorption and (re)emission of photons causes them to have a mean free path of about a few millimeters which in turn leads to enormously long traveling times across this region. Estimates based on random walk analysis indicate that it takes roughly 170,000 years for photons to diffuse through the radiative zone \citep{1992ApJ...401..759M}. Now, once these photons reach the convection zone, the temperature starts to drop rapidly which increases the opacity due to formation of neutral H and He (most of the H and He exist in the fully ionized state in the radiative zone). At about 0.7~\radsun\ the temperature gradient becomes so large that the plasma cannot remain in stable equilibrium thereby triggering convective instabilities that mark the onset of the convection zone \citep{2014masu.book.....P}. The opacity becomes so high in this region that most of the energy is transported by moving blobs of plasma via the process of convection. 

\subsection{Photosphere}
\label{subsection:photosphere_description}
\begin{figure*}[!h]
    \centering
    \includegraphics[width=\textwidth]{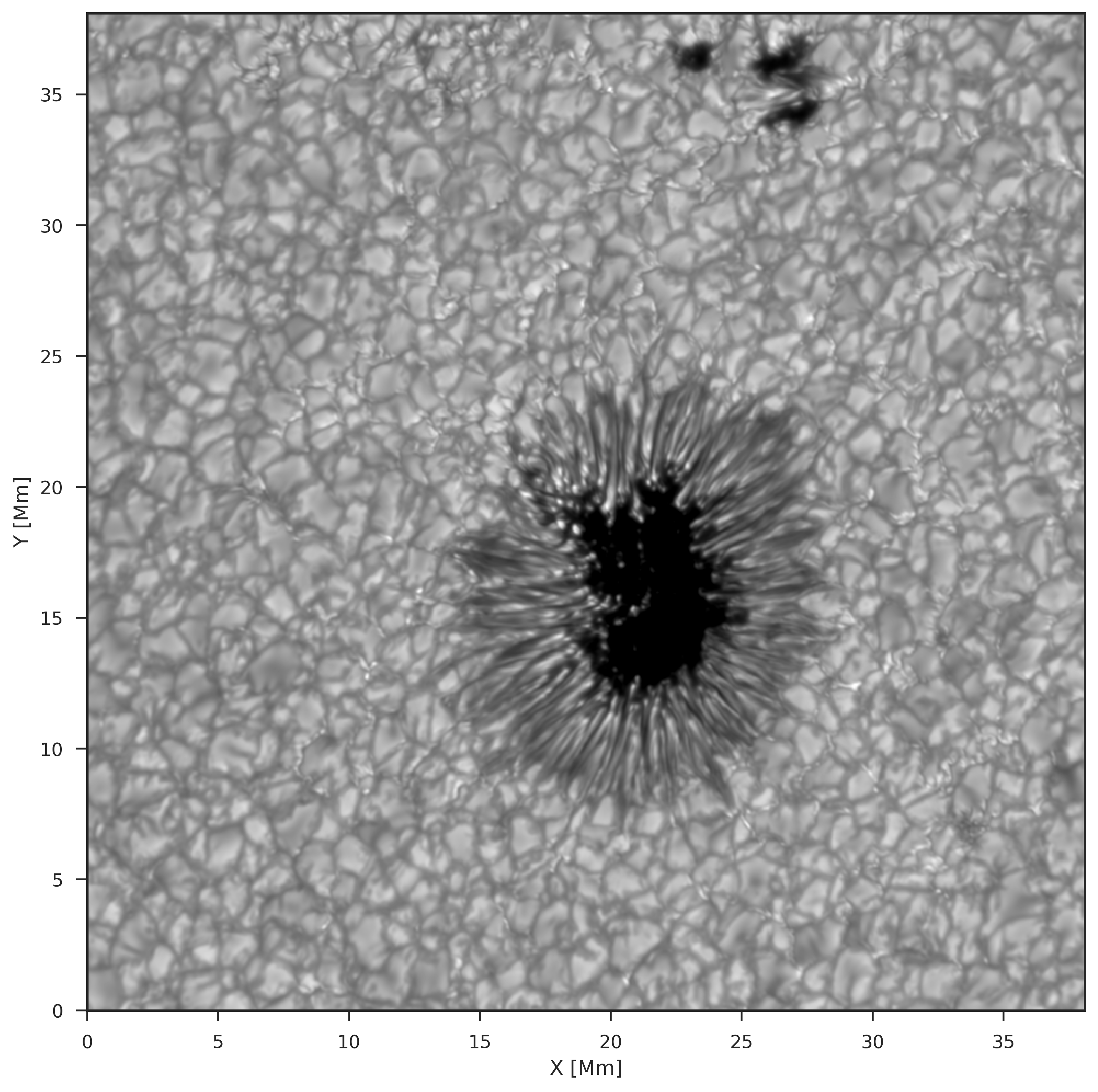}
    \caption{High resolution wide band image observed with the Swedish 1-m Solar Telescope on the island of La Palma, Spain showing a distinct photospheric scene with a sunspot, granulation pattern and several pores (dark structure at the top), on 7 August, 2020. The solar north is upwards. Courtesy of Luc Rouppe van der Voort. }
    \label{figure_ch-1:sunspot_context}
\end{figure*}

\begin{figure}[!h]
    \centering
    \includegraphics[width=\textwidth]{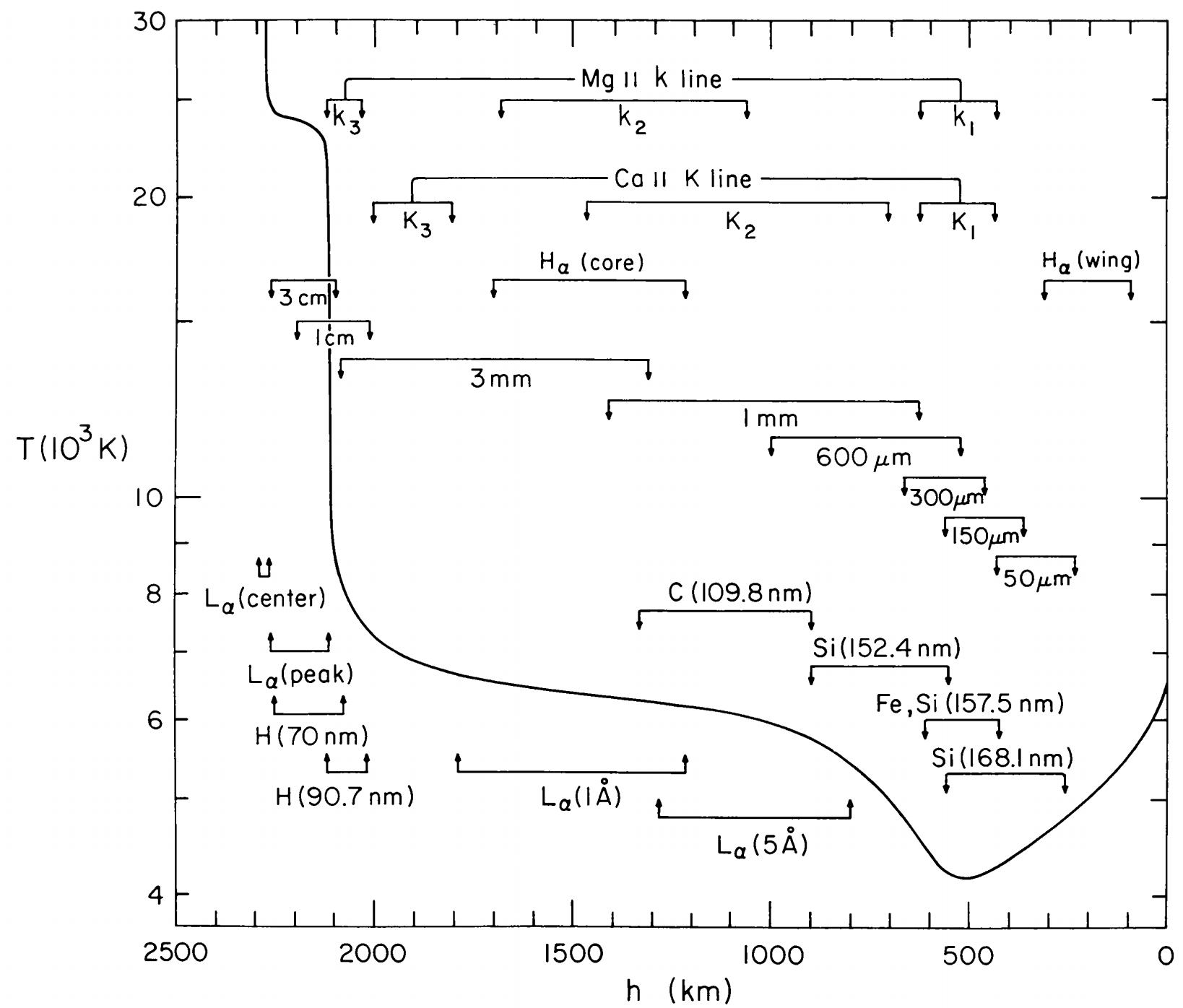}
    \caption{Stratification of temperature as a function of height in the solar atmosphere from the well-known VAL semi-empirical model. The region up to a height (h) of about 500~km forms the photosphere. The chromosphere is marked by an increase in the temperature beyond 500~km up to about 2000~km. The rapid increase in temperature beyond 2000~km represents the so-called transition region which separates the chromosphere from the million degree corona. The different atomic species indicate the spectral lines along with their ionization stages and average formation height. Image credit: \cite{1981ApJS...45..635V}. \textcopyright\ AAS. Reproduced with permission. }
    \label{figure_ch-1:VAL_model}
\end{figure}

The hot plasma which rises from the convective zone reaches a state where the opacity is insufficient to prevent any escape of the radiation. Eventually, the plasma expands, undergoes radiative cooling and returns to convective stability again \citep{2014masu.book.....P}. This generally happens in the layer which is termed as the photosphere $-$ i.e. the layer visible to the naked eye. The photosphere has a distinct appearance that shows the presence of a large number of granular cells. These are basically the tops of hot plasma blobs that arise from the convection zone and cover most of the visible area of the solar disk. In addition to the granular cells, strong magnetic flux concentrations, such as sunspots, pores, and plages are also seen in the photosphere which can alter the granulation pattern (as shown in \cref{figure_ch-1:sunspot_context}). The photosphere is the densest layer of the solar atmosphere which forms an interface between the solar interior and its outer atmospheres. According to the temperature stratification of the VAL semi-empirical model \citep{1981ApJS...45..635V} shown in \cref{figure_ch-1:VAL_model}, it is only a few hundred kilometers thick and continues until the temperature reaches a minimum at roughly 525~km above the surface, which is defined as the height where the optical depth ($\tau$) is unity at 500~nm.

\subsection{Chromosphere}
\label{subsection:chromosphere_description}

\begin{figure}[!h]
    \centering
    \includegraphics[width=\textwidth]{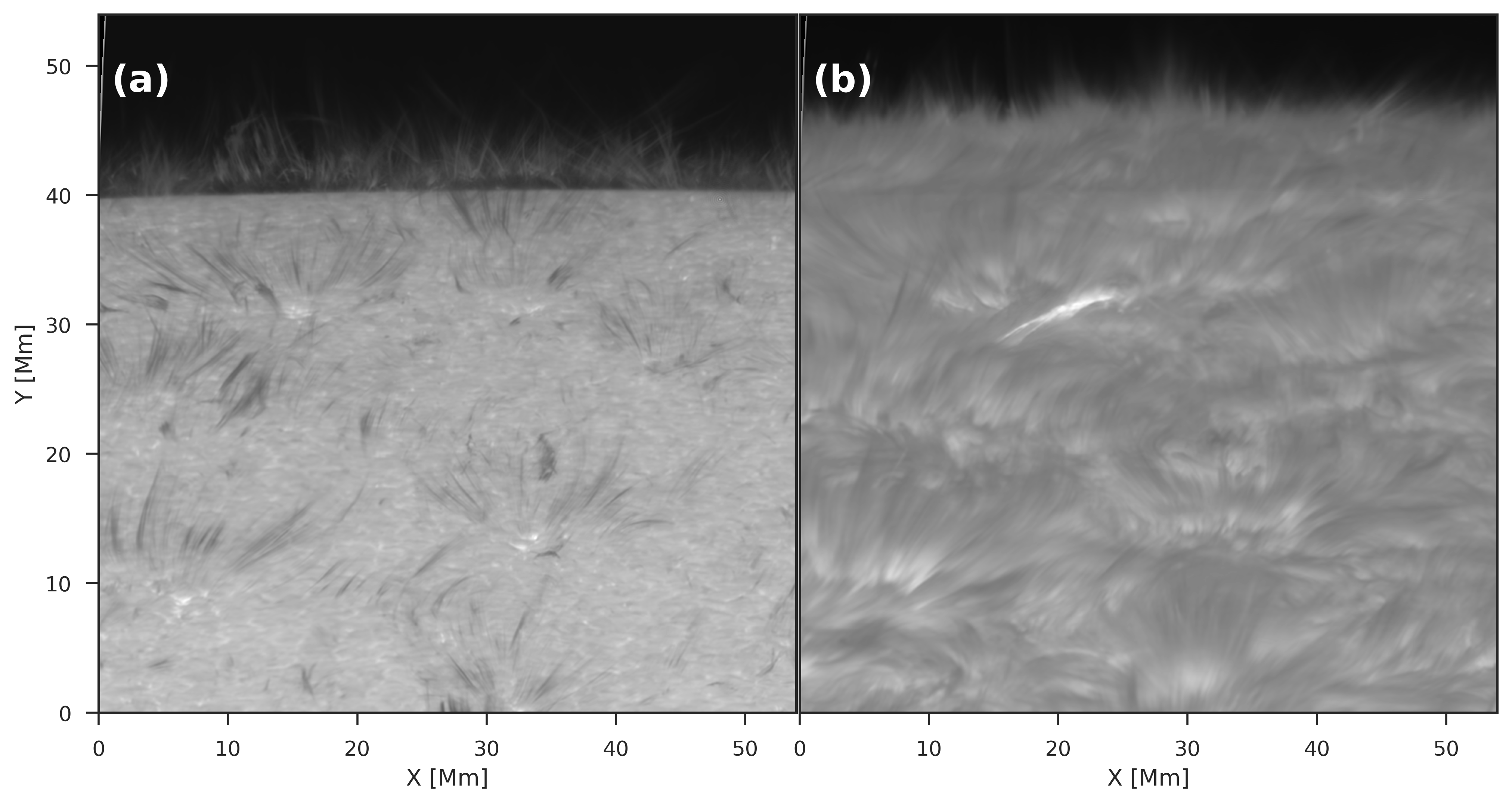}
    \caption{High resolution images of the north solar limb observed from the Swedish 1-m Solar Telescope on 20 August 2020, using the narrow band CRisp Imaging SpectroPolarimeter \halpha{} filter. Panel~(a) shows an image in the blue wing of the \halpha{} spectral line corresponding to a Doppler position of 45~\kms{} ($+$0.98~\AA\ from the line core) and panel~(b) shows the line core of \halpha{}. The field-of-view covers roughly 55$\times$55~Mm$^{2}$. Data reduced by Luc Rouppe van der Voort. }
    \label{figure_ch-1:Chromosphere_context}
\end{figure}

The chromosphere is the layer of the solar atmosphere that sits immediately above the photosphere. According to the semi-empirical VAL model shown in \cref{figure_ch-1:VAL_model}, the onset of the chromosphere is marked by an increase in the temperature beyond the temperature minimum. The layer derives it name from the Greek word \textit{chroma} (color), which is mainly attributed to the reddish colored observations of this layer visible during solar eclipses \citep{1868RSPS...17..131L}. The color is mainly due to the radiation emitted by neutral hydrogen atoms at 656~nm and the corresponding absorption line, called the \halpha{} spectral line, is one of the strongest lines observed in the chromosphere. It is quite unique from the perspective of observations, where we see a myriad of different features such as dynamic fibrils, filaments, mottles, and spicules, to name a few, that are not visible in the photosphere. A representative example is shown in \cref{figure_ch-1:Chromosphere_context}, where panels~(a) and (b) show high-resolution off-limb images of the solar chromosphere observed in the blue wing and line core of the \halpha{} spectral line from the Swedish 1-m Solar Telescope \citep[SST,][]{2003SPIE.4853..341S}. Panel~(b) shows the characteristic rich fibrillar appearance associated with dense canopies that are visible all over the field-of-view (FOV). This is in stark contrast to panel~(a) that appears predominantly photospheric with some thin, slender, and thread-like structures that are visible all over the limb and scattered on the disk. The chromosphere is also known to play a crucial role in heating and mass loading of the million-degree solar corona, since most of the nonthermal energy that drives these mechanisms propagates through this layer. The VAL model suggests that this layer is roughly 1500~km thick, and is more or less homogeneous. However, based on abundant observations in the past \citep[see][]{2007ASPC..368...27R}, and from the high-resolution observations, such as the one in \cref{figure_ch-1:Chromosphere_context}, it is clear that the chromosphere is not only far from homogeneous, but it is also very dynamic. Therefore, as suggested by \cite{2019ARA&A..57..189C}, current numerical modeling attempts rely on physics-based definitions of the solar chromosphere rather than purely geometrical arguments.
%\todo[inline]{Images!}

\subsection{Corona}
\label{subsection:corona_description}
\begin{figure}[!p] %change to !hbp
    \centering
    \includegraphics[width=\textwidth]{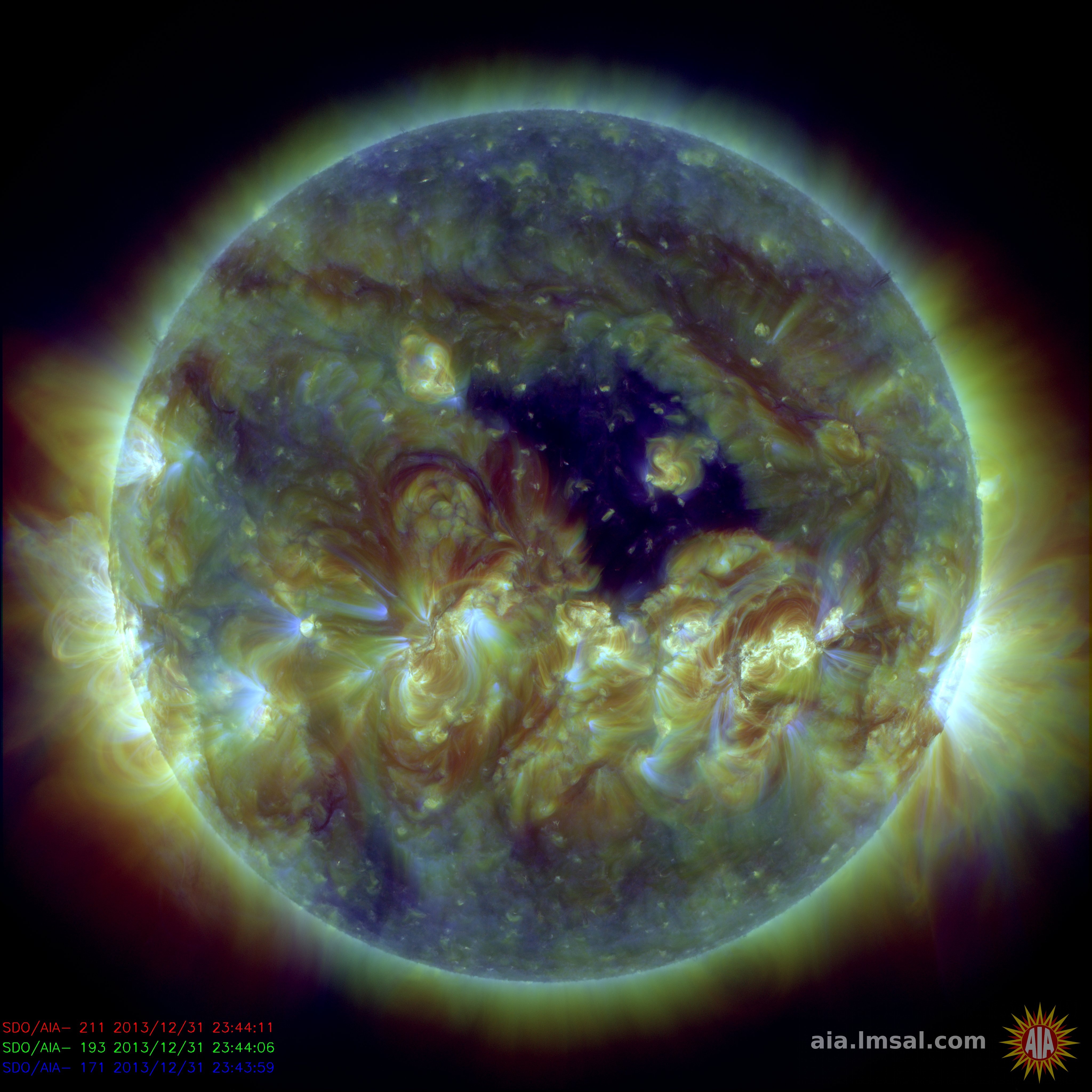}
    \caption{RGB composite image of the solar corona seen in three different passbands: 21.1~nm (formed at temperatures around 2 million Kelvin) is shown in red, followed by 19.3~nm (roughly 1.3 million Kelvin) in green and 17.1~nm (formed around 800,000~Kelvin) in blue, with NASA's SDO/AIA instrument. The brightest areas correspond to active regions which also show the presence of coronal loops that connects magnetic islands of opposite polarities. The dark region close to the center of the disc is a coronal hole which acts as a potential driver of solar wind. Image credit: Heliophysics Events Knowledgebase, Lockheed Martin Solar and Astrophysics Lab, Palo Alto.}
    \label{figure_ch-1:Corona_context}
\end{figure}
The solar corona is the outermost layer of the solar atmosphere that is visible to the naked eye only during a solar eclipse. The temperature of this layer is much hotter (several orders of magnitude) than the average temperature of the photosphere. In terms of its thickness, it can be said that it extends from the top of the chromosphere and out into the solar wind, which fills the entire heliosphere. Despite its enormous temperature, the radiation from the solar corona is roughly a million times less bright than that of the photosphere. This is attributed to the extremely low plasma density prevailing in this region. Many emission spectral lines observed in the corona are attributed to high ionization state of known elements, thereby implying that the temperature in this region can be at least a million Kelvin \citep{2014masu.book.....P}. The representative image in \cref{figure_ch-1:Corona_context} shows the solar corona in three different wavelength channels, two of which (i.e. the 19.3 and 21.1~nm) sample temperature regimes that are higher than 1~million K. Lighter nuclei of elements, such as hydrogen, are completely ionized whereas heavier elements, such as iron, can have an ionization state as high as $+$15. The amount of energy (and the temperature) required to ionize these elements indicate that the corona is only visible in the extreme ultraviolet wavelengths that can only be observed from space \citep{2014masu.book.....P}. The corona is also responsible for the generation of solar wind, which is a stream of charged particles consisting mainly of electrons, protons and $\alpha$ particles. The solar wind interacts with the inter-planetary atmospheres in the solar system and can impact the space-weather around the Earth and other planets. Since the density of the plasma is extremely low, magnetic fields dictate the environment of this layer. Various features like loops \citep{2002ESASP.505..191A}, coronal holes \citep{2009LRSP....6....3C}, and coronal X-ray jets \citep[][to name a few]{1992PASJ...44L.173S,2015Natur.523..437S}, can be observed in this region. We refer to \cref{figure_ch-1:Corona_context} for representative examples of coronal loops (mostly seen in the close vicinity of active regions) and holes.

\subsection{The Transition Region}
\label{subsection:TR_description}

\begin{figure}[!h] %change to !hbp
    \centering
    \includegraphics[width=\textwidth]{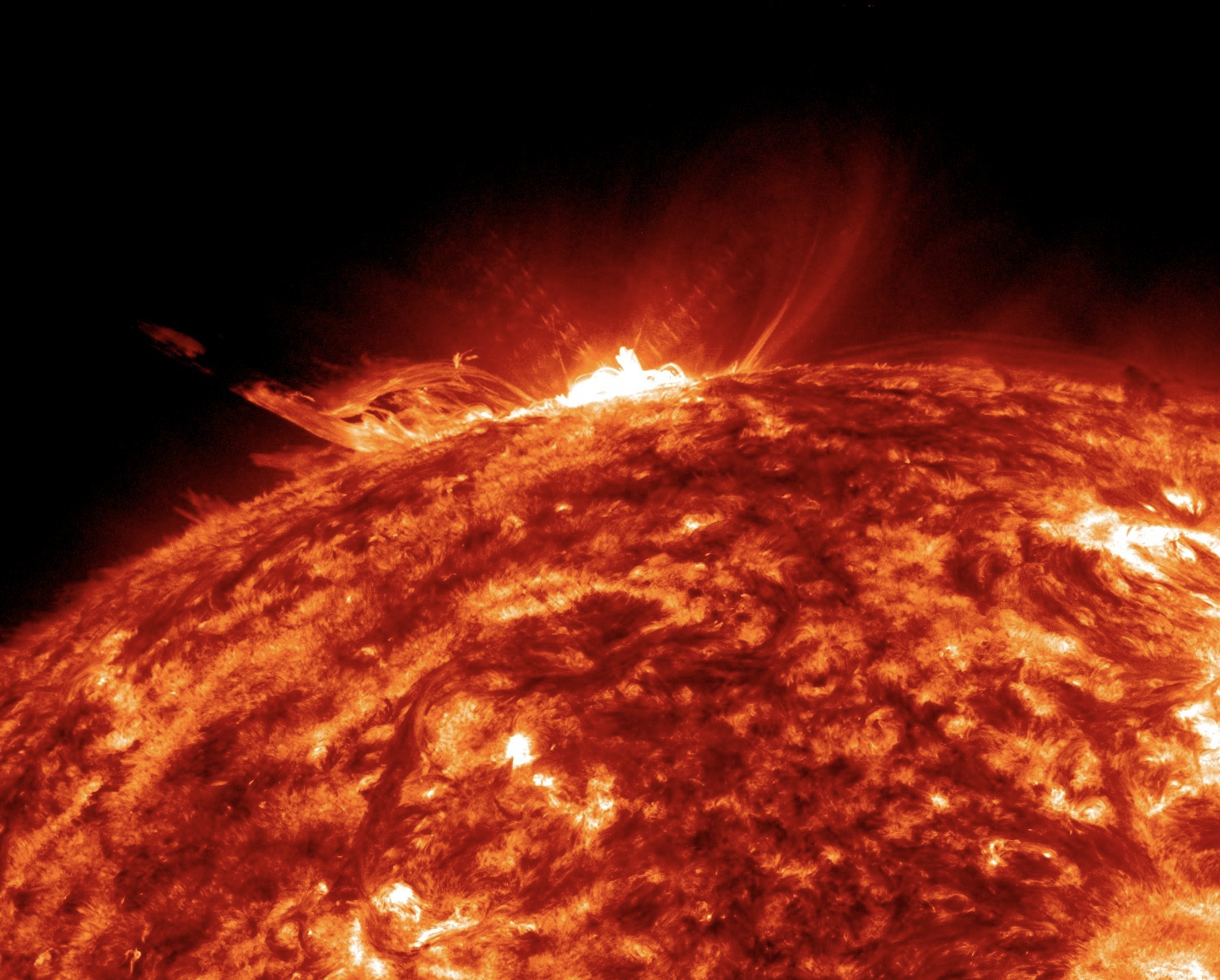}
    \caption{Transition region image of a flaring active region (observed on 10 April, 2013) close to the solar limb through NASA's SDO/AIA 30.4~nm channel. Image credit: Heliophysics Events Knowledgebase, Lockheed Martin Solar and Astrophysics Lab, Palo Alto.}
    \label{figure_ch-1:TR_context}
\end{figure}

The temperature stratification shown in \cref{figure_ch-1:VAL_model}, indicates a very rapid rise from about 8000~K to 30,000~K within a span of roughly 100~km. Here, the solar plasma undergoes a drastic "transition" from chromospheric to coronal temperatures. This (relatively) thin layer of the solar atmosphere, sandwiched between the chromosphere and the corona, is known as the transition region (TR). Due to high temperatures (albeit not as high as the solar corona) most of the hydrogen is ionized and, like the corona, it is dominated by emission spectral lines ions of heavier elements such as carbon, oxygen, silicon and magnesium with ionization states that can range between $+1$ and $+8$, indicating temperatures that can be as high as 100,000~K \citep{2004ApJ...617L..85P}. These ions emit radiation almost entirely in the ultraviolet part of the electromagnetic spectrum and therefore, like the corona, it is only possible to study them from space. With the increase of temperature and the decrease in the plasma density, the solar atmosphere changes rapidly from being optically thick in the upper chromosphere to optically thin in the TR. While an optically thin assumption is generally a good approximation in the relatively quieter areas of the TR (and also to a large extent in the solar corona), some studies such as \cite{2019ApJ...871...23K}, suggest that it may not always be the case especially during highly energetic and impulsive events such as flares. \cref{figure_ch-1:TR_context} shows a TR image of a flaring active region close to the limb observed through the Atmospheric Imaging Assembly's \citep[AIA,][]{2012SoPh..275...17L} 30.4~nm wavelength channel onboard the Solar Dynamics Observatory \citep[SDO,][]{2012SoPh..275....3P}. 
%A closer look at this image clearly indicates that the dense fibrillar appearance, which is commonly observed in the optically thick solar chromosphere, is missing here. Rather, the Lyman-$\alpha$ image shows the presence of a large number of loop-like structures (similar to coronal images) connecting the strong magnetic patches in and around the vicinity of the active region.
% \todo[inline]{Images!}

\section{The \textit{Interface} region and its importance}
\label{sec:interface_region}

 The \textit{interface} region is a term used by many solar physicists to describe the combination of the solar chromosphere and the TR \citep{2021SoPh..296...84D}. As discussed in Sects.~\ref{subsection:chromosphere_description} and \ref{subsection:TR_description}, it is through this region that all the non-thermal energy responsible for driving the solar corona (and also solar wind) is propagated. Characterizing and interpreting the observations in these regions can be extremely complex due to a multitude of reasons, such as, transitioning from a high to low plasma-$\beta$ (the ratio between the gas pressure and magnetic pressure) environment, partially ionized plasma in the chromosphere to fully ionized plasma in the corona, and the influence of non-local thermodynamic equilibrium (non-LTE) in radiation transport, to name a few \citep{2019ARA&A..57..189C}. As a result, a detailed study of the interface region can lead to a better understanding of the physical processes like magnetic reconnection, equilibrium and non-equilibrium ionization of atomic species, ion-neutral interactions and so on. 
 
 \begin{figure}[!h]
    \centering
    \begin{adjustbox}{minipage=\linewidth,scale=1.}
    \includegraphics[width=\textwidth]{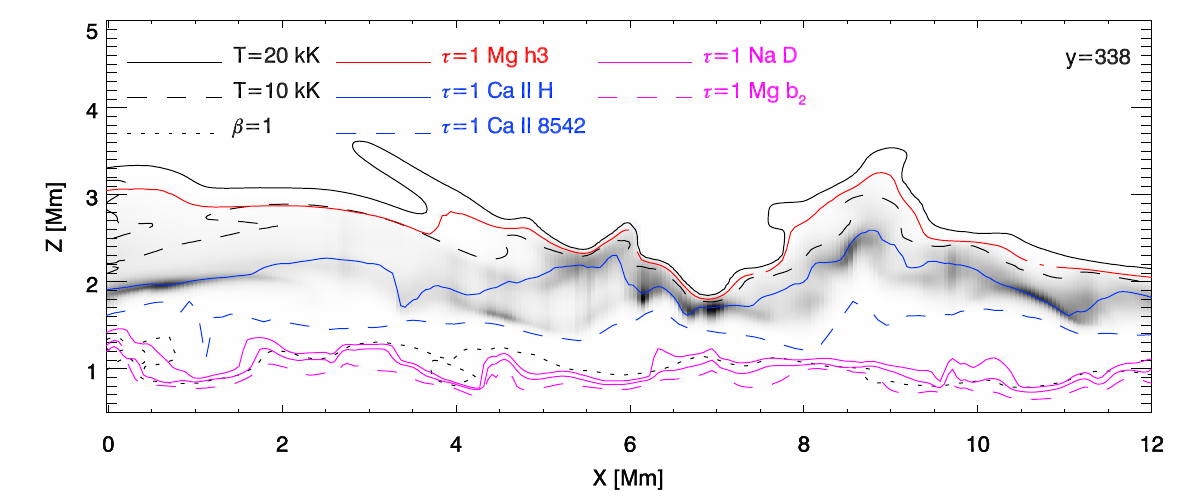}
    \end{adjustbox}
    \caption{Formation height of spectral lines corresponding to various atomic species in an $XZ$ slice of a stratified \textit{Bifrost} model atmosphere of an enhanced network computed by \cite{2016A&A...585A...4C}. The height where the optical depth is unity is used as a representative of the formation height of the optically thick lines, and for the optically thin \ion{He}{i}~1083~nm line, contribution function to the intensity is shown in gray scale. The simulation is publicly available at \url{http://sdc.uio.no/search/simulations}. Image credit: Mats Carlsson, Rosseland Center for Solar Physics, Oslo.  }
    \label{figure_ch-1:stratification_bifrost}
\end{figure}

 Classical semi-empirical modeling attempts give a clue about the temperature stratification in this region, indicating a rise in the temperature beyond the photosphere, but models such as the well-known VAL atmosphere, do not paint an entirely accurate picture. Although it does successfully reproduce a large number of spectral lines of different atomic species, like the ones indicated in \cref{figure_ch-1:VAL_model}, it provides a very simplistic scenario of the stratification of the solar atmosphere, indicating only the mean properties. In reality, the solar atmosphere (in particular the interface region) is highly time-dependent, structured and consists of non-homogeneous plasma that is mostly not in equilibrium. Depending on the associated local heating or cooling processes, the plasma density and temperature are continuously changing. \cref{figure_ch-1:stratification_bifrost} exemplifies the complex stratification in the solar atmosphere from the perspective of a more realistic numerical simulation. It can be seen that the temperature stratification is different in different regions in the horizontal domain, which in-turn affects the formation heights of the different spectral lines. This is in contrast to the VAL model which shows the formation of each spectral feature of interest at its own fixed height for the different spectral lines of interest. Nevertheless, the VAL model has been of quintessential importance in the past that serves as a reference atmosphere for exploring the chromopsheric line formation even today.
 
The rapid rise in temperature past the temperature minimum (for example as in \cref{figure_ch-1:VAL_model}) is rather unexpected and has been puzzling solar physicists for many decades. This is because the temperature is expected to continuously decrease with an increase in the distance from the solar core (source). However, as we see from numerous observations (mainly space-based), the reality is completely opposite. This increase in temperature with height is commonly referred to as the coronal (also chromospheric) heating problem that can be considered as the holy grail in solar physics \citep{2006SoPh..234...41K}. Despite abundant observational and numerical modeling efforts, the major cause of such heating remains a mystery till today. Classical consensus to explain such a controversial observation can broadly be divided into two categories. One school of thought is supportive of the idea that a large number of magnetic reconnection processes goes on continuously in the solar atmosphere at various spatial and temporal scales, that eventually leads to the heating of the outer atmospheres -- either via the production of a large number of nanoflares \citep{1988ApJ...330..474P,2002ApJ...576..533P} or braiding of the magnetic field lines via photospheric motions \citep{1972ApJ...174..499P,2013Natur.493..501C,2015ApJ...811..106H}. On the other hand, the alternative scenario is based on the argument that the heating is dominated by the damping/dissipation of magnetohydrodynamic (MHD) waves that permeate in the solar atmosphere \citep{1947MNRAS.107..211A,2020SSRv..216..140V}. Today it is known that that the solar atmosphere is highly structured, time varying and dynamic. Therefore, it is likely that different mechanisms act simultaneously to heat the upper atmospheres of the Sun \citep{2012RSPTA.370.3217P}.

\section{Coupling between the solar chromosphere and the transition region/corona}
\label{section:coupling}

\begin{figure}[!h]
    \centering
    \begin{adjustbox}{minipage=\linewidth,scale=1.0}
    \includegraphics[width=\textwidth]{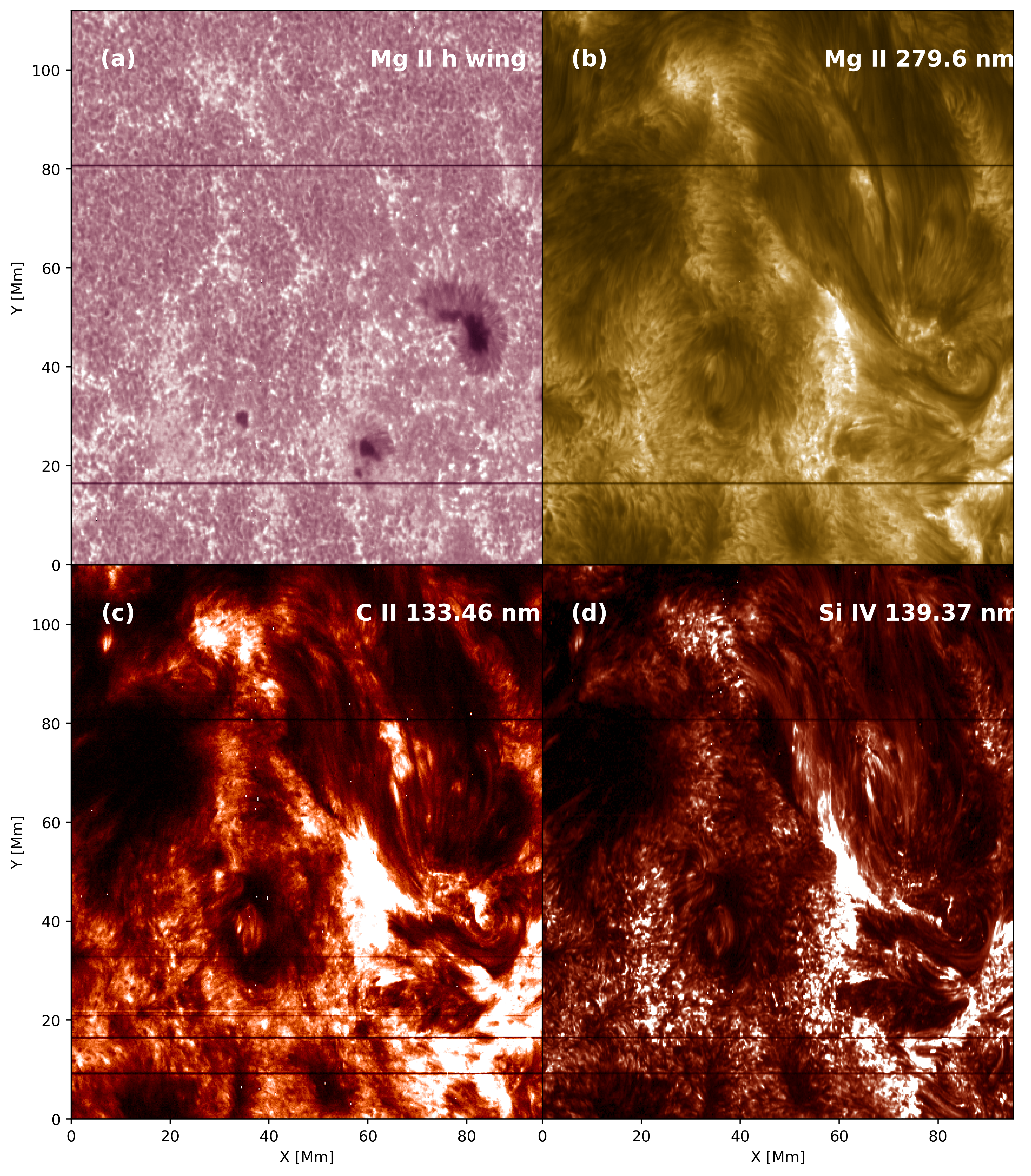}
    \end{adjustbox}
    \caption{IRIS spectroheliograms of active region (AR) 12056 observed on 15 May 2014 showing the coupling between the solar features observed in (a)~photosphere, (b)~upper chromosphere, (c)~lower TR, and (d)~middle to upper TR. The bright structures seen in panels~(c) and (d) correspond to the footpoints of AR loops. Large fibrils are seen above the AR, especially in panels~(b) and (c), with a distinct canopy connecting ($X$,$Y$)=(40,100)~Mm and (80,40)~Mm.  }
    \label{figure_ch-1:coupling_IRIS}
\end{figure}
The dynamic interaction between the chromosphere and the TR was not fully understood in older studies focusing on these regions owing to the lack of adequate observations. Although, it was known that large-scale events, for example jets, are largely multithermal and could be observed across multiple layers, the intricate connection between the cooler chromospheric and hotter TR/coronal channels among the relatively small-scaled features, such as spicules, microjets in sunspots, and low-lying loops, were unexplored. The scenario changed significantly after the advent of NASA's Interface Region Imaging Spectrograph \citep[IRIS,][]{Bart2014}. With the help of its high-resolution imaging and spectroscopic observations (see \cref{figure_ch-1:coupling_IRIS} for an example), the coupling between the cooler photosphere, chromosphere and the hotter layers above it became much more apparent \citep[see][for more detailed discussions of studies focusing on this connection]{2019ARA&A..57..189C,2021SoPh..296...84D}.

Coordination with ground-based instruments, in particular SST, has provided an unprecedented view of the small-scaled dynamic interactions taking place across the different layers of the solar atmosphere, ranging from the photosphere to the corona. These observations have indicated that any layer of the Sun's atmosphere, for example the TR or the solar corona, cannot be investigated independently without accounting for the observations in the upper chromosphere. As we will see in \cref{chap:spicules}, upward propagating chromospheric spicules (for example) have been linked with mass-loading and heating of the solar corona. A similar relationship also holds in the opposite direction where events in the corona, like nanoflares, can cause short-lived brightenings in the TR and upper chromosphere, via the release of non-thermal particles \citep[such as electron beams, see][]{2014Sci...346B.315T}. This thesis further exploits the coupling of the solar atmosphere by not only making use of the coordination between different space-based instruments, but also between the ground and space-based observations, leading to a comprehensive and a detailed coverage of the solar atmosphere. This is further detailed in \cref{chap:Instruments_simulations}.

\section{Outline of the thesis}
\label{section:Outline}

The description of the solar atmosphere presented in this chapter paints a highly dynamic and a structured scenario where the different layers of the atmosphere are highly coupled to one another. The advent of modern instruments with high spatial, spectral and temporal resolution, made it possible to study the evolution of various small-scaled dynamic structures that can potentially unlock many mysteries of the solar atmosphere. The aim of this thesis focuses on the study and analysis of spicules -- thin, jet-like, highly dynamic, small-scaled structures observed ubiquitously in the solar chromosphere, their characterization in different chromospheric and TR spectral lines, and their role in mass balance and heating of the solar atmosphere. The breakdown of the rest of the thesis is described below. 

\cref{chap:Instruments_simulations} describes the details on the numerical simulation of spicules (that has been used for the purpose of comparing with the observations), instruments from the different observatories, their data products, the processes and challenges of co-aligning the different datasets along with their importance. This is followed by \cref{chap:spicules} where an introduction to spicules and their associated mass flows, have been described extensively starting from a historical perspective. Also, the developments in terms of understanding spicules from the point of view of modern instrumentation and observations, their types, properties and possible origin mechanisms have been addressed as well. Furthermore, the studies undertaken as a part of this thesis and their motivation have also been described in the above chapter in the context of our current understanding of spicules and their associated mass flows. Finally, \cref{chap:paper_summary} provides a summary of the three papers that form the basis of this thesis along with concluding remarks and possibilities for future investigations.
    \chapter{Instruments and simulations}
\label{chap:Instruments_simulations}
%the intricately complex...nature of the sun and its atmospjere has beene possible onlu because  bla bla
The ever-increasing amount of information about the Sun and its atmosphere has been possible with the rapid developments in sophisticated instrumentation techniques, such as large aperture telescopes, that aim not only to record the highest possible resolution images of the Sun, but also to capture an unprecedented amount of light by acting as a "photon bucket". Depending on the local physical processes at play, it may be necessary to record observations in the ultraviolet (UV) or extreme-UV (EUV) part of the solar spectrum, in addition to optical frequencies. This is only possible from space since the Earth's atmosphere is mostly opaque at those wavelengths. Furthermore, the turbulence in the atmosphere around us plays a detrimental role in degrading solar observations recorded from ground-based telescopes. Observing directly from space can overcome this effect but the cost of operating space-based telescopes is significantly higher in comparison to ground-based telescopes. Moreover, there is a limit to the size of telescopes that can be placed off-Earth primarily due to their high cost and limited telemetry. Therefore, astronomers have turned to a method called adaptive optics that uses sophisticated deformable mirrors that can correct (lower-order) atmospheric distortion effects in almost real-time. This is generally followed by post-processing image restoration techniques such as speckle interferometry \citep{1992A&A...264L..24D} or Multi-Frame Blind Deconvolution \citep[MFBD,][]{1993JOSAA..10.1064S,2002SPIE.4792..146L} which minimizes the remaining higher-order seeing effects. The datasets that have been used in this thesis make use of both space and ground-based telescopes (including adaptive optics and post-facto image restoration techniques), which leads to both advantages and challenges that we address in this chapter.

The use of numerical simulations in solar physics has been as important as observations, especially in recent times. Observations have revealed that the solar atmosphere is far from static, and is hardly ever in equilibrium which makes interpretation challenging. Therefore, simulations come in to the picture which helps with the understanding of the physics behind the observations. However, depending on the geometry (1D, 2D or 3D) and also the physical processes involved, simulating the solar atmosphere (or at least a part of it) can be very computationally demanding or can even be beyond our current computational capabilities \citep{2020LRSP...17....3L}. Nevertheless, advanced simulations play a pivotal part in interpreting (and also sometimes predicting new) observations. The last two decades have witnessed the development of a plethora of different codes, such as \textit{Stagger} \citep{1998ApJ...499..914S}, MURaM \citep{2005A&A...429..335V}, \textit{Bifrost} \citep{2011A&A...531A.154G} and COBOLD \citep{2012JCoPh.231..919F} to name a few, that aim to model solar and/or stellar atmospheres in 3D geometry. In this thesis (i.e. in Paper~III), we have made use of an existing 2.5D MHD simulation of spicules by \citet{2017Sci...356.1269M}, computed with the \textit{Bifrost} code for the purpose of comparing with the observations. 

\section{The Swedish 1-m Solar Telescope}
\label{subsection:SST}

\begin{figure}[!htp] %change to !hbp
    \centering
    \includegraphics[width=\textwidth]{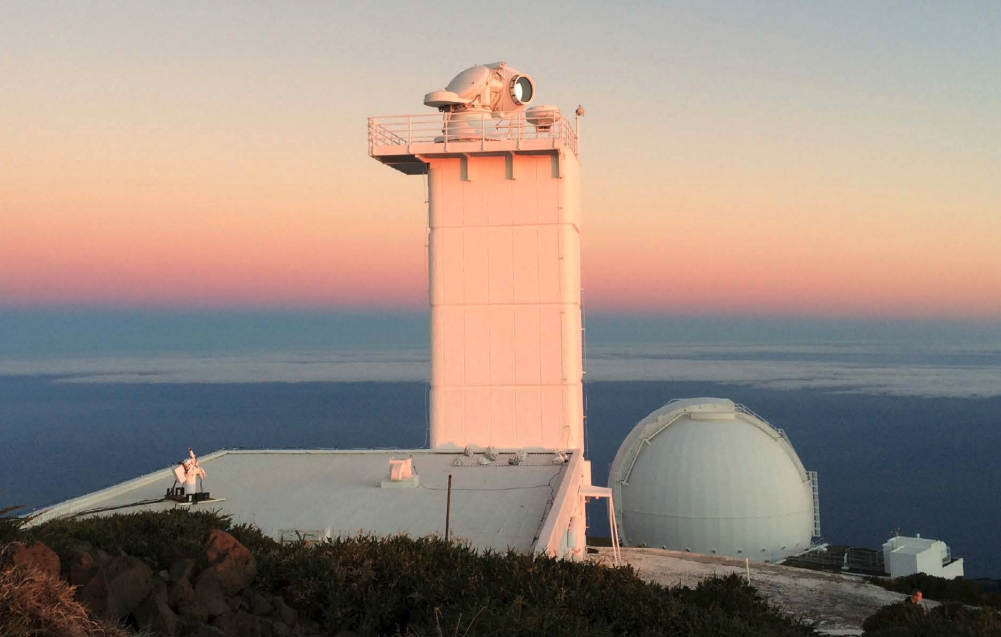}
    \caption{The Swedish 1-m Solar Telescope with the distinct white tower at the  the Observatorio Roque de los Muchachos on the island of La Palma. The primary imaging element, the 1~m lens which is one of the largest pieces of optical glass in the world, can be seen gazing at the rising sun at the top. Image credit: Rosseland Center for Solar Physics, Oslo.  }
    \label{figure_ch-2:SST_tower}
\end{figure}

The SST is a refracting telescope with an effective aperture diameter of 97~cm, and is used for high-resolution observations of the solar photosphere and chromosphere. It is located on the island of La Palma, which is a part of the Canary Islands, Spain. The telescope is shown in \cref{figure_ch-2:SST_tower} with the main objective lens situated on top of the tower. Besides being the primary imaging element, the main objective also serves as the entry window to the vacuum tower. This dual function of the 1-m lens also reduces the number of optical elements to a minimum inside the telescope, which would otherwise lead to a reduction in the photon efficiency and add to optical aberrations in the main science data. The tower contains a vacuum tube that prevents the generation of turbulence due to the high (nearly 700~W) heat from the beam by maintaining a constant atmospheric pressure below 5~mbar. This results in preventing the internal seeing inside the tower which would otherwise affect the image quality (in the same way as atmospheric seeing). The main objective lens is polished (and designed) in such a way that it is able to withstand the pressure difference on its two sides, which further limits the size of the lens that can be used for these applications. The telescope is optimized for multi-wavelength observations between 350--1000~nm. It consists of two main instruments based on dual-etalon Fabry-P{\'e}rot interferometers (FPIs) that are tuned to observe in the blueward (between 380--500~nm) and redward (beyond 500~nm) part of the visible spectrum. The SST makes use of an adaptive optics system with both tip-tilt and a deformable mirror (having 85 movable parts) that reduces the effect of aberrations in the wavefront due to atmospheric turbulence. In this thesis, we make use of high-resolution observations from both the FPIs, and we focus on the \cak{}, \ion{Fe}{I}~630.2~nm, \ion{Ca}{ii}~854.2~nm, and \halpha{} spectral lines that are described below.   

\subsection{CRIsp Imaging Spectropolarimeter}
\label{subsection:CRISP}

The CRIsp Imaging Spectropolarimeter \citep[CRISP,][]{Crisp_2008} is a dual-etalon FPI tuned to observe in the red to near-infrared part of the electromagnetic spectrum. The two etalons are used in tandem along with a pre-filter that limits the spectral coverage of the light that passes through the optical system by eliminating multiple transmission peaks \citep{1998A&A...340..569K}. Moreover, CRISP is also mounted with a pair of liquid crystal retarders in sequence which makes full Stokes spectropolarimetric observations possible by allowing four different combinations of polarization analysis. Both spectropolarimetric and spectroscopic images recorded using CRISP have a pixel scale of 0\farcs058. We refer to \cite{2015A&A...573A..40D} and \cite{2018arXiv180403030L} for further details on the optical setup and the imaging systems.

\subsection{CHROMospheric Imaging Spectrometer}
\label{subsection:CHROMIS}
The CHROMOspheric Imaging Spectrometer \citep[CHROMIS,][]{2017psio.confE..85S} is also a dual-etalon FPI, similar to CRISP, but is used for observations in the wavelength range 380--500~nm and was installed at the SST in 2016. Unlike CRISP, CHROMIS does not have polarimetric capabilities at the moment, and it is particularly optimized for observations in the chromospheric \ion{Ca}{ii}~H and K and the H-$\beta$ spectral line (centered at 486.22~nm). The current set of pre-filter configuration of CHROMIS allows simultaneous chromospheric imaging observations in either the \ion{Ca}{ii}~H or K wavelength regions and photospheric observations with help of a wide band filter tuned around 395~nm, between the H and K lines (Bose et al. in prep). Another configuration allows observations in the chromospheric H-$\beta$ spectral line, with the wide band collected in the far blue wing centered at 484.55~nm. I refer to \cite{2018arXiv180403030L} for more details on the pre-filter arrangements and the passbands of the different filters. It is to be noted that with the advent of CHROMIS, it has been possible to observe the solar atmosphere with an unprecedented high-resolution that is better than 0\farcs1 (or \textasciitilde70~km) at wavelengths shortward of 400~nm, with a pixel scale of 0\farcs037.

In summary, both CRISP and CHROMIS are dual-etalon, FPI-based instruments capable of performing high spatial and spectral resolution observations ranging from the photosphere to the chromosphere. CRISP has spectropolarimetric (with the photospheric \ion{Fe}{i}~617.3 and 630.2~nm, and chromospheric \ion{Ca}{ii}~854.2~nm lines being most widely used for this purpose) and spectroscopic capabilities at wavelengths beyond 500~nm (including \halpha{}), while CHROMIS is primarily used for imaging spectroscopic observations shortward of 500~nm with \ion{Ca}{ii}~H and K lines being the most common choice in addition to H-$\beta$.

This thesis made use of the CRISP full Stokes imaging spectropolarimetric \ion{Fe}{i}~630.2~nm data from the 25th of May 2017 campaign and \halpha{} imaging spectroscopic data from both the 25th of May, 2017 (henceforth dataset~1) and 19th September 2020 (henceforth dataset~2) campaigns. Dataset~1 and 2 targeted an enhanced network with a cadence of 19.6~s, and quiet Sun region with a cadence of 23.6~s, respectively, close to the disk center. The \ion{Fe}{i}~630.2~nm data was sampled at 16 wavelength positions with a variable wavelength sampling ranging between 40--80~m\AA\ depending on the position with respect to the 630.2~nm line center. The \halpha{} spectral line was sampled at 32 wavelength positions for dataset~1, and 31 wavelength positions for dataset~2 with 100~m\AA\ wavelength steps. In addition, CRISP was also used to record spectroscopic \ion{Ca}{ii}~854.2~nm and spectropolarimetric \ion{Fe}{i}~617.3~nm data for dataset~2. 

\begin{figure}[!htp]
    \centering
    \includegraphics[width=\textwidth]{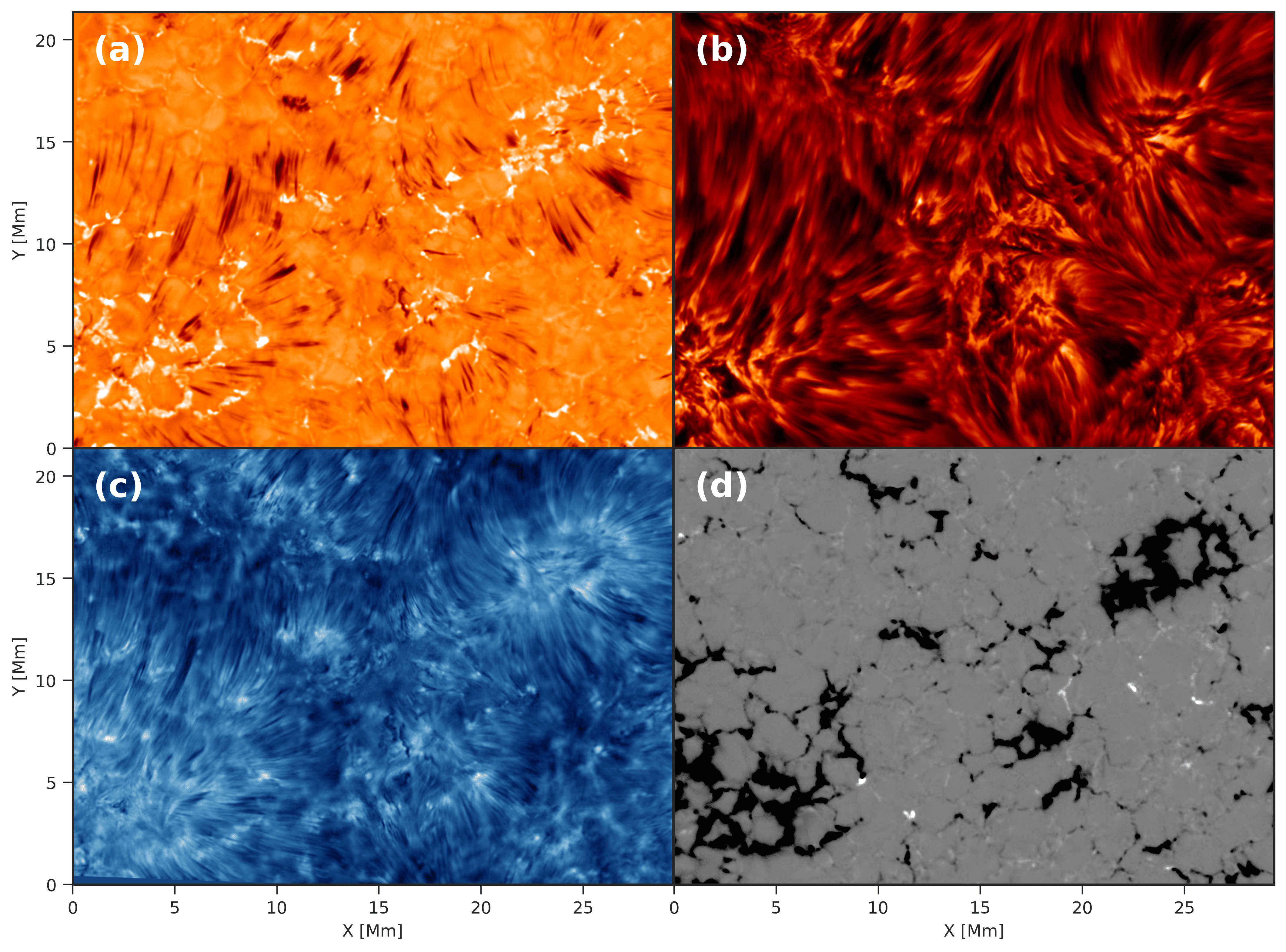}
    \caption{Overview of dataset~1 targeting an enhanced network region observed with the SST on 25 May 2017. Panels~a and b, show images observed in the blue wing (at a Doppler offset of $-40$~\kms{} with respect to the line center) and the line core of the \halpha{} spectral line. Panel~c shows the image observed at the inner blue-wing of the \cak{}~393.4~nm. Panel~d shows the photospheric line-of-sight magnetic field saturated between $\pm$~500~G derived from the \ion{Fe}{i}~630.2~nm line using a Milne-Eddington inversion scheme based on the implementation by \cite{2019A&A...631A.153D}. This dataset has been described and analyzed extensively by the author in several publications, such as Paper~I, Paper~II, and Paper~III, that form a part of this thesis. }
    \label{figure_ch-2:data_overview}
\end{figure}

In addition to CRISP, I also made use of the CHROMIS \cak{} (centered around 393.37~nm) imaging spectroscopic data corresponding to the enhanced network target of dataset~1. CHROMIS sampled the \cak{} spectral line at 41 wavelength positions with 63.5~m\AA\ steps within $\pm$1.28~\AA\ of the line-center. Moreover, images from a continuum wavelength position at 400~nm, and CHROMIS 395~nm wide band filter (that lies between the H and K lines) were also sampled. The cadence of this dataset was 13.6~s. Dataset~2 also has CHROMIS co-observations but they were not used as a part of this thesis. An overview of the observations from dataset~1 is shown in \cref{figure_ch-2:data_overview}.

%The two datasets described here have been used in Paper~\ref{pap:Paper1}, Paper~\ref{pap:Paper2}, and Paper~\ref{pap:Paper3} of this thesis. 

Despite excellent atmospheric conditions, corrections for atmospheric seeing effects were necessary to achieve high quality science data. As mentioned earlier, lower order seeing effects were removed by means of a tip-tilt and a deformable mirror, that forms a part of the upgraded version of the SST adaptive optics system \citep[described in][]{2019A&A...626A..55S}. The Multi-Object Multi-Frame Blind Deconvolution method \citep[MOMFBD,][]{vannoort2005MOMFBD} was further used as a post-facto image reconstruction technique to minimize the residual higher-order seeing effects, which the adaptive optics system alone is generally unable to correct for. Moreover, adaptive optics is also unable to fully compensate for the spatial variation of seeing (blurring) all over the FOV, that is attributed to the fact that different lines of sight are affected differently as they pass through the Earth's atmosphere (also known as anisoplanatism). Advanced image restoration techniques, like MOMFBD, have the capability efficiently to correct for this anisoplanatism as described in \cite{vannoort2005MOMFBD} \citep[also see,][]{2012A&A...548A.114H}. MOMFBD is an extension of the MFBD technique that includes simultaneous reduction of Multiple-Object (MO) data recorded using filters that are sufficiently close in wavelength. The use of MO has facilitated improved image restorations, by providing better constrains in estimating the seeing point spread function, in comparison to single-object restorations. For CRISP and CHROMIS, the wide-band and the sequentially recorded narrow-band channels serve as MOs necessary for the MOMFBD process, ultimately leading to observations at the highest possible spatial resolution down to the diffraction limit of the telescope (given by 1.22$\lambda$/$D$, where $\lambda$ is the wavelength and $D$ represents the effective diameter of the telescope) which are 0\farcs1 for \cak{} and 0\farcs16 for \halpha{}.

The SSTRED data reduction pipeline \citep{2015A&A...573A..40D,2018arXiv180403030L} was used for further reduction of the data post MOMFBD. The observations from CRISP were then internally co-aligned and destrechted to remove the warping due to seeing effects across the FOV. The internal co-alignment was done by cross-correlating the wide band images of \ion{Fe}{i}~630.2~nm with the corresponding \halpha{} wide band images (with the latter serving as reference) for dataset~1, since both wide band images predominantly show photospheric features. Furthermore, the alignment between CRISP and CHROMIS was done by expanding CRISP to CHROMIS pixel scale, followed by cross-correlating the corresponding wide band \halpha{} images with CHROMIS 395~nm wide band images. Finally, the CRISP FOV was adequately cropped so as to limit it to the CHROMIS FOV. \cref{figure_ch-2:data_overview} shows CRISP and CHROMIS images from dataset~1 that have been co-aligned using the method described in this section. For dataset~2, I did not make use of the CHROMIS data but the internal CRISP alignment was done in exactly the same way as dataset~1, but with \ion{Ca}{ii}~854.2~nm wide band serving as the reference. 

\section{Interface Region Imaging Spectrograph}
\label{section:IRIS}

\begin{figure}[!htp]
    \centering
    \includegraphics[width=\textwidth]{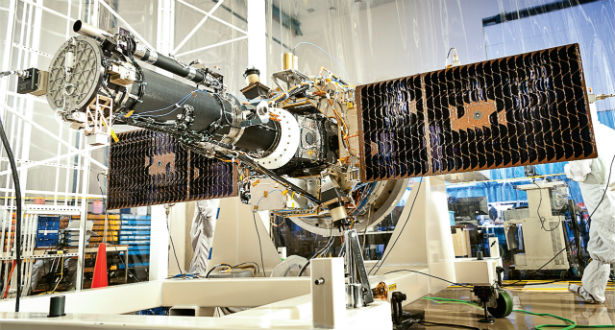}
    \caption{The Interface Region Imaging Satellite being prepared for launch in the clean room facility of NASA's Ames Research Center. Image credit: Jim Dowdall, Lockheed Martin.}
    \label{figure_ch-2:iris_satellite}
\end{figure}

Launched by NASA in June 2013, IRIS (see \cref{figure_ch-2:iris_satellite}) is a small explorer space mission studying the interface region (refer to \cref{sec:interface_region}), that has led to significant advancements in our understanding of the solar chromosphere and the TR \citep{2021SoPh..296...84D}. IRIS consist of a 20~cm telescope that feeds near UV (NUV) and far-UV (FUV) light into a slit-based high-resolution spectrograph and it has opened up a possibility of obtaining an unprecedented spatial (0\farcs33 in the FUV, and 0\farcs4 in the NUV), and spectral (2.5~\kms{} pixel$^{-1}$) resolution, with a cadence down to $\approx$1~s. It obtains spectra in three major passbands between 133.2--135.8~nm (FUV1), 138.9--140.7~nm (FUV2) and 278.3--283.4~nm (NUV), with simultaneous spectral observations from the photosphere to the transition region. Spectral lines that are most widely used by the community (and also this thesis) are the chromospheric \ion{Mg}{ii}~h (280.3~nm) and k (279.6~nm) lines \citep{2013ApJ...772...89L,2013ApJ...772...90L}, the upper chromospheric/transition region \ion{C}{ii}~133.4/133.5~nm lines \citep{2015ApJ...811...81R} and the TR \ion{Si}{iv}~139.4/140.3~nm lines. The spectrograph slit is roughly 0\farcs33 wide and has a length of 175\arcs. It can be displaced in the direction parallel to the plane of the sky (perpendicular to the direction of dispersion) to build up a raster that can have an area of up to 130\arcs~$\times$~175\arcs (roughly 94~Mm~$\times$~123~Mm) in the solar-$X$ and $Y$ directions. 
In addition, IRIS can also record slit-jaw images (SJIs) in four different wavelength bands i.e. \ion{C}{ii}~133, \ion{Si}{iv}~140, \ion{Mg}{ii}~279.6 (dominated by the core and inner wings of the \Mgk{} line) and \ion{Mg}{ii}~283~nm (that serves to provide a valid context of the region observed on the solar surface). The SJIs also serve as a means to co-align IRIS with other ground or space-based observations. A database of coordinated and co-aligned IRIS and SST datasets has been recently made available by \cite{2020A&A...641A.146R}.

This thesis made use of the coordination between IRIS and SST for both datasets~1 and 2. Each IRIS observation run has a unique id associated with it which is known as the OBS-ID. An OBS-ID is a unique 10-digit number that essentially encodes the details of the observation including the number of raster steps, observed spectral lines, cadence, exposure time etc. On the 25th of May, 2017 (dataset~1) IRIS was selected to run in a so-called medium dense 8-step raster mode (with an OBS-ID \verb|3633105426|) with an exposure time of 2~s and the spectral raster covering a FOV of 2\farcs8~$\times$~62\arcs centered around the enhanced network target. For dataset~2, observed on 19 September 2020, IRIS ran in an large dense 4-step raster mode (OBS-ID \verb|3633109417|) with an 8~s exposure time covering a FOV of 1\farcs36~$\times$~120\arcs co-observing the quiet Sun target with SST. 

\subsection{Co-aligning IRIS and SST}
\label{subsection:co_align_IRIS_SST}
The co-alignment between the IRIS and the SST datasets were made by employing a cross-correlation technique between the pair of images that were morphologically most similar with respect to one another. For dataset~1, this meant IRIS 279.6~nm \ion{Mg}{ii}~SJI and CHROMIS \cak{}. I first began by expanding the SJIs up to the CHROMIS pixel scale by employing the standard cubic interpolation method of \cite{PARK1983258}. Then the SJIs (that were closest in time to CHROMIS observations) were cross-correlated with the \cak{} images obtained by computing the spectral average between $\pm$20~\kms{}. Before the cross-correlation, we employ a Fourier filtering technique using a 1$^{\mathrm{st}}$ order Butterworth filter that helped to remove the high-frequency (noisy) pixels from the expanded IRIS SJIs. This step proved to be crucial in achieving a better correlation between SST (which was rotated to account for the orientation of the solar north prior to cross-correlation) and IRIS, thereby leading to a co-aligned sequence with a cadence of 13.6~s. Co-aligning the IRIS dataset by expanding its dimensions to CHROMIS pixels is a relatively uncommon choice, which was performed in order to maximize the latter's novel science potential at the highest possible resolution. 

Dataset~2 focused on a quiet-Sun region, and from the perspective of CRISP, \ion{Ca}{ii}~854.2~nm data bore the closest morphological resemblance to IRIS~279.6~nm SJI. As a result, the two sets of data were co-aligned by means of cross-correlation in a way similar to dataset~1. However, co-aligning a quiet region on the Sun with minimal (visual) activity across the different wavelength channels proved to be far more challenging than dataset~1. This was circumvented by integrating the \ion{Ca}{ii}~854.2~nm data over the entire wavelength dimension (between $\pm$28~\kms{}) for each time step, which was followed up by cross-correlating the resultant image with an expanded (to CRISP pixel scale) 279.6~nm SJI that was closest in time to the CRISP counterpart. Once again, the SJI was filtered with the same Butterworth filter (as in dataset~1) to get rid of the high-frequency noise pixels before performing the cross-correlation. Since CRISP \ion{Ca}{ii}~854.2 data served as the reference, the cadence of the resulting co-aligned sequence was 23.6~s. Furthermore, both the co-aligned datasets were visualized with CRISPEX \footnote{\url{https://github.com/grviss/crispex}\label{fnote:crispex}} \citep{2012ApJ...750...22V}, an Interactive Data Language (IDL) widget-based tool. Panels~(d) and (e) of \cref{figure_ch-2:co_aligned_all} show the \ion{Mg}{ii}~279.6~nm and \ion{Si}{iv}~140.0~nm SJI co-aligned to the corresponding SST observations for dataset~2 shown in panels~(a) and (b). These, along with the co-aligned SST and IRIS observations for dataset~1, have been used in Paper~I and Paper~III of this thesis. 

%Co-aligning quiet Sun datasets can be challenging because, unlike a plage or an enhanced network region that has a distinct emission pattern, quiet Sun regions often do not have such distinct signatures across multiple wavelength bands. Therefore, for co-aligning the SST and IRIS sequences for dataset~2, we first integrated the \ion{Ca}{ii}~854.2~nm data over the entire wavelength dimension for each time step and the resultant image was cross-correlated with the corresponding 279.6~nm SJI in a nearest neighbor fashion. Panels~(d) and (e) of \cref{figure_ch-2:co_aligned_all} show the magnesium and silicon SJI co-aligned to the corresponding SST observations for dataset~2. These, along with the co-aligned SST and IRIS observations for dataset~1, have been used in \cref{pap:Paper1} and \cref{pap:Paper3} of this thesis. 

\section{Solar Dynamics Observatory}
\label{section:SDO}

The Solar Dynamics Observatory (SDO) has been one of the flagship missions of NASA's Heliophysics Systems Observatory \citep{2016AGUFMSH21F..06C}. Launched in 2010 under the Living With a Star (LWS) program, the SDO observes the Sun in a quasi-simultaneous fashion in the visible and EUV wavelengths and has been instrumental in monitoring the activity of the Sun for over a decade now. With its high cadence (\textasciitilde12~s) full disk observations, one of the major objectives of SDO is to understand the dynamic physical processes taking place in the chromosphere and the corona and relate them with the manifestations of the magnetic field in the photosphere. It consists of three different instruments:
\begin{itemize}
  \item The Atmospheric Imaging Assembly \citep[AIA,][]{2012SoPh..275...17L}
  \item The Helioseismic and Magnetic Imager \citep[HMI,][]{2012SoPh..275..229S}
  \item The Extreme ultraviolet Variability Experiment \citep[EVE,][]{2012SoPh..275..115W}
\end{itemize}
In this thesis, we primarily use observations from AIA for scientific analysis and HMI continuum images for the purpose of co-alignment with the SST images using the method described below. The co-aligned datasets are used in the analysis presented in Paper~III.

\subsection{AIA}
\label{subsection:AIA}
AIA records full disk images of the Sun with a spatial resolution of \textasciitilde1\farcs0 and a temporal resolution of \textasciitilde24~s in two UV and \textasciitilde12~s in seven EUV wavelengths with a pixel scale of 0\farcs6. The nine (E)UV wavelength channels are observed with four telescopes that aim to provide a complete thermal coverage of the solar atmosphere ranging from 6~$\times$~10$^{4}$~K to roughly 20~$\times$~10$^{7}$~K \citep{2012SoPh..275...41B}. 

\subsection{HMI}
\label{subsection:HMI}
HMI is used for the measurement of photospheric velocity and magnetic field maps via spectropolarimetric measurements of the \ion{Fe}{i}~617.3~nm line. In addition to the full disk line-of-sight magnetic field and Doppler maps that is available at a cadence of 45~s, HMI also provides full Stokes vector magnetograms with a cadence of 720~s and a spatial resolution of 1\farcs11. The pixel scale is 0\farcs5.

%In this thesis, we used SDO/AIA and HMI data co-aligned to SST datasets for the analysis presented in \cref{pap:Paper3}.

\subsection{Co-alignment between SDO/AIA and SST datasets}
\label{subsection:Co-alignment_SDO_SST}

\begin{figure*}[!hbp]
\centering
\begin{adjustbox}{minipage=0.85\linewidth,scale=1.2}
 \begin{subfigure}[b]{0.5\textwidth}
    \includegraphics[width=\textwidth]{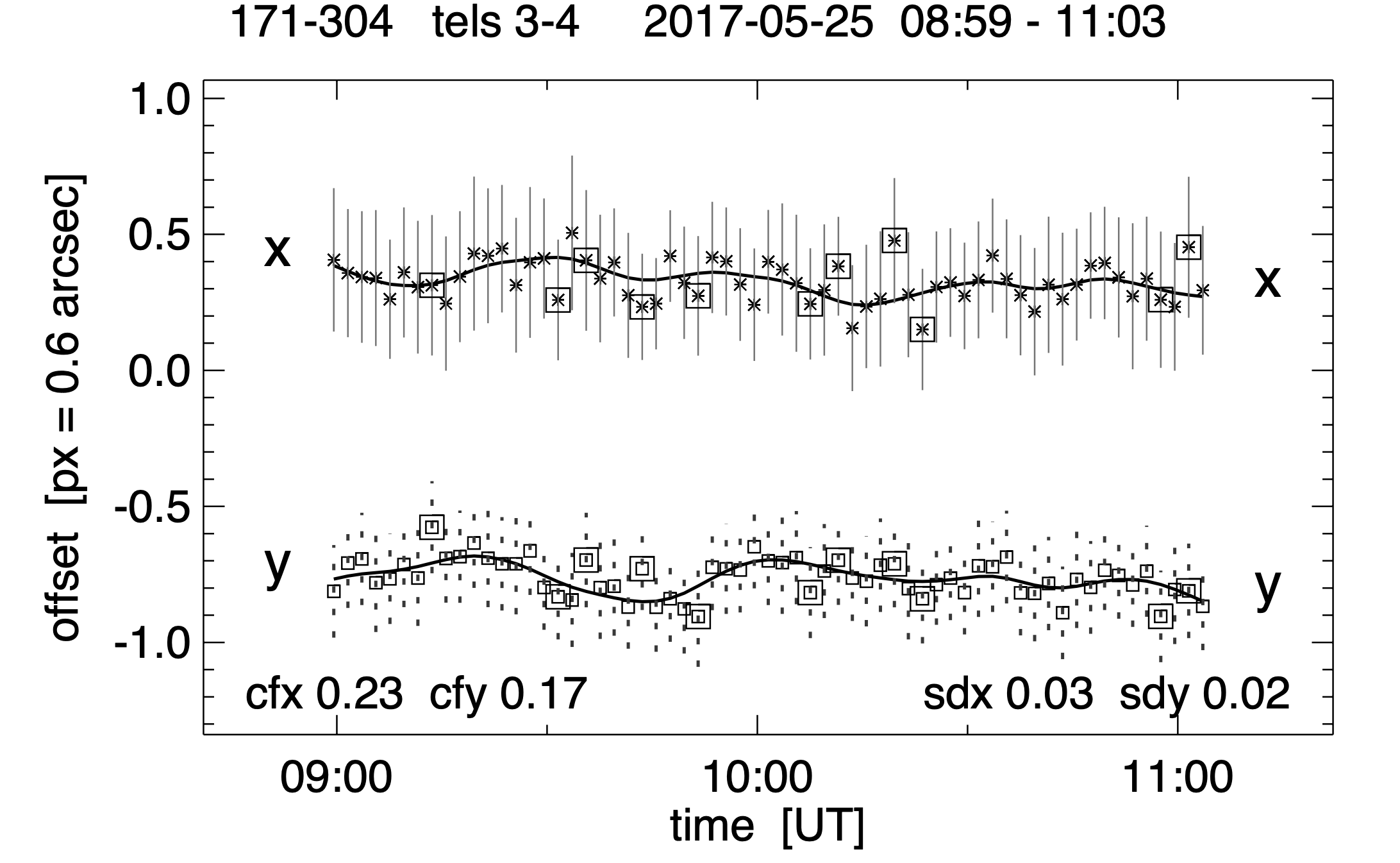} 
    %\caption{Flower one.}
    %\label{fig:f1}
  \end{subfigure}
  \hfill
  \begin{subfigure}[b]{0.5\textwidth}
    \includegraphics[width=\textwidth]{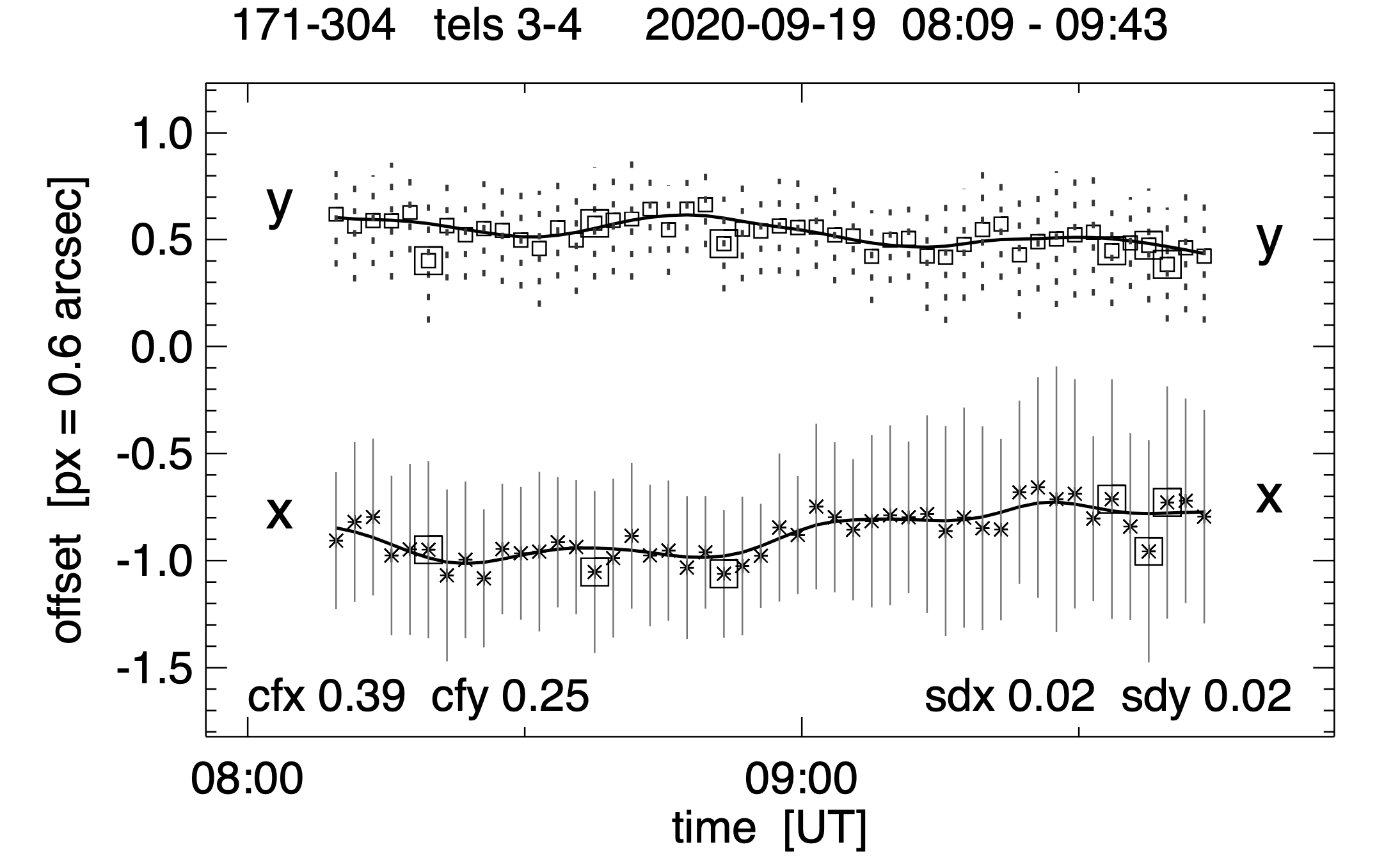}
    %\caption{Flower two.}
    %\label{fig:f2}
  \end{subfigure}
  \end{adjustbox}
  \caption{\label{figure_ch2:AIA_cross_alignment}%
  Horizontal~($x$, asterisks) and vertical~($y$, squares) offset plots, generated by the automatic alignment routine, between the co-aligned AIA 171~\AA\ (17.1~nm) and 304~\AA\ (30.4~nm) channels for dataset~1 (left panel) and dataset~2 (right panel) as a function of time. The $y$-axis indicates the offsets in terms of AIA pixels. The error bars denote 96\% (2-$\sigma$) confidence limits per time step. Their mean values are indicated by $cfx$ and $cfy$ in the lower left corner of each plot. Outliers in the $x$ and $y$ shifts are indicated by bounding boxes around asterisks or squares, and are removed while obtaining an iterative spline fit (solid black curves). The spline fits are saved by the pipeline and are used to co-align the target data. The values of $sdx$ and $sdy$ in the lower right corner represent the 1-$\sigma$ uncertainty in the spline fits, which suggests an accuracy of the co-alignment that is better than an AIA pixel scale (1~px). The term $tels$ in the title of each plot indicates the telescope number assigned by the pipeline.   }
\end{figure*}

The analysis presented in Paper~III made use of the co-aligned SDO/AIA and SST data sequences. I haved used the publicly available\footnote{\url{https://webspace.science.uu.nl/~rutte101/rridl/00-README/sdo-manual.html}} IDL based automated alignment routines developed by Rob Rutten for this purpose. The co-alignment procedure can be summarized in four major steps. The SDO cutout sequences corresponding to both datasets~1 and 2 were first downloaded. All the AIA channels were then co-aligned to HMI continuum/magnetogram images (for our analysis AIA 30.4~nm served as the anchor channel and it was aligned to HMI magnetogram), the HMI images were then expanded (resampled) from their original pixel scale (0\farcs5) to SST pixel scale (0\farcs037/0\farcs058 for CHROMIS/CRISP) and aligned to the SST wideband images via an iterative cross-correlation procedure. Finally, the SDO sequences were cropped to have the same FOV as the SST wide band FOVs. 

A major challenge in using the cut-out sequences from SDO/AIA is the internal co-alignment among the different AIA channels (especially the (E)UV passbands). This is because the temperature sensitivities of the different (E)UV passbands can range from a few tens of thousands to more than a few million Kelvin, depending on the channels of interest (refer to \cref{figure_ch-1:Corona_context} and \cref{figure_ch-1:TR_context}). To overcome this, the alignment pipeline downloads two sets of SDO cutout sequences. One at full (12~s) temporal cadence that targets the SST FOV, and the other set covers a FOV of 700\arcs~$\times$~700\arcs around the disk center at a lower cadence. The latter sequence is further used to find offsets between different AIA channels in an iterative fashion, by cross-correlating smaller sub-FOVs of roughly 30\arcs~$\times$~30\arcs, and applying the height of formation differences between the different channels. The two panels of \cref{figure_ch2:AIA_cross_alignment} show the offsets between two AIA channels (30.4 and 17.1~nm) after applying the co-alignment procedure described above. Based on the 1-$\sigma$ error estimates of the spline fit (denoted by $sdx$ and $sdy$) that is used to co-align the actual target data, it can be said that the accuracy of the alignment between the two channels is better than 1 AIA pixel (or 0\farcs6) for both the datasets. We refer to \cite{2020arXiv200900376R} for more details on the alignment procedure. It is important to note here that since the two datasets under investigation have a heliocentric angle $\mu$ $\approx$1 (i.e. they are close to the disk-center), the effects of projection are rather small. The accuracy of co-alignment is expected to be much lesser for datasets targeting further away from the disk center (i.e. at a lower $\mu$). 

\begin{figure}[!p]
    \centering
    \begin{adjustbox}{minipage=\linewidth,scale=0.7}
    \includegraphics[width=\textwidth]{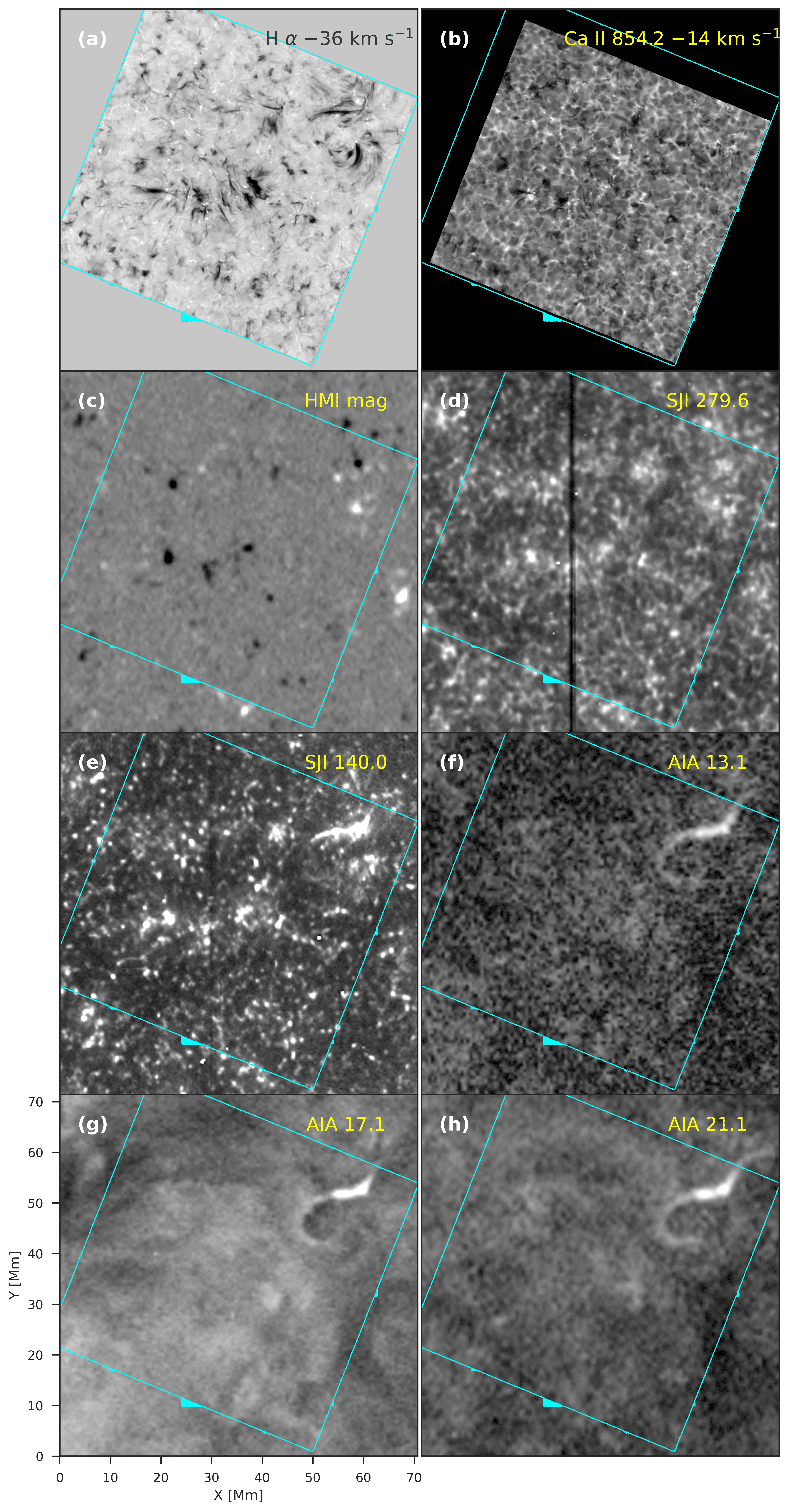}
    \end{adjustbox}
    \caption{Coordinated SST, IRIS and SDO observations of the quiet Sun disk center target from dataset~2. The images from the different telescopes are nearly co-temporal and are co-aligned to each other using the methods described in this chapter. Panels~(a)--(h) show the target observed through different passbands (from different telescopes) that are indicated in the top right corner in each case. The cyan colored box in each panel indicates the FOV corresponding to the \halpha{} blue wing image (panel~(a)) which serves as a reference. The HMI magnetogram shown in panel~(c) is clipped at $\pm$~1~kG.}
    \label{figure_ch-2:co_aligned_all}
\end{figure}

The co-alignment procedure described above generates eleven (nine AIA and two HMI) SDO image cut out sequences that have the same pixel scale as SST-CRISP/CHROMIS, and the images are chosen in such a way that they are closest in time to their SST counterparts. The resulting co-aligned SDO datasets were also visualized with CRISPEX in tandem with SST and IRIS counterparts, which revealed a good match among them. Panels~(c), and (f)--(h) in \cref{figure_ch-2:co_aligned_all} show SDO/HMI and AIA images corresponding to dataset~2 co-aligned to SST observations (panels~(a) and (b)). In Paper~III of this thesis, I have used images from AIA 30.4, 17.1 and 13.1~nm channels.

\section{Why do we need coordinated observations?}
\label{section:Need_for_coordinated_datasets}
From the discussions presented in the preceding sections, it is clear that combining observations from multiple telescopes is far from trivial. It is therefore valid to ask why is it necessary to coordinate between multiple telescopes in the first place. The answer lies in the fact that all telescopes, whether ground or space-based, have limitations in terms of their temporal and spatial resolution, wavelength (passbands) of observations, and range of temperature coverage. It is clear from the description presented in \cref{section:coupling}, and also from the observations presented in this chapter, that the solar atmosphere is highly non-homogeneous, time-dependent and structured. Furthermore, many small (and large) scale dynamic events often show indications of heating and can have an impact over a range of atmospheric layers. It is impossible for a single telescope to fulfill all the above criteria. Therefore, it is necessary to combine observations from different instruments that are complementary to one another in order to have a comprehensive understanding of the coupling between the layers of the solar atmosphere. 

% For example, an event that is visible in the solar corona almost always has a relationship with the dynamics happening at relatively deeper layers such as the TR or the chromosphere\footnote{In reality, it can often be challenging to tie the observations in the corona directly to the lower atmosphere.}.
\section{Numerical simulation of spicules}
\label{section:simulation_description}
\begin{figure}[!ht]
    \centering
    \includegraphics[width=\textwidth]{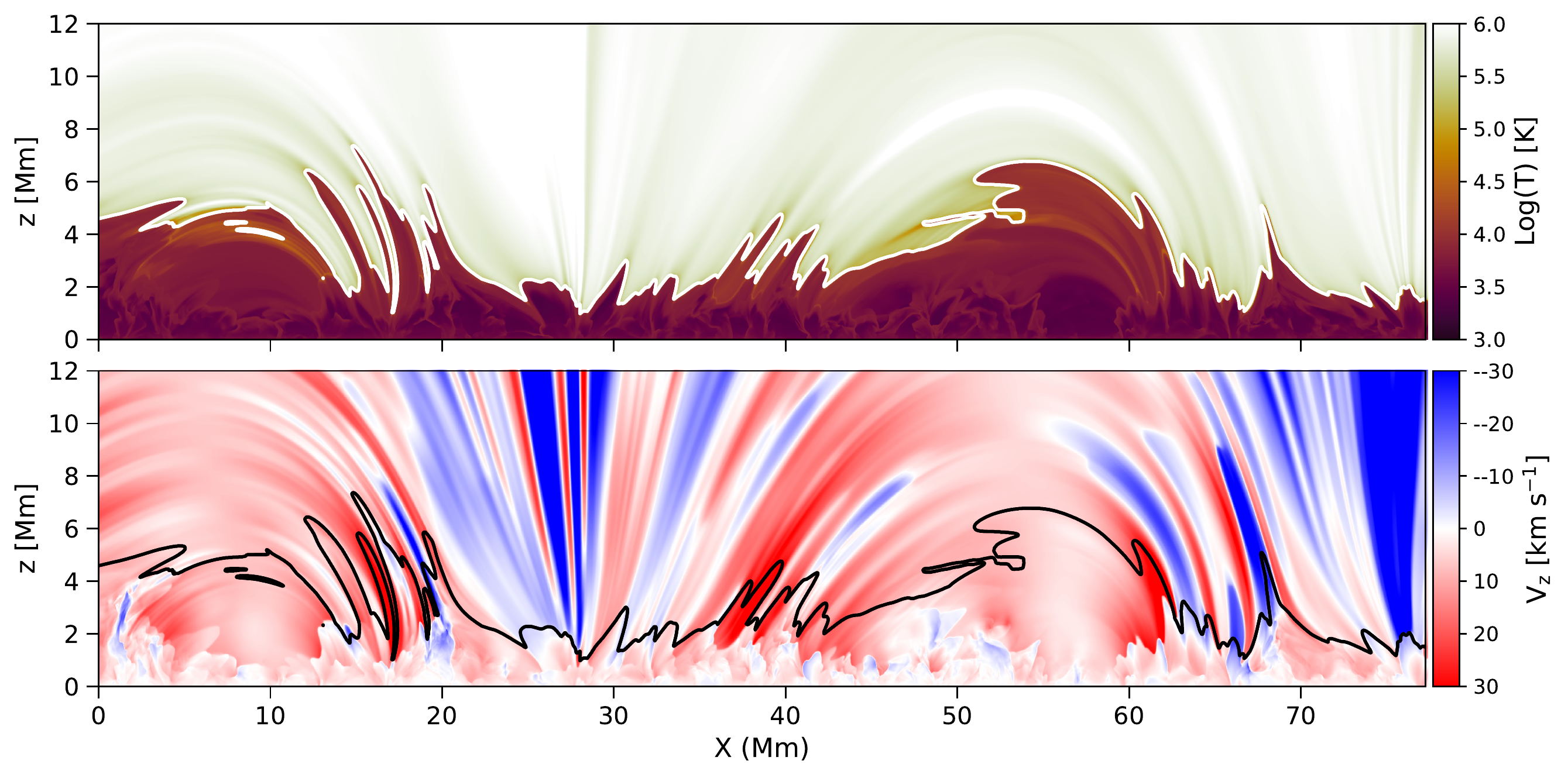}
    \caption{Outline of the 2.5D MHD simulation of spicules computed using the \textit{Bifrost} code. The top row shows a map of the temperature in log scale while the bottom row shows the corresponding signed vertical velocity (positive is downflow) clipped to $\pm$~30~\kms{}. Spicules appear as thin, finger-like intrusions in the upper solar atmosphere that are cooler than the surroundings. The black contour in the bottom row overlaps the region where the temperature equals 30,000~K. Analysis of this simulation is performed in Paper~III.}
    \label{figure_ch-2:simulation_context}
\end{figure}

This thesis makes use of an existing 2.5-dimensional (or 2.5D) MHD simulation of spicules computed by \cite{2017Sci...356.1269M} and \cite{2017ApJ...847...36M} using the \textit{Bifrost} code. The simulation includes the effect of ion-neutral interactions which is also known as ambipolar diffusion. Ambipolar diffusion can be interpreted as the process where the neutral particles (such as hydrogen) decouple from the plasma whereas the ions (charged species such as the chromospheric \ion{Ca}{ii}, \ion{Mg}{ii}, or TR ions such as \ion{Si}{iv}) remain tied to it. The so-called type-II spicules occurred naturally in this simulation as cool, thin, finger-like protrusions of plasma in the hotter corona (see \cref{figure_ch-2:simulation_context}). This was a significant advancement over older \textit{Bifrost} simulations \citep[such as,][]{2013ApJ...771...66M} which produced only one spicule after many solar hours of computation, indicating that physical processes like ambipolar diffusion, along with large-scale magnetic field configuration and high spatial resolution are the key ingredients to simulate spicular ubiquity. A discussion on the relevance of this simulation in the context of driving mechanism behind spicules is given in \cref{section:driving_mechanism} of \cref{chap:spicules}. The simulation domain spans from the upper convection zone (roughly 2.5~Mm below the visible photosphere or $z$=0~Mm) to about 40~Mm into the solar corona with a non-uniform resolution ranging from 12~km to 60~km in the vertical direction. In the horizontal domain the simulation covers 96~Mm with a uniform resolution of \textasciitilde12~km. \cref{figure_ch-2:simulation_context} shows an outline of the 2.5D simulation used in Paper~III, where the top row shows the temperature map (in logarithmic scale) and the bottom row shows the corresponding vertical velocity. 

The publicly available RH1.5D code\footnote{Available at: \url{https://github.com/ITA-Solar/rh}} \citep{Tiago_RH_2015} was used to perform optically thick radiative transfer computations from the above MHD simulation, which led to a synthesis of the chromospheric \cak{} and \Mgk{} lines. A comparison of the synthetic and observed spectra facilitates an approach based on which simulations can be compared reliably with observations. The RH1.5D code is a massively parallelized version of the former RH radiative transfer code developed by \cite{Uitenbroek_2001}, and is capable of performing multi-atom, multi-level, non-LTE radiation transport under both complete and partial frequency redistribution (CRD and PRD) of photons. Both the \cak{} and \Mgk{} spectral lines were synthesized under PRD (i.e. in Paper~I and Paper~III) as their line source function is dependent on frequency \citep[see,][]{1974ApJ...192..769M,1974SoPh...38..367V}. Furthermore, a detailed analysis of the optically thick spectral line formation mechanism in spicules was performed with the help of contribution functions. Synthetic \Si{}~139.4~nm intensity was computed assuming optically thin approximation and under ionization equilibrium using the CHIANTI database \citep{2009A&A...498..915D} in a way similar to \cite{2016ApJ...817...46M}, for example. The details of the analysis has been described extensively in Paper~III.

    \chapter{Spicules and mass flows in the solar atmosphere}
\label{chap:spicules}

\section{Historical perspective}
\label{section:history_of_spicules}
\begin{figure}[!h]
    \centering
    \begin{adjustbox}{minipage=\linewidth,scale=1.}
    \includegraphics[width=\textwidth]{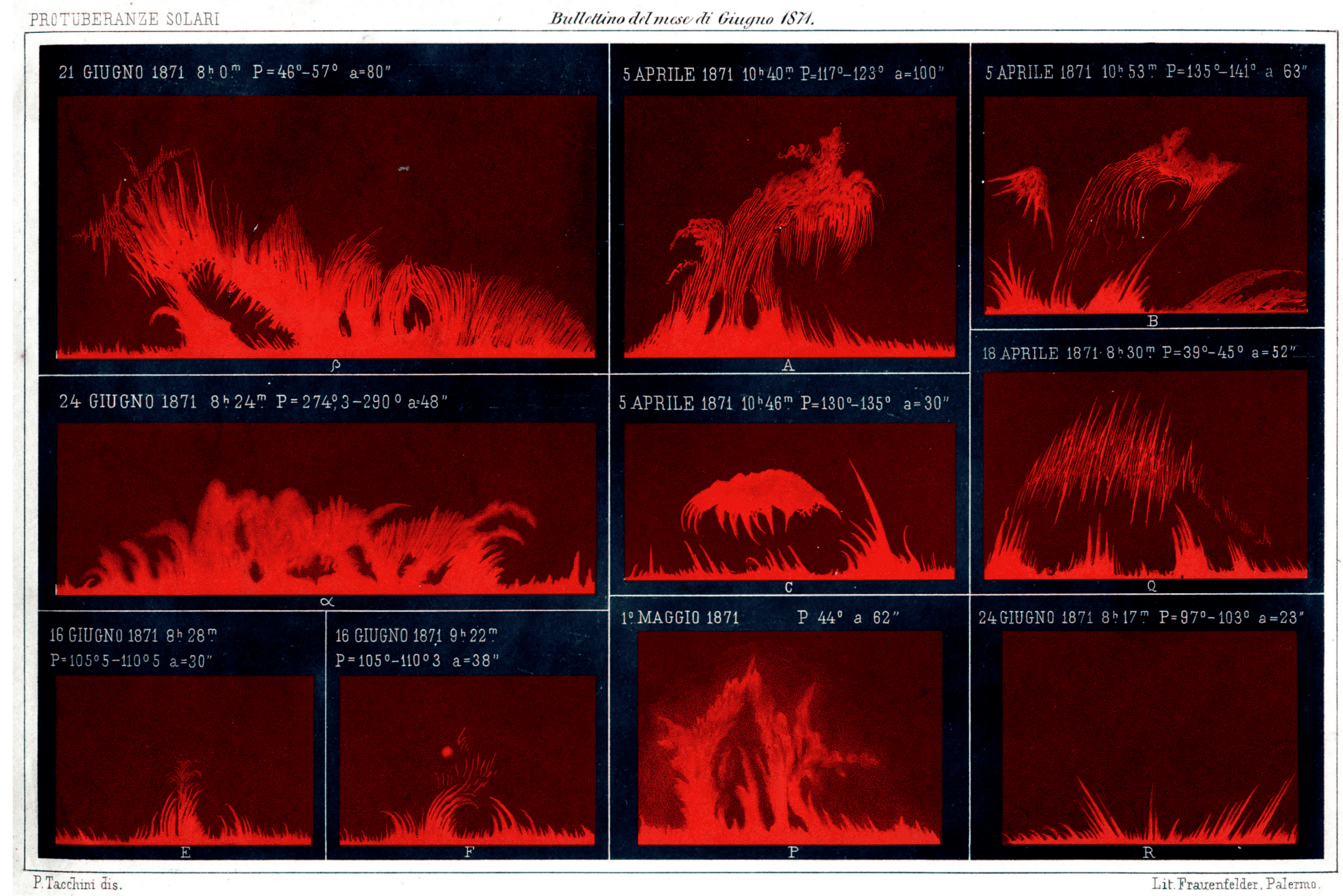}
    \end{adjustbox}
    \caption{First lithographed drawings of spicules and solar prominences observed by Pietro Tachinni in 1871. Thin hair-like protrusions resembling spicules are seen in almost all the drawings. Image Credit: \cite{2006MSAIS...9...28C}. Reproduced with permission from Ileana Chinicci, INAF, Rome.}
    \label{figure_ch-3:Tachinni_spicules}
\end{figure}

High resolution off-limb observations of the solar chromosphere reveal rapidly changing hair-like structures, that protrude outwards from the limb. These structures are known as \textit{spicules} and they form the major theme of this thesis. To the best of our knowledge, the first lithographed drawings of spicules (along with other off-limb features such as prominences) could be jointly attributed to the 19$^{\mathrm{th}}$ century Italian astronomer Pietro Tachinni \citep[refer to][for a modern account on the achievements of Tachinni]{2006MSAIS...9...28C} and the Italian priest cum astronomer Angelo Sechhi \citep{1871AdPd}, both in April, 1871. Sechhi later described them formally in the second edition of \textit{Le Soleil} which was published in 1877 \citep{1877arnp.book.....S}. I refer to \cref{figure_ch-3:Tachinni_spicules} which shows some of the drawings of spicules and prominences by Pietro Tachinni in 1871. 

\begin{figure}[!h]
    \centering
    \begin{adjustbox}{minipage=\linewidth,scale=1.}
    \includegraphics[width=\textwidth]{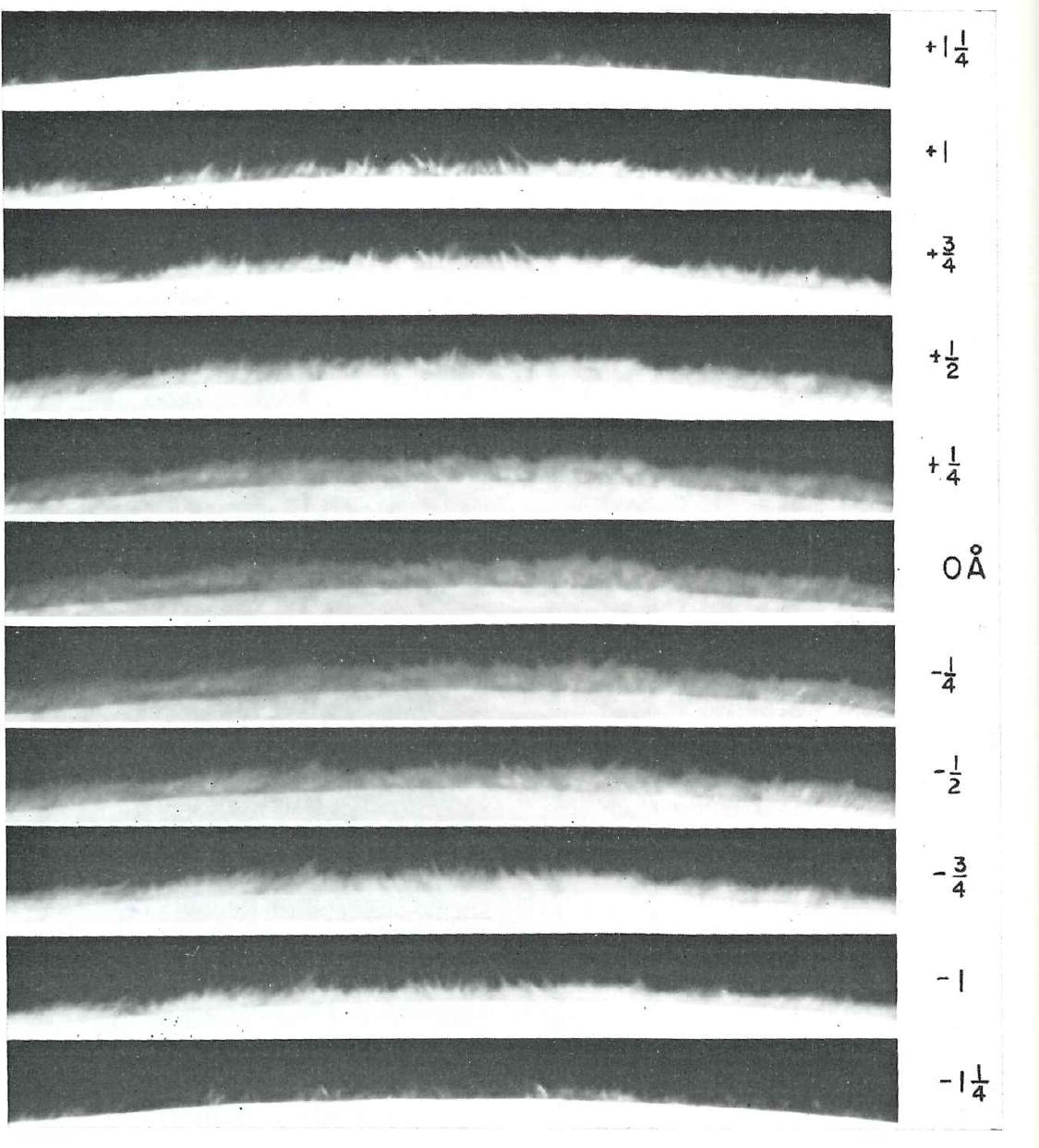}
    \end{adjustbox}
    \caption{Filtergram images of off-limb spicules observed by R.B.~Dunn at different wavelength positions between $\pm$~1.25~\AA\ with respect to the line core of \halpha{}. Image credit: \cite{1972ARA&A..10...73B}. Reproduced with permission from the Annual Review of Astronomy and Astrophysics, Volume 10 \textcopyright\ 1972 by Annual Reviews, \url{http://www.annualreviews.org}.}
    \label{figure_ch-3:Dunn_spicules}
\end{figure}

The term spicules was coined around 75 years later by \cite{1945ApJ...101..136R}, who also observed "spikes" of chromospheric plasma that appeared all over the north pole of the Sun. Most of the early observations of spicules, including those of \cite{1945ApJ...101..136R}, were based on filtergrams obtained primarily in \halpha{}. One of the best known early observation of spicules in \halpha{} is credited to R.B. Dunn \citep{1971BAAS....3..267D}, where he shows filtergrams obtained across various wavelength positions in and around the \halpha{} line core using the Sacramento Peak vacuum telescope. \cref{figure_ch-3:Dunn_spicules} is a reproduction of the spicules observed by Dunn through a tunable Zeiss \halpha{} filter.
In addition, \cite{1973SoPh...32..337B} observed off-limb spicules through a K filter centered on the line core of \cak{} spectral line and noted that the solar chromosphere appears more diffused and the spicules are not as well resolved as it is with an \halpha{} filter. However, other parameters such as the lifetime and height of spicules were well in line with the preceding studies by \cite{1968SoPh....3..367B,1971BAAS....3..267D} and \cite{1972ARA&A..10...73B}.

\section{Developments since the 1960s}
\label{section:Development_modern}
As described in the preceding section, the first examples of spicules were observed and recorded as early as in the 19$^{\mathrm{th}}$ century by Sechhi and Tachinni in 1871. More recent decades saw a much increased interest, driven by novel instrumentation and also due to the realization that these are truly ubiquitous features in the Sun.
%Since their discovery by Sechhi and Tachinni in 1871, spicules have been a topic of great interest in solar physics. Their ubiquity in the solar chromosphere made them particularly interesting for the observers in second half of the twentieth century. 
Since the 1960s there have been several studies that aimed at determining some of the basic properties of spicules, such as, their maximum height that can be between 5--10~Mm above the photosphere, their propagation speed which can be between 20--50~\kms, and their lifetime that ranges between 5--10~min \citep[see][and references therein for extensive reviews on classical and modern understanding of spicules and their properties]{1968SoPh....3..367B,1972ARA&A..10...73B,2000SoPh..196...79S,2012SSRv..169..181T}.

Spicules have also been thought to play a potential role in supplying mass and energy to the solar corona. Studies, such as \cite{1977A&A....55..305P,1978SoPh...57...49P}, estimated that the amount of mass flux carried by spicules propagating upwards is nearly 100 times larger than the mass needed to balance the solar wind. This means that even if a small fraction of this spicular mass is carried by the solar wind (and the rest eventually falls back to the chromosphere), it would be enough to balance the mass losses sustained by the solar corona. Shortly after this, \cite{1982ApJ...255..743A,1984ApJ...287..412A} studied the energetics of spicules and reasoned that they might play an important role in heating and transferring energy from the deeper layers of the solar atmosphere to outer atmospheres such as the chromosphere and the corona. This idea was however challenged first by \cite{1983ApJ...267..825W} and later by \cite{1987ApJ...319..465M}, who found no evidences of spicules in the EUV emission lines formed in the TR and corona.

 Despite a multitude of studies since the pioneering review of \cite{1968SoPh....3..367B}, spicules still remain one of the most elusive and least understood phenomena in the solar atmosphere. This could be mainly attributed to the lack of high spatial and temporal resolution observations, particularly during the second half of the twentieth century which failed to provide proper constraints in the development of satisfactory numerical spicule models that could explain all of their observed properties \citep{2000SoPh..196...79S}. With the recent developments in advanced instrumentation and numerical modeling capabilities over the past decade or so, we have witnessed significant breakthroughs in our attempt to understand this "beast", that happens to appear almost everywhere on the solar limb. However, we still lack definitive models that could explain their role in heating the solar corona or reproduce all of their observed properties, including their fine structure \citep{2019ARA&A..57..189C,2019PASJ...71R...1H}.
 
 \section{Types of spicules and their dynamics}
\label{section:spicule_types_dynamics}

Until the dawn of modern instrumentation techniques that resulted in the possibility of obtaining high-resolution observations of the Sun, spicules were mostly observed against the limb. With the help of high-resolution ground-based observations from the 65~cm aperture solar telescope at the Big Bear Solar Observatory, \cite{1995ApJ...450..411S} observed on-disk counterparts of spicules in an enhanced network region. The above set of authors used the term \textit{fibrils} to describe any generic fibrous structure that appear either in the wing or line-core images of \halpha{}. 

\begin{figure*}[!h]
\centering
\begin{adjustbox}{minipage=0.85\linewidth,scale=1.2}
 \begin{subfigure}[b]{0.5\textwidth}
    \includegraphics[width=\textwidth]{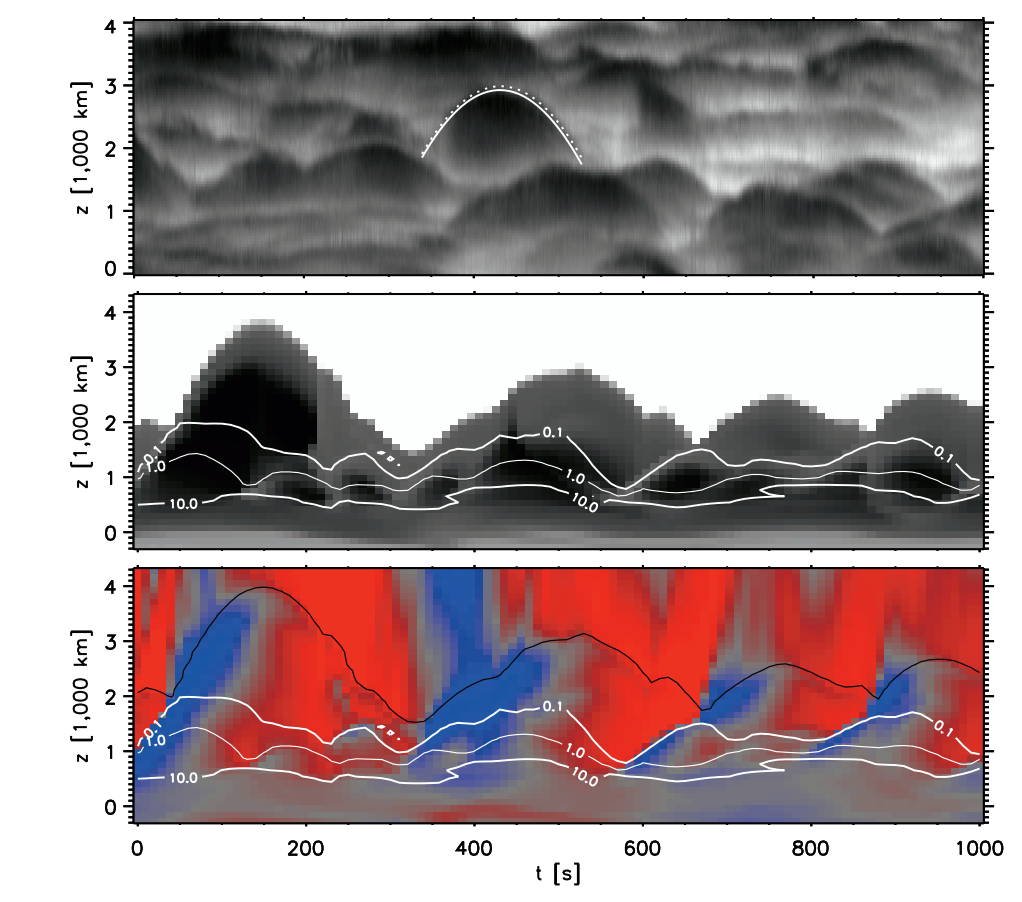} 
    %\caption{Flower one.}
    %\label{fig:f1}
  \end{subfigure}
  \hfill
  \begin{subfigure}[b]{0.5\textwidth}
    \includegraphics[width=\textwidth,height=0.21\textheight]{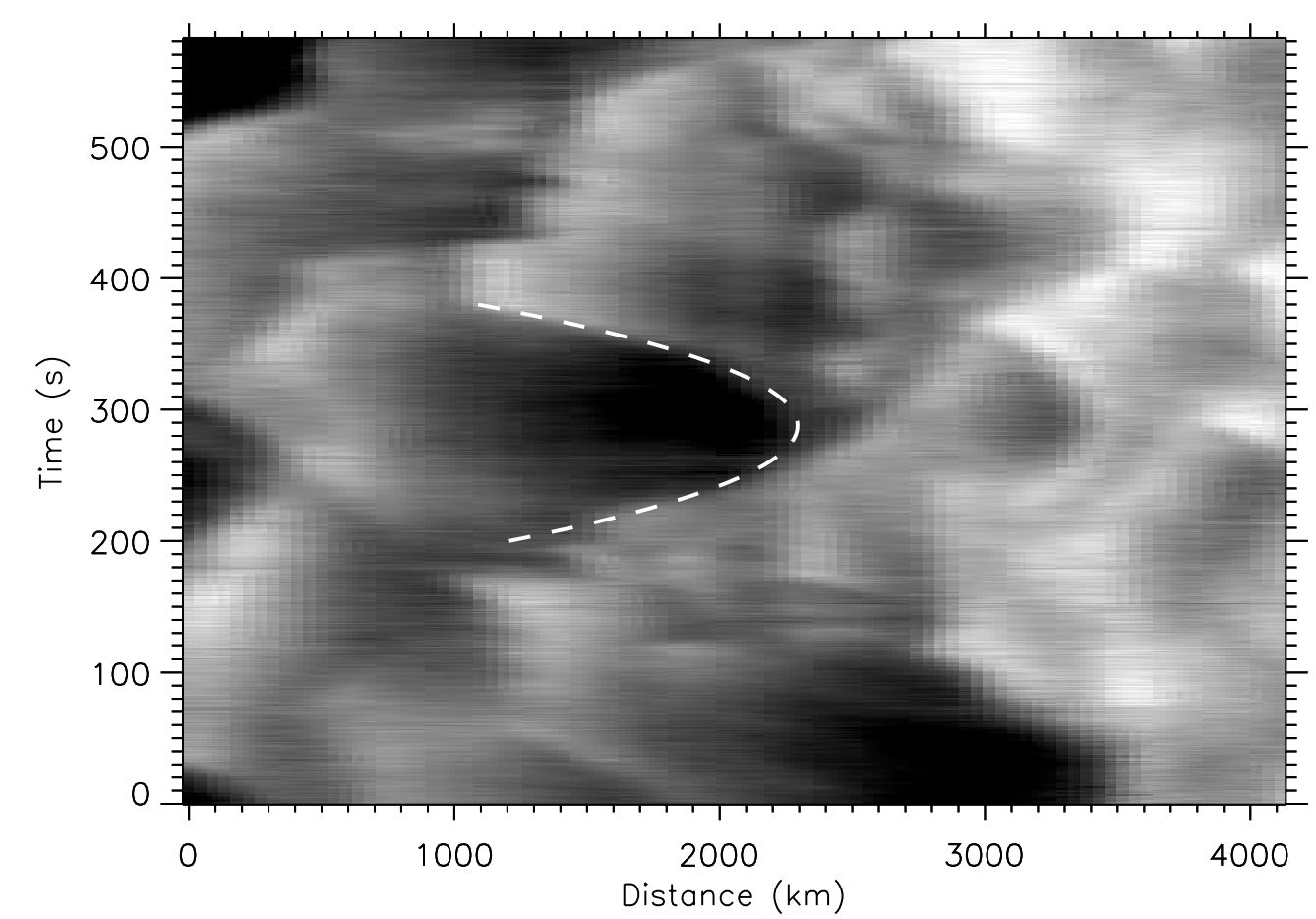}
    %\caption{Flower two.}
    %\label{fig:f2}
  \end{subfigure}
  \end{adjustbox}
  \caption{\label{figure_ch3:DFs}%
  Evolution of on-disk dynamic fibrils in high resolution observation and simulations. The left panel shows the spatio-temporal evolution of dynamic fibrils from SST (\halpha{} line core) observations (top row), plasma temperature and vertical velocity (middle and bottom rows) from the MHD simulation. The right panel shows another example of the space-time evolution undergone by a dynamic fibril. Distinct parabolic paths are seen in both the panels, including the simulation. Upward velocities (in blue) mark the ascending phase of the fibrils indicating a shock-like behavior. Image credit: \cite{2006ApJ...647L..73H,2007ApJ...655..624D}. \textcopyright\ AAS. Reproduced with permission.}
\end{figure*}

About a decade later, studies such as \cite{2006ApJ...647L..73H} and \cite{2007ApJ...655..624D}, revealed the existence and the spatio-temporal evolution of short, highly dynamic fibrils in and around the close vicinity of an active region plage. High quality time series observations from SST, in tandem with sophisticated adaptive optics and advanced post-processing image restoration techniques like MOMFBD (see \cref{subsection:CHROMIS}), enabled the possibility of tracking these spicular features for the very first time. The results from these studies indicated that dynamic fibrils undergo typical parabolic up-down motion when observed in the line-core images of \halpha{}, propagate with velocities that are in the range of 10--40~\kms, show compelling correlation between maximum velocity and deceleration, last between 3--10~min, and can be between 1.5--3~Mm in length. Many of these values seemed to correlate well with the low resolution off-limb observations by \cite{1968SoPh....3..367B,1972ARA&A..10...73B}. Furthermore, a detailed comparison with advanced 2D radiative MHD simulations with the \textit{Bifrost} code revealed striking similarities with the observed properties of dynamic fibrils, suggesting that fibrils are most likely driven by acoustic shock waves that are leaked into the solar chromosphere by convective motions or $p$-modes (see \cref{figure_ch3:DFs}). The leakage of photospheric oscillations have also been shown to produce 5-min periodicities in dynamic fibrils \citep{2004Natur.430..536D}. 
The quiet Sun counterpart of active region dynamic fibrils were reported by \cite{2007ApJ...660L.169R} and were linked to chromospheric mottles that were historically observed in \halpha{} \citep[see also,][]{1995ApJ...450..411S}. These mottles showed similar properties as the active region fibrils, including the characteristic parabolic up-down motion. This was suggestive of the fact that mottles were driven in a way similar to the dynamic fibrils.

At the limb however, the dynamic fibrils are not seen outside of active regions. This is because of the rapid expansion of the magnetic field with height in weaker field environments, such as quiet Sun or coronal holes, which dampens the energy carried by the shock waves in fibrils. As a result, they produce only minor protrusions in the limb observations and remain mostly hidden among the taller (stronger) features due to superposition along the line of sight \citep{2019ARA&A..57..189C}. However, this explanation caused confusion among the solar physicists because traditionally spicules have been observed everywhere against the limb regardless of active regions. 

\begin{figure}[!h]
    \centering
    \begin{adjustbox}{minipage=\linewidth,scale=1.}
    \includegraphics[width=\textwidth]{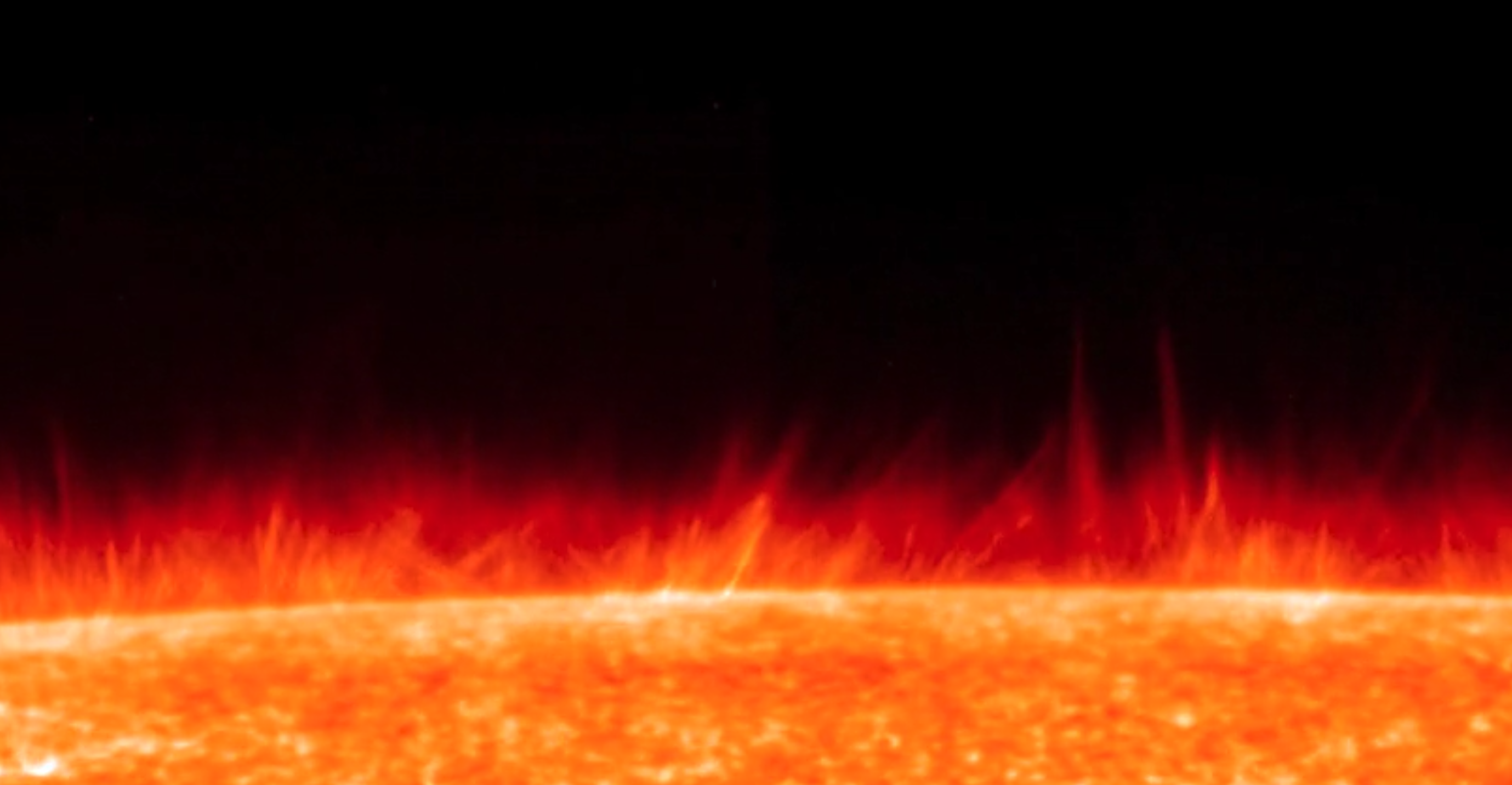}
    \end{adjustbox}
    \caption{Spicules observed against the solar limb with Hinode \ion{Ca}{ii}~H filter that has a FWHM of 0.22~nm. The off-limb part of this image has been enhanced to make the faint spicules clearer. Image credit: Mats Carlsson, Rosseland Center for Solar Physics, Oslo.}
    \label{figure_ch-3:Hinode_spicules}
\end{figure}

The advent of high-resolution, seeing-free, space-based \ion{Ca}{ii}~H observations that were made available by the Hinode satellite \citep[][]{2007SoPh..243....3K}, provided an explanation regarding the above confusion which led to a paradigm shift in our understanding of spicules. Targeting quiet Sun and coronal hole limb observations, \cite{2007PASJ...59S.655D} found evidence of a potentially new category of spicules that, contrary to previously observed fibrils or mottles, have much faster velocities (30--150~\kms), typically last less than 2~min, and show a rapid apparent upward motion without the characteristic parabolic motion prevalent in fibrils. These properties differed significantly from the canonical spicular observations by \cite{1968SoPh....3..367B,1972ARA&A..10...73B,1973SoPh...32..337B}. Therefore, based on these differences, and also the fact that these features were observed predominantly in the weaker magnetic field environments, \cite{2007PASJ...59S.655D} described them as "type-II" spicules to distinguish them with previously observed active region dynamic fibrils or quiet Sun mottles (which were later termed as "type-Is"). It was also interpreted that the fading or the lack of a distinct downward trajectory in the type-IIs could be a sign of vigorous heating associated with type-II spicules. \cref{figure_ch-3:Hinode_spicules} shows a representative example of spicules observed against the limb with the Hinode Solar Optical Telescope in \ion{Ca}{ii}~H filter.

The discovery of type-II spicules sparked a massive debate in the scientific community, where many questioned the existence of this new category, and its role in upper atmospheric heating. Based on their numerical modeling efforts, \cite{2012JGRA..11712102K} found that most of the newly discovered type-II spicules were only moderately heated to temperatures of the order of 10$^{5}$~K (weak emission in AIA 30.4~nm channel) and therefore challenged their role in heating the solar corona. \cite{2012ApJ...750...16Z}, in particular, analyzed the same sets of data as \cite{2007PASJ...59S.655D}, and concluded that there was no credible evidence of type-II spicules, since they argued that more than 60\% of their detected spicules showed a characteristic type-I like behavior. Consequently, \cite{2012ApJ...750...16Z} advised caution while investigating the role of spicules in coronal heating. Immediately after this, \cite{Tiago_2012} performed a statistical study on over several hundreds of spicules detected from the same dataset as \cite{2012ApJ...750...16Z} and convincingly demonstrated that type-IIs were not just real but they are more abundant in the solar chromosphere, except in the active regions (where type-Is or dynamic fibrils were found to be dominating). Furthermore, their statistical distribution also showed distinct differences in the lifetimes, velocities and maximum heights attained between the two categories as was originally suggested by \cite{2007PASJ...59S.655D}. Earlier, studies conducted by \cite{2010ApJ...714L...1S} also with Hinode/SOT revealed the presence of type-II spicules in a coronal hole close to the north pole of the Sun with no apparent downward motion. In addition, the above set of authors also found brightenings close to the footpoints associated with some of these spicules when seen on the disk. 

Observations from Hinode certainly paved way for significant improvements in our understanding of spicules, which helped to clear the air regarding the omnipresence of spicules in classical observations. However, it posed a new question. Since type-IIs were more dominant in the Sun, then how come the classical observations (by \cite{1968SoPh....3..367B}, for example) revealed properties that were more correlated with type-Is? A likely scenario was painted by \cite{2013ApJ...764...69P} who found that, if the Hinode observations were artificially degraded to imitate older ground-based images, the deduced properties of spicules resembled type-Is. In other words, the measured velocities and lifetimes were significantly affected by degrading the resolution which led the above set of authors to conclude that classical observations of spicules did detect type-IIs, but owing to the lower spatial and temporal resolution they remained unnoticed.

\section{Evidence of heating in spicules}
\label{section:heating_with_spicules}

\begin{figure}[!h]
    \centering
    \begin{adjustbox}{minipage=\linewidth,scale=1.}
    \includegraphics[width=\textwidth]{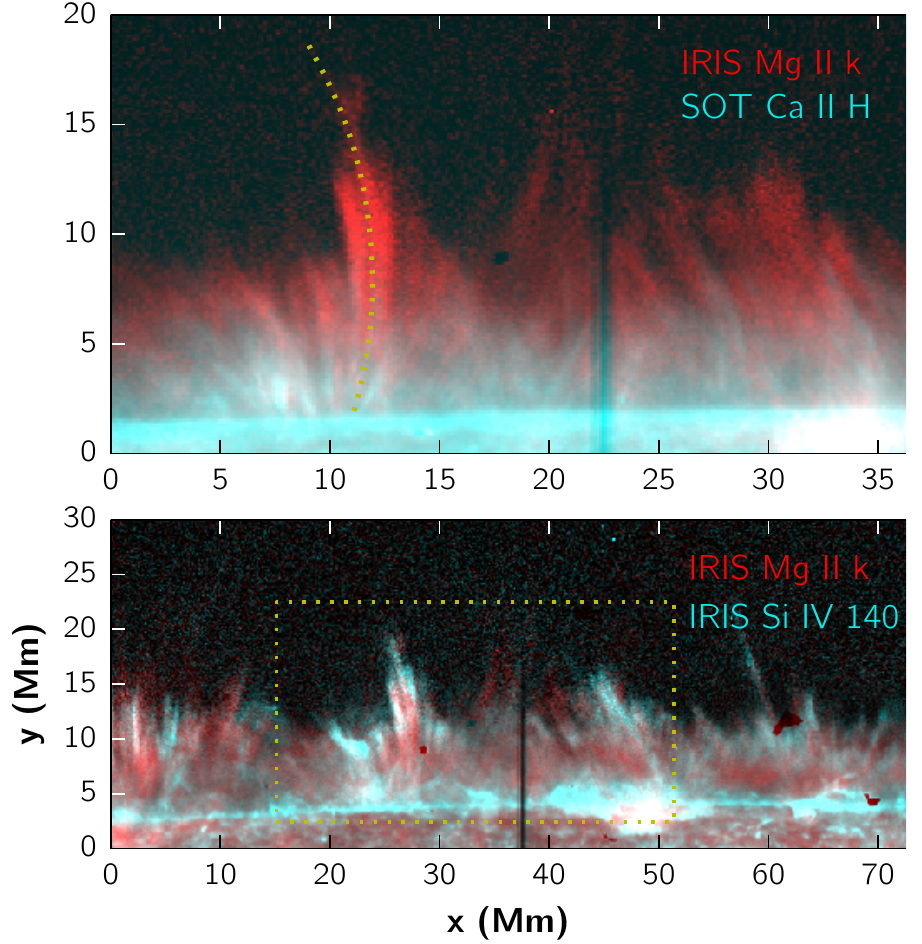}
    \end{adjustbox}
    \caption{Red-Blue composite images of spicules observed by combining Hinode SOT and IRIS \ion{Mg}{ii} and \ion{Si}{iv} SJIs. The top panel shows \ion{Mg}{ii}~SJI in red and Hinode \ion{Ca}{ii}~H filter in blue and the bottom row shows the IRIS \ion{Mg}{ii} and \ion{Si}{iv} SJIs in red and blue respectively, at the same time as the top panel. The yellow-dotted rectangle in the bottom panel indicates the (Hinode) FOV of the top panel. Image credit: Tiago~M.~D.~Pereira, Rosseland Center for Solar Physics, Oslo }
    \label{figure_ch-3:Pereira_spicules}
\end{figure}

The fading of type-II spicules in the Hinode \ion{Ca}{ii}~H passbands were speculated to be associated with rapid heating which causes the spicules to lose opacity as they evolve. This idea, along with the speculation that spicules could contribute to the heating of the outer solar atmosphere, had been around for a few decades well before the advent of Hinode or high-resolution space-based observations \citep[see][for example]{1983ApJ...267..825W,1984ApJ...285..843S}. One of the first clear observational evidence of spicules being heated, and responsible for supplying hot plasma to the solar corona was exemplified by \cite{2011Sci...331...55D} with coordinated observations from the Hinode and SDO space missions. However, this aspect was made astoundingly clear only after the launch of IRIS by \cite{Tiago_2014_heat}, where the authors combined Hinode and IRIS observations and unambiguously found that the fading \ion{Ca}{ii}~H spicule continue to evolve in passbands that are sensitive to higher temperatures, such \ion{Mg}{ii} and \ion{Si}{iv}, sampling the upper chromosphere and TR. \cref{figure_ch-3:Pereira_spicules} describes the scenario where we see that spicules that fade out of the Hinode passbands continue to evolve in the hotter IRIS channels. 
Furthermore, IRIS observations by the above set of authors also revealed that these heated spicules eventually fall back in the lower atmosphere, after attaining a certain maximum height. The falling phase, however, was not distinctly visible in the \ion{Ca}{ii}~H passbands. This was also confirmed a year later in a more detailed study conducted by \cite{2015ApJ...806..170S}. Further evidence of spicules being associated with heating beyond the TR temperatures is discussed later in \cref{sectiom:mass_flows}.

Since the IRIS observations revealed that the rapidly heated type-II spicules also undergo a parabolic path, similar to type-I spicules, it prompted the question again whether a segregation between the two types of spicules is justified or not. It is to be noted here, that despite similar trajectories, fundamental differences do exist between the two categories. \cite{2016ApJ...817..124S} found evidence of grain-like brightenings in the TR (in the \ion{Si}{iv} passband) above an active region plage. These brightenings were found to be due to the shocks propagating from the chromosphere to the (less denser) TR associated with the tops of dynamic fibrils (or on-disk counterparts of type-Is). A spatio-temporal analysis of these grains revealed parabolic trajectories in the TR, but the brightenings were only restricted to the tops of those spicules. This is in contradiction to the earlier analysis by \cite{Tiago_2014_heat}, where the spicules appeared to be emission along their entire lengths. Moreover, the propagation speed of type-II spicules observed by \cite{Tiago_2014_heat} \citep[and also by][]{2015ApJ...806..170S} were substantially higher than type-Is, which adds more strength in favor of the differences between the two types.

\section{On-disk counterparts of type-II spicules}
\label{section:RBEs_RREs}

\begin{figure}[!h]
    \centering
    \begin{adjustbox}{minipage=\linewidth,scale=1.}
    \includegraphics[width=\textwidth]{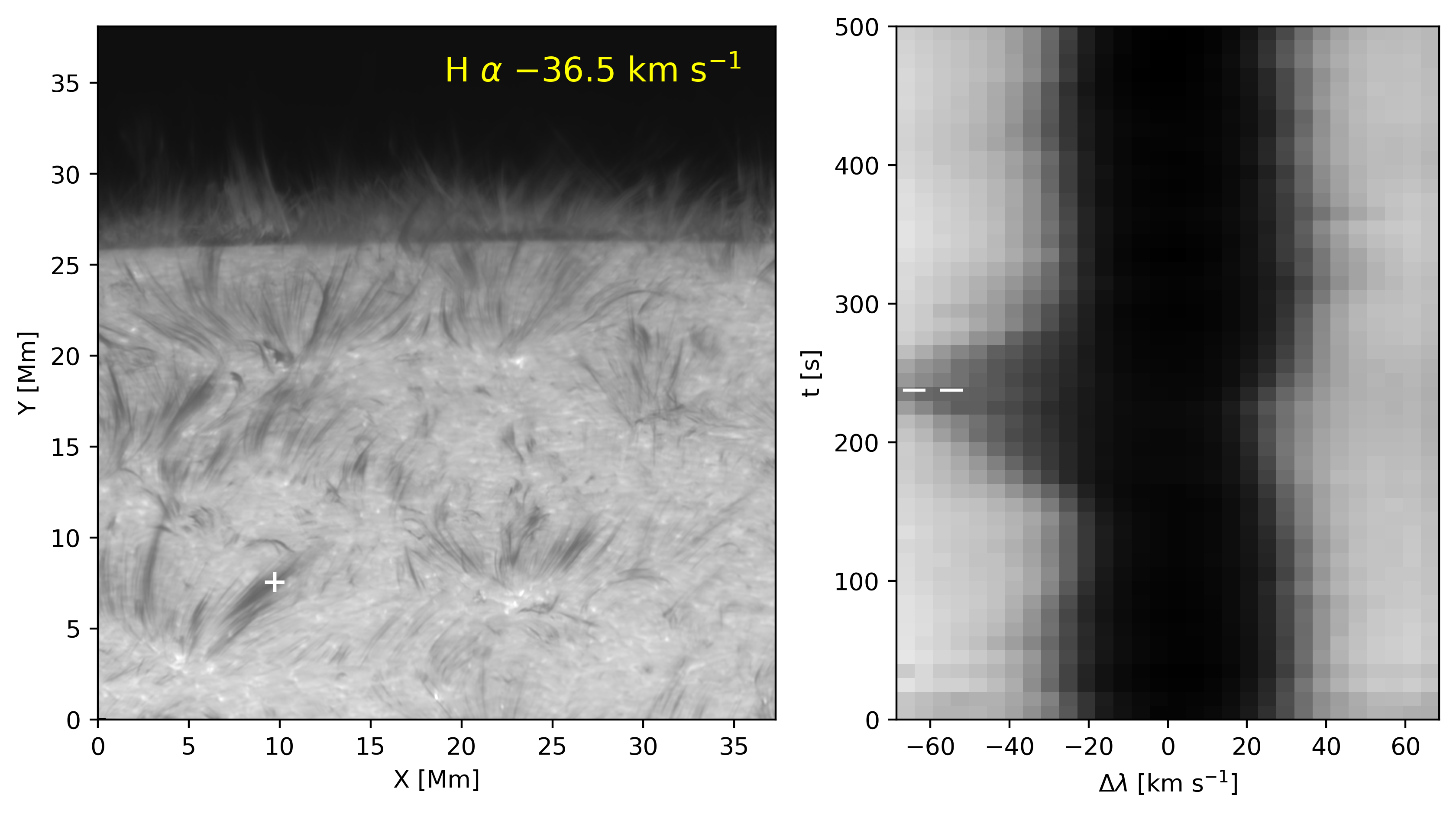}
    \end{adjustbox}
    \caption{Rapid blueshifted excursions (RBEs) observed with SST CRISP on 20 August 2020. Left panel: \halpha{} blue wing image (at $-36.5$~\kms or $-$0.8~\AA\ from the line center) showing ubiquitous presence of dark, threaded, hair-like structures (RBEs) both on-disk and off-limb. Right panel: \halpha{} spectral--time ($\lambda t$) diagram corresponding to the spicule indicated by the white marker (plus sign) in the left panel, showing the rapid development of a blueward excursion asymmetry between $t$=200 and 300~s. The dashed horizontal marker in the right panel indicates the time at which the image in the left panel is shown. Data reduction by Luc Rouppe van der Voort}
    \label{figure_ch-3:RBEs_SST}
\end{figure}
%Further evidence in this direction was provided by \cite{2011Sci...331...55D} who were able to connect the rapidly propagating on-disk spicules in the blue wing of the narrow band \halpha{} filter on-board Hinode/SOT  with the corresponding emission signatures in the hotter SDO/AIA channels. 
As noted in \cref{section:spicule_types_dynamics}, fibrils were observed to be the on-disk counterparts of type-I spicules. They show expansion and contraction in the inner wing images of \halpha{} with velocities between 10--30~\kms, last for several minutes and show strong periodicities between 3--7~min \citep{2006ApJ...647L..73H,2007ApJ...655..624D}. After the discovery of type-II spicules by \cite{2007PASJ...59S.655D}, there was a search for the on-disk counterparts of this new category of spicules. With the help of imaging spectroscopic observations with the Interferometric BIdimensional Spectrometer \citep[IBIS,][]{2006SoPh..236..415C}, \cite{2008ApJ...679L.167L} found short-lived (\textasciitilde40~s), high speed events (15--30~\kms) that appeared as blueshifted excursions in the \ion{Ca}{ii}~854.2~nm line without any following redshift. These were described as rapid blueshifted events (RBEs). Later, with higher temporal, and spatial resolution observations from the SST, \cite{Luc_2009} not only confirmed the presence of RBEs by observing excursions in the blue wings of the \ion{Ca}{ii}~854.2~nm and \halpha{}, but also described the appearance of RBEs as dark-threaded elongated structures that are visible in the far wing images of the chromospheric spectral lines. They performed a detailed statistical analysis showing that RBEs have lifetimes, appearance and morphology that are similar to earlier reported properties of type-II spicules by \cite{2007PASJ...59S.655D}. Moreover, RBEs were also found to mostly appear in the close vicinity of strong field network regions and rapidly propagate outwards with speeds in excess of 50~\kms. Later, \cite{Sekse_2012} expanded the statistics by developing an automated detection procedure that enabled them to detect a large number of RBEs in their datasets, further confirming that RBEs are indeed the on-disk manifestations of type-II spicules. In addition, their analysis revealed that RBEs observed in \ion{Ca}{ii}~854.2 are linked to \halpha{} RBEs, and the former have the lower part of RBEs closer to the network regions in comparison to the \halpha{} RBEs.

\begin{figure}[!h]
    \centering
    \begin{adjustbox}{minipage=\linewidth,scale=1.}
    \includegraphics[width=\textwidth]{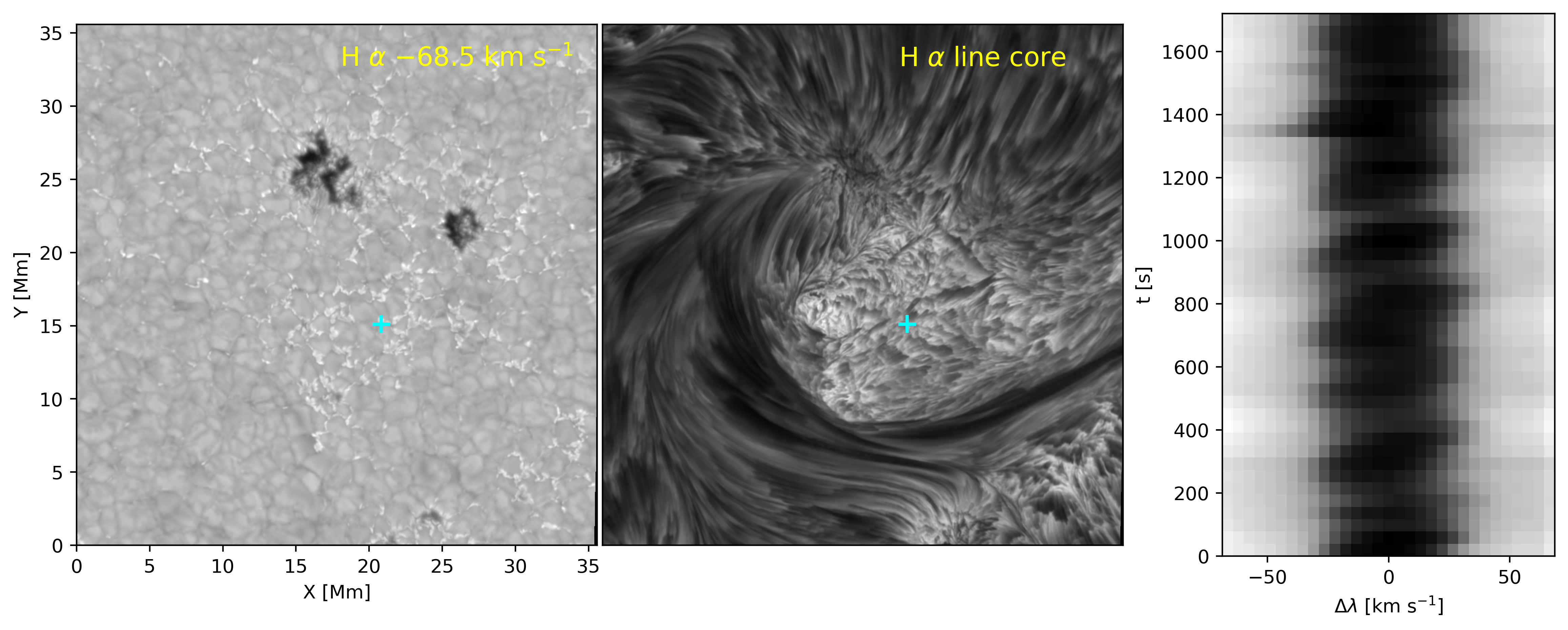}
    \end{adjustbox}
    \caption{Active region dynamic fibrils (type-I spicules) observed with SST CRISP on 11 August 2020. Left panel: Far wing photospheric \halpha{} image ($-68.5$~\kms{} or $-$1.5~\AA\ from the line center) showing the darker pores and brighter plage regions. Middle panel: Purely chromospheric \halpha{} line core image showing a predominant fibrilar scene with a large canopy extending from one end of the image to the other; right panel: \halpha{} $\lambda t$ diagram corresponding to the location of a dynamic fibril indicated by a cyan marker (plus sign) in the left and middle panels. The dynamic fibrils are mostly found in the vicinity of the central plage region. Data reduction by Luc Rouppe van der Voort}
    \label{figure_ch-3:DF_sawtooth_SST}
\end{figure}

The right panel of \cref{figure_ch-3:RBEs_SST} shows an example of the blueshifted excursion associated with an RBE observed in the blue wing of \halpha{} with the help of a spectral-time ($\lambda t$) diagram. RBEs appear as dark threaded structures on the disk and they are morphologically similar to the bright hair-like features that protrude outwards from the limb (left panel). The $\lambda t$ diagram clearly illustrates the rapid blueward asymmetry in the \halpha{} spectral line, corresponding to the RBE indicated by the marker in the left panel, between $t$=200 and 300~s without any following redward excursion. This is in tandem with the first observations of RBEs presented by \cite{2008ApJ...679L.167L}, \cite{Luc_2009} and \cite{Sekse_2012}, and is contradictory to the behavior observed in dynamic fibrils or type-Is that show a characteristic saw-tooth pattern in the $\lambda t$ plots as shown in \cref{figure_ch-3:DF_sawtooth_SST} \citep[also see examples in][]{2008ApJ...673.1194L,2016ApJ...817..124S}. 

In addition to the radially outward flows, RBEs were also found to undergo characteristic swaying motion of the order of 10--20~\kms that were interpreted as a sign of \alfvenic{} waves propagating in the solar chromosphere \citep{2007Sci...318.1574D}. Spatio-temporal analysis of the wave propagation led \cite{2007Sci...318.1574D} to conclude that the observed \alfvenic{} waves could be responsible for driving the solar wind. A detailed study of these waves extending into the TR and corona by \cite{2011Natur.475..477M}, led to the suggestion that spicules could not only drive the fast solar wind, but are energetic enough to balance the losses suffered in the quiet corona.

%The statistical properties of the \alfvenic{} waves were further expanded by Okamoto and De Pontieu 2011 who upward and downward propagating standing waves in spicules with Hinode observations.
\begin{figure}[!h]
    \centering
    \begin{adjustbox}{minipage=\linewidth,scale=1.}
    \includegraphics[width=\textwidth]{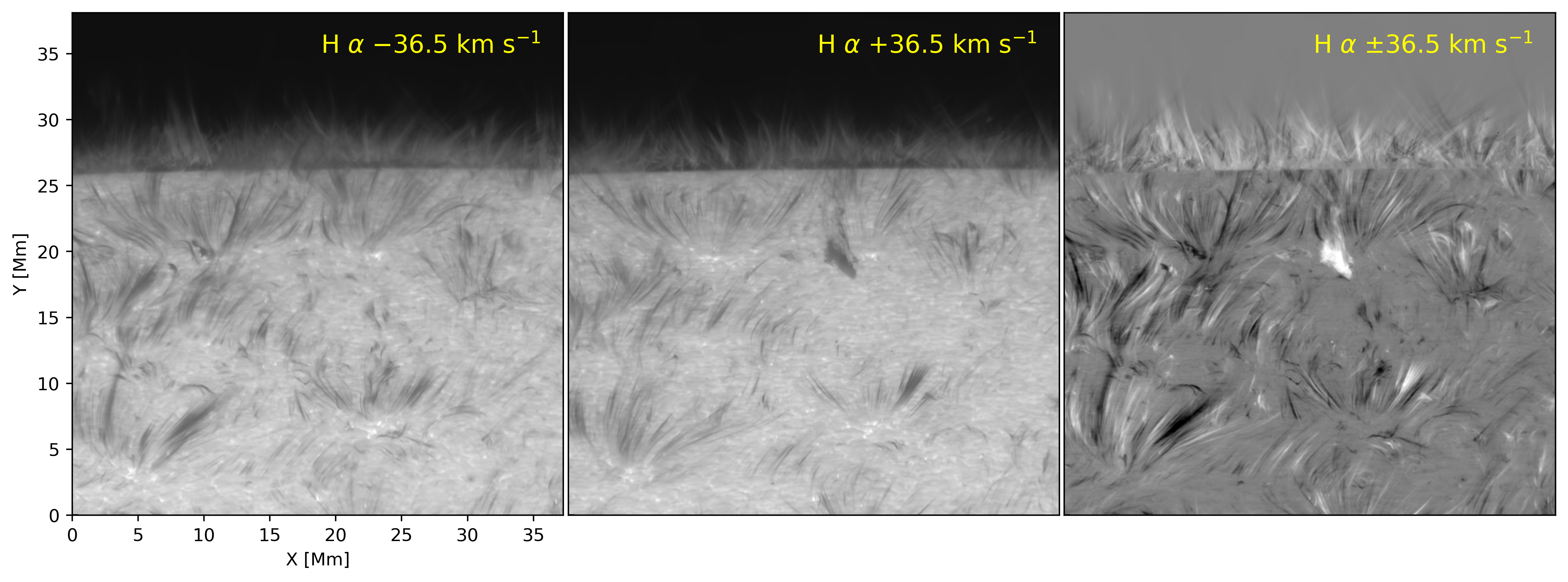}
    \end{adjustbox}
    \caption{Observation of spicules and their associated twists in \halpha{}. The left and middle panels respectively show spicules in \halpha{} blue and red wing images at Doppler offsets of $-$36.5 ($-$0.8~\AA) and $+$36.5~\kms ($+$0.8~\AA) with respect to the line center. The right panel show the corresponding Dopplergram (refer to text for details) at $\pm$36.5~\kms\ constructed from the left and middle panels. The existence of adjacent bright and dark threaded structures illustrate twists associated with spicules. Data reduction by Luc Rouppe van der Voort.}
    \label{figure_ch-3:Dopplergram_SST_spicule}
\end{figure}

Studies in the second half of the twentieth century led by \cite{1967SvA....10..570L} and \cite{1968SoPh....5..131P} provided indications that spicules could undergo rotational motions around their axis. Later with Hinode observations, \cite{2008ASPC..397...27S} provided stronger evidences that spicules can not only show rotational oscillations around their axis, but can also be multi-threaded. However, this aspect of torsional (or twisting) motions associated with spicules were not definitive until the study by \cite{Bart_3_motions}. With the help of high-resolution SST observations the above set of authors concluded that spicules exhibit ubiquitous twisting motions in addition to the radial field aligned upflows and swaying motions. They further interpreted that the appearance of the so-called spicular bushes in the both red and blue wings of \ion{Ca}{ii}~854.2 and \halpha{} images could be due to a combination of all the three kinds of motions as described above. These results were immediately expanded by \cite{2013ApJ...769...44S} which led to the first systematic observation of the red wing counterparts of RBEs, termed as rapid redshifted excursions (RREs), that were also confirmed later by \cite{2015ApJ...802...26K}. 
%The advent of IRIS observations revealed that the twists associated with spicules are far more common than previously thought and they appear ubiquitously in the chromosphere and the TR \citep{2014Sci...346D.315D}. 

Dopplergrams, constructed by subtracting red wing images from blue wing images that are symmetrical with respect to the line center, is an efficient way to visualize twists associated with spicules. Essentially, a Dopplergram captures features that have strong Doppler shifts associated with them, such as in spicules \citep{2013ApJ...769...44S,2016ApJ...824...65P}. \cref{figure_ch-3:Dopplergram_SST_spicule} shows an \halpha{} Dopplergram at $\pm$36.5~\kms\ corresponding to the same time as \cref{figure_ch-3:RBEs_SST}. RREs appear as bright streaks on-disk while RBEs appear darker. The scenario is opposite for off-limb spicules, since \halpha{} is an absorption line on-disk while it appears as an emission line off the limb. A glance at \cref{figure_ch-3:Dopplergram_SST_spicule} immediately suggests that the chromosphere is replete with spicules, with many appearing adjacent to each other. The presence of adjacent bright-dark features, such as the one seen close to to ($X$,$Y$) =(26,10) in \cref{figure_ch-3:Dopplergram_SST_spicule}, implies the existence of torsional motions in spicules. Also in some cases, RBEs evolve into RREs and vice-versa suggesting the presence of propagating torsional motions that can have a phase speed between 100--300~\kms \citep{Bart_3_motions}. With coordinated SST and IRIS observations \cite{2014Sci...346D.315D} showed that twists (associated predominantly with spicules) are ubiquitous in the solar atmosphere, regardless of the magnetic activity, and they are seen even in the upper chromospheric and TR channels which suggesting rapid heating. As a result, it is speculated that the non-thermal energy required to balance the losses suffered in the chromosphere and the TR could be attributed to these ubiquitous twists in the solar atmosphere. 

\section{Spectral signatures of RBEs and RREs in \cak{}}
\label{section:CHROMIS_spicule_spectra}

\begin{figure}[!p]
    \centering
    \begin{adjustbox}{minipage=\linewidth,scale=1.}
    \includegraphics[width=\textwidth]{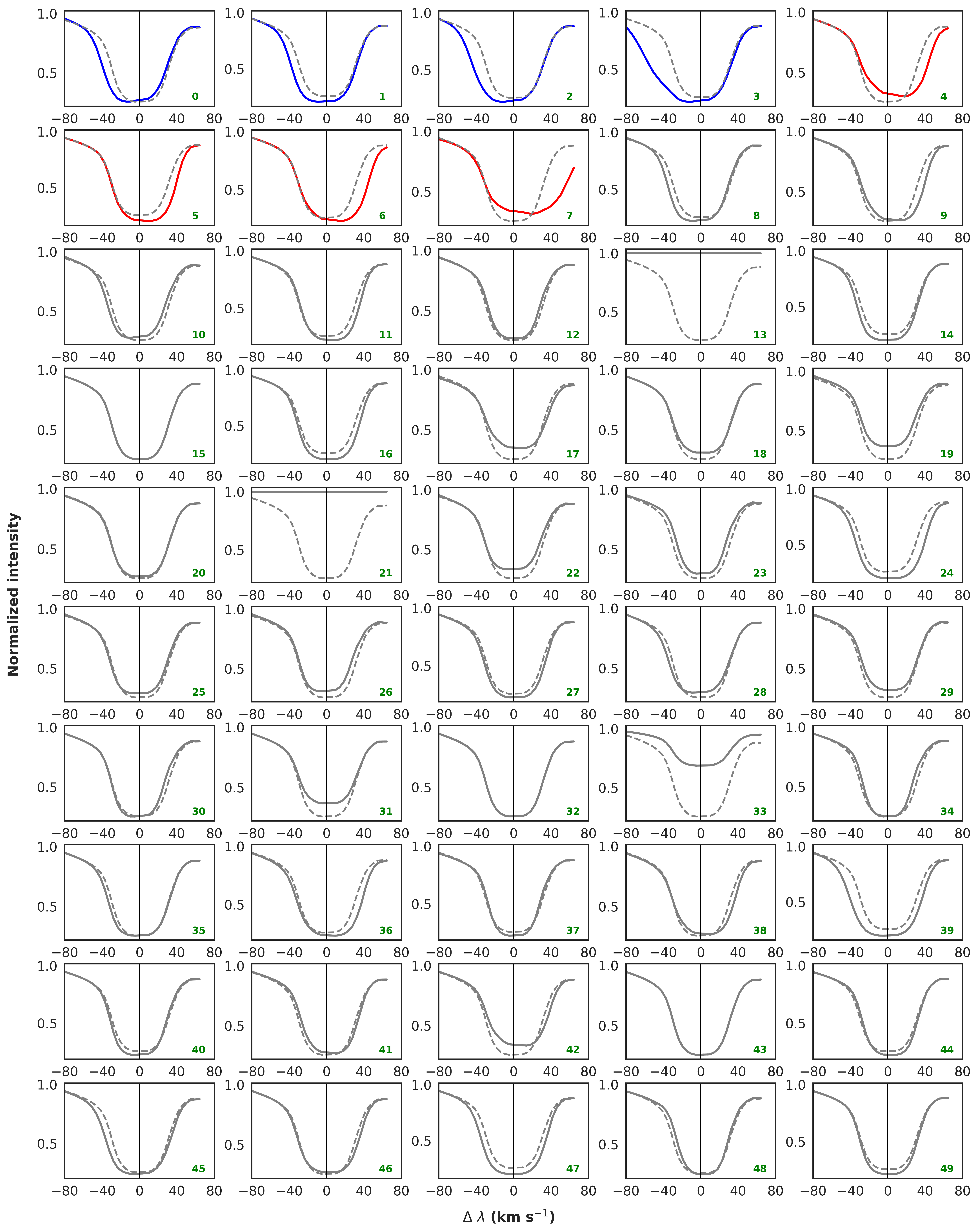}
    \end{adjustbox}
    \caption{Set of 50 \halpha{} RPs resulting from the $k$-means clustering algorithm from dataset~1, and presented in Paper~II. The RP index is indicated in the bottom right corner in each plot. RPs 0--3 show absorption asymmetries in the blue wing of \halpha{} and are considered as RBE-like spectral profiles (shown in blue color), whereas RPs 4--7 show absorption asymmetries in the red wing of \halpha{} and correspond to RRE-like spectral profiles (shown in red color). Rest of the RPs (8--49) are shown in gray. The dashed-gray \halpha{} profile in each panel represents the spatial and temporal average spectral profile over the whole dataset. The solid vertical line in each panel indicates the position with a Doppler offset of 0~\kms. }
    \label{figure_ch-3:RP_H_alpha}
\end{figure}

The spectral signatures of on-disk counterparts of type-II spicules have been identified in the IRIS \ion{Mg}{ii}~h and k resonance lines for the first time by \cite{Luc_2015}. The installation of the CHROMIS instrument at the SST in 2016 enabled the possibility of observing the solar atmosphere at an unprecedented spatial-resolution that is better than 70~km at wavelengths shorter than 400~nm. Observations at such spatial resolution have the potential to unravel spicule properties at even smaller temporal and spatial scales that eventually could lead to a better understanding of the physical processes at play. Moreover, it could provide better observational constraints which in-turn could lead to better numerical modeling efforts. Paper~I \citep{2019A&A...631L...5B} of this thesis, provides the very first characterization of RBEs and RREs in \cak{}.
%with the help of the robust $k$-means clustering algorithm. 

As discussed in Paper~I, identifying RBE/RRE spectral signatures in \cak{} is not a straightforward task since chromospheric images appear more diffuse in \cak{} in comparison to \halpha{} due to a lack of opacity gap between the line core and the wings in the former. This was also pointed out by \cite{1973SoPh...32..337B} when he compared off-limb spicule images observed in the K line with R.B.~Dunn's \halpha{} spicules. Therefore, a systematic spectral characterization was performed by applying the robust $k$-means clustering algorithm on the combined \halpha{} and \cak{} dataset (dataset~1 in \cref{chap:Instruments_simulations}), so that the relatively well-known spectral signatures of RBEs and RREs in \halpha{} could serve as a reference. 

The $k$-means method \citep{everitt_1972,macqueen1967some} has been extremely useful of late to characterize a variety of solar phenomena and observations such as flares, spicules and penumbral Ellerman bombs \citep[see,][for applications outside solar spicules]{2018ApJ...861...62P,2020A&A...641L...5J,2021A&A...648A..54R,2019ApJ...875L..18S}. The details behind this clustering algorithm have been described extensively in Paper~I, and also in \cref{section:appendix_kmeans} to this thesis. In this section, the focus is mainly on the results from the clustering algorithm, and spectra of RBEs and RREs in \cak{}. Basically, the $k$-means algorithm partitions the dataset into $k$ clusters and uniquely assigns each pixel in the co-aligned FOV of \halpha{} and \cak{} images to one representative profile (henceforth RP), that is the mean over millions of spectral profiles belonging to that cluster (see panel~(d) of \cref{figure_ch-3:Spicule_detection}). Analysis based on the mean of the standard deviation within each cluster suggested that dataset~1 could be optimally partitioned into 50 clusters, each with their own RP (more details in \cref{section:appendix_kmeans}). 

\begin{figure}[!p]
    \centering
    \begin{adjustbox}{minipage=\linewidth,scale=1.}
    \includegraphics[width=\textwidth]{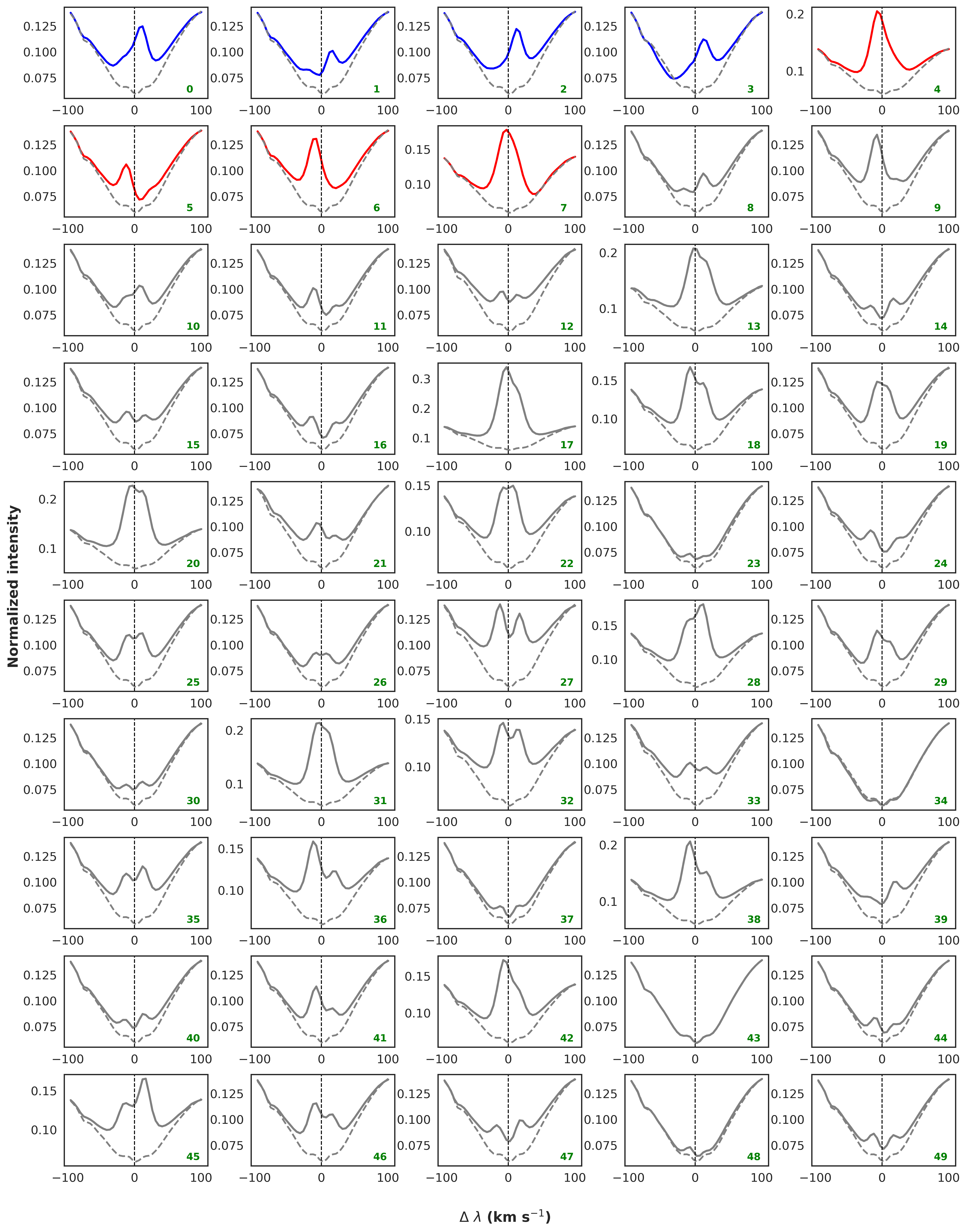}
    \end{adjustbox}
    \caption{Set of 50 \cak{} RPs corresponding to the \halpha{} RPs resulting from the $k$-means clustering algorithm shown in the same format as \cref{figure_ch-3:RP_H_alpha}. RBE-like RPs are indicated in blue whereas RRE-like RPs are indicated in red. Rest of the RPs are shown in gray.}
    \label{figure_ch-3:RP_Ca_K}
\end{figure}

\cref{figure_ch-3:RP_H_alpha} and \cref{figure_ch-3:RP_Ca_K} show the \halpha{} and \cak{} RPs corresponding to the 50 clusters obtained from the clustering algorithm. Since $k$-means was performed on the combined (and co-aligned) \halpha{} and \cak{} data, a given RP index is attributed to the same feature in the data. In other words, RP~0 in \halpha{} (for example) corresponds to the same feature in the solar atmosphere as RP~0 in \cak{}. I attribute RPs~0--3 (RPs~4--7) to RBEs (RREs) in the two figures. This attribution is motivated based on previous studies \citep[such as,][]{Luc_2009,Sekse_2012,2015ApJ...802...26K} where RBEs (RREs) were primarily identified based on strong absorption asymmetries in the blue (red) wings of the \halpha{} spectral line. Further details that forms the basis of the selection of the above RPs is briefly discussed in \cref{section:spicule_detection}, and a more thorough description is given in Paper~II. The remainder of the RPs, i.e. 8--49, belong to features that are neither RBEs nor RREs. 

A close look at the RBE RPs in the two figures suggest that Doppler shifts associated with the calcium K$_{3}$ cores \citep[I follow the standard nomenclature associated with \cak{} as illustrated in][]{1991SoPh..134...15R} is in tandem with the blueward excursion asymmetries in the \halpha{} RPs. This has also been shown in Paper~I with the help of \cak{} and \halpha{} $\lambda t$ diagrams for one typical example of an RBE. We also notice that, the blue K$_{2\mathrm{v}}$ peak is either reduced or completely suppressed by the Doppler shift of K$_{3}$. A similar but opposite shift (towards the redward side) is seen for the RRE RPs, where the red K$_{2\mathrm{r}}$ is affected in the same way. This causes an asymmetry in the \cak{} spectral line suggesting the presence of velocity gradients along the LOS. Moreover, the K$_{2}$ peak that lies in the wavelengths opposite to the flow show an enhancement in their intensity with respect to the average background because of the differential removal of upper layer opacities. This in turn leads to an increased contribution from the deeper layers of the solar atmosphere which causes the observed enhancement of the (opposite) K$_{2}$ peak. Based on these observations a systematic characterization of the \cak{} spicule spectra was performed in Paper~I, which led to the conclusion that the \cak{} line formation in spicules is dominated by opacity shifts with strong gradients in the LOS velocity, and the Doppler shift of the line minimum (K$_{3}$) is a true representative of the mass velocity associated with spicules. 

It is to be noted here that similar arguments in favor of Doppler removal of upper layer opacities have also been made by \cite{1997ApJ...481..500C} to explain the formation of internetwork bright grains observed in \ion{Ca}{ii}~H and K, \cite{2015ApJ...810..145D} to explain the source function enhancement in the red wing of \ion{Ca}{ii}~854.2 due to a granular-sized flux emergence, by \cite{2017ApJ...845..102H} and \cite{2019A&A...627A..46B} to explain the enhanced emission in the blue wing of \ion{Ca}{ii}~854.2 spectrum and \ion{Mg}{ii}~k$_{2\mathrm{v}}$ peak as a consequence of downflows in a flashing umbra.

\section{Spectral signatures of RBEs and RREs in \Mgk{} and \Si{} }
\label{section:IRIS_spicule_spectra}

\begin{figure}[!h]
    \centering
    \begin{adjustbox}{minipage=\linewidth,scale=1.}
    \includegraphics[width=\textwidth]{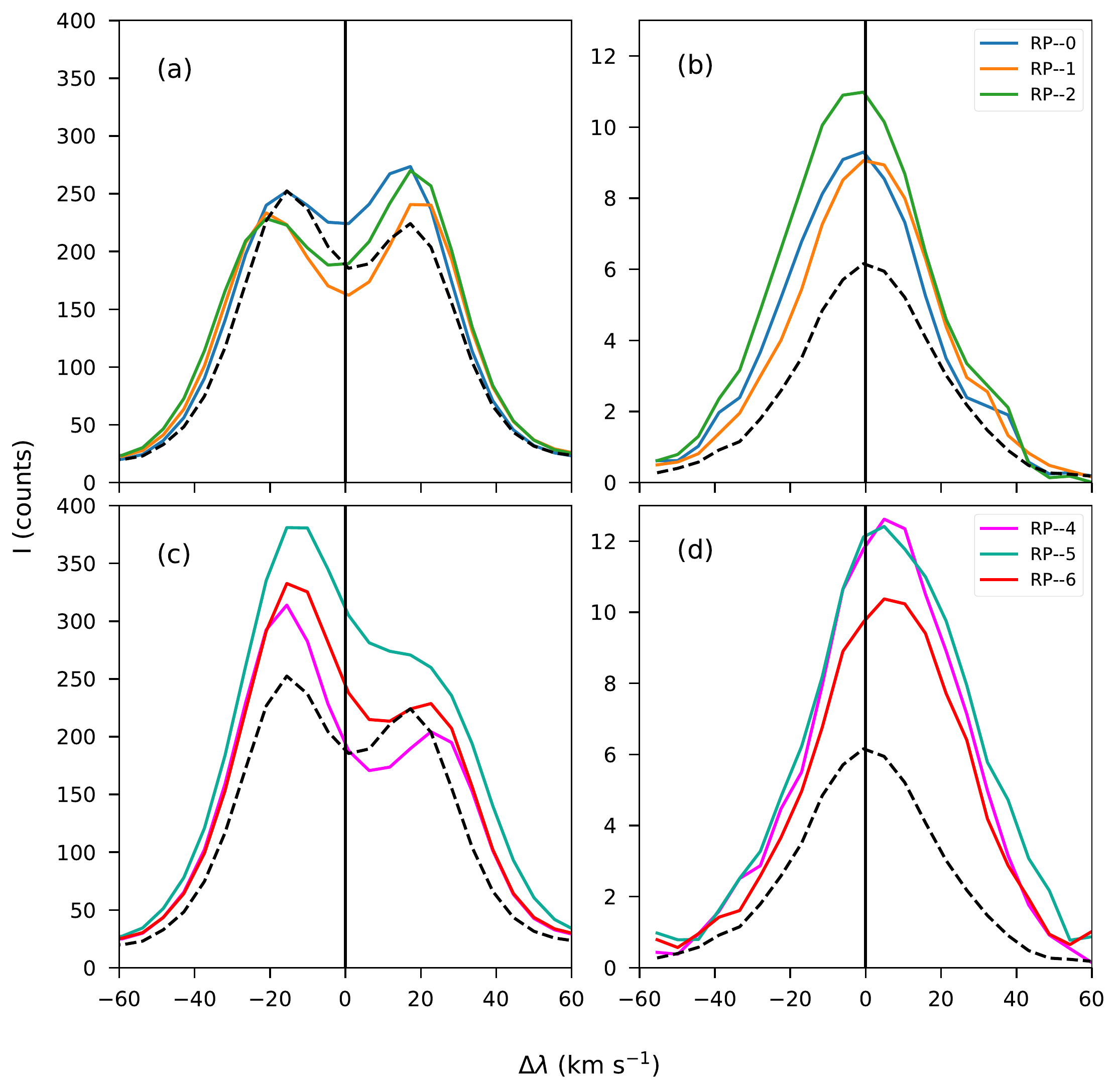}
    \end{adjustbox}
    \caption{Upper chromospheric and TR RPs associated with rapid blue and redshifted excursions (RBEs and RREs). Top row shows (RBE) RPs~0--2, while the bottom row shows (RRE) RPs~4--6 in IRIS \Mgk{} (panel~(a) and (c)) and \Si{}~139.4~nm (panel~(b) and (d)) spectral line. The dashed spectral line in each panel corresponds to the \Mgk{} and \Si{}~139.4~nm spectra obtained after averaging the spectral spatially and temporally. The solid black vertical line denotes the location corresponding to a Doppler offset of 0~\kms. }
    \label{figure_ch-3:RP_Mg_Si}
\end{figure}

Signatures of RBEs and RREs in \ion{Mg}{ii}, \ion{C}{ii}, and \Si{} spectra have been reported in the past with IRIS observations by \cite{2014Sci...346A.315T}, and coordinated SST and IRIS observations by \cite{Luc_2015}, with the former focusing primarily on the counterparts of on-disk network jets using SJIs. However, the above studies were limited to a handful of RBE/RRE examples and a systematic characterization over hundreds of thousands of spicule spectra, in a way similar to \cak{}, was found to be lacking in the IRIS spectra. The analysis presented in Paper~I provides a fertile ground in this direction since IRIS co-observed the same target as the SST. However due to the limited spatial coverage of the IRIS raster, its spectra were not included while performing the $k$-means clustering analysis. Rather, I relied on the clusters overlapping the raster FOV, and the RPs corresponding to the IRIS spectra were obtained by spatially and temporally averaging the profiles in each cluster separately for both \Mgk{} and \Si{}. As a result, I found that the RPs corresponding to the strongest K$_{3}$ absorption in both RBEs and RREs (i.e. RPs 3 and 7) were absent in the IRIS data owing to the limited spatial coverage of the narrow rasters. Nonetheless, a comparison of the remaining RBE and RRE RPs with \cak{} spectra revealed that the \Mgk{} line formation in spicules is also affected by differential Doppler removal of the upper layer opacities in the same way as in \cak{}, and the Doppler shift of k$_{3}$ reduces or suppresses the k$_{2}$ peaks in the direction of the flow (though not as distinctly as in \cak{} possibly due to its higher formation height, relatively broader spectral profile and lower resolution of the IRIS data, further aggravated by 2-pixel binning). This is rather unsurprising since both \cak{} and \Mgk{} are chromospheric resonance lines that have similar formation mechanisms \citep{2013ApJ...772...90L,2018A&A...611A..62B}, with the latter forming relatively higher up in the chromosphere owing to higher abundance of Mg. The above interpretation is also well correlated with the spectral signatures observed by \cite{Luc_2015}. Panels~(a) and (c) in \cref{figure_ch-3:RP_Mg_Si} shows the RPs corresponding to the \Mgk{} RBE and RRE spectra respectively\footnote{The \Mgk{} RPs has also been shown in Paper~I.}, and a close look immediately reveals that the RRE RPs (particularly the k$_{2}$ peaks) are enhanced in comparison to the RBE RPs. Since the intensity of the k$_{2}$ emission peaks is found to be well correlated with the temperature at their formation heights \citep{2013ApJ...772...90L}, an enhancement in their specific intensity suggests the possibility of heating based mechanisms that could be prevalent among RREs. This aspect has been studied elaborately with the help of optically thick radiative transfer diagnostics in Paper~III. 

The FUV \Si{} TR spectral counterpart of RBEs and RREs can be quite challenging to find in the IRIS rasters due to relatively low signal-to-noise in comparison to \ion{Mg}{ii}. In this thesis introduction, I extend the analysis performed in Paper~I, and obtain the \Si{}~139.4~nm RPs corresponding to the clusters overlapping the IRIS raster in a way similar to \Mgk{}. Panels~(b) and (d) of \cref{figure_ch-3:RP_Mg_Si} show the \Si{} RPs corresponding to RBEs and RREs, respectively. We see a clear signal in \Si{} spectra that is well beyond the noise level and also note that the shift of the emission peak is in concert with the Doppler shift of k$_{3}$ in \Mgk{} seen in panels~(a) and (c). This is complementary to the examples of RBE and RRE spectra in \Si{} obtained by \cite{Luc_2015}. Moreover, we also find that the \Si{} RPs are broadened in comparison to the average \Si{} spectra (that is obtained by spatial and temporal average over the whole time series), which was also reported in the context of TR network jets by \cite{2014Sci...346A.315T}. The spectral signatures shown in \cref{figure_ch-3:RP_Mg_Si}, together with the robustness of the $k$-means clustering algorithm, unquestionably confirms that the on-disk counterparts of type-II spicules can be heated to at least TR temperatures (80,000~K under equilibrium ionization conditions).

%----------Maybe MENTION THE HEATING SIGNATURES ASSOCIATED WITH RBES AND RRES IN THE PAST LITERATURE?

\section{Detection of on-disk spicules}
\label{section:spicule_detection}
The analysis and discussions presented in the preceding sections (i.e. \cref{section:CHROMIS_spicule_spectra} and \cref{section:IRIS_spicule_spectra}) provide a basis for identifying the spectra corresponding to on-disk spicules. However, detection of spicules in an automated fashion requires a few more steps that are briefly discussed in the following paragraph. As discussed earlier, spicules can show complex dynamic behavior owing to their field aligned flows, torsional and swaying motions \citep{Bart_3_motions,2013ApJ...769...44S}. Moreover, they also exhibit variations in the Doppler offsets along their length and breadth as they evolve. This in turn affects the spectral profiles which can show an enhancement in its line widths \citep[e.g. in \halpha{}][]{2016ApJ...824...65P}. From the perspective of RPs, this would mean that an RBE (or an RRE) can be associated with multiple spectral profiles with different Doppler offsets that would contribute to their overall morphpology. These variations in the Doppler offsets along the length and breadth of spicules are of crucial importance while developing any automated detection procedure. If ignored, it would lead to an underestimation of the number of successful detections that could further lead to a bias in the derived statistical properties. Moreover, inclusion of multiple RPs also facilitates the detection of spicules which evolve in groups. Evolution of spicules in groups is a commonly observed behavior since the observations of \cite{1957SCoA....2...15L}, who described \textit{porcupine} and \textit{wheat field} like patterns associated with them. Later \cite{1968SoPh....3..367B} suggested that both porcupines and wheat fields grow and disappear as one unit which is a characteristic of the commonly observed group behavior in spicules.

\begin{figure}[!ht]
    \centering
    \begin{adjustbox}{minipage=\linewidth,scale=1.0}
    \includegraphics[width=\textwidth]{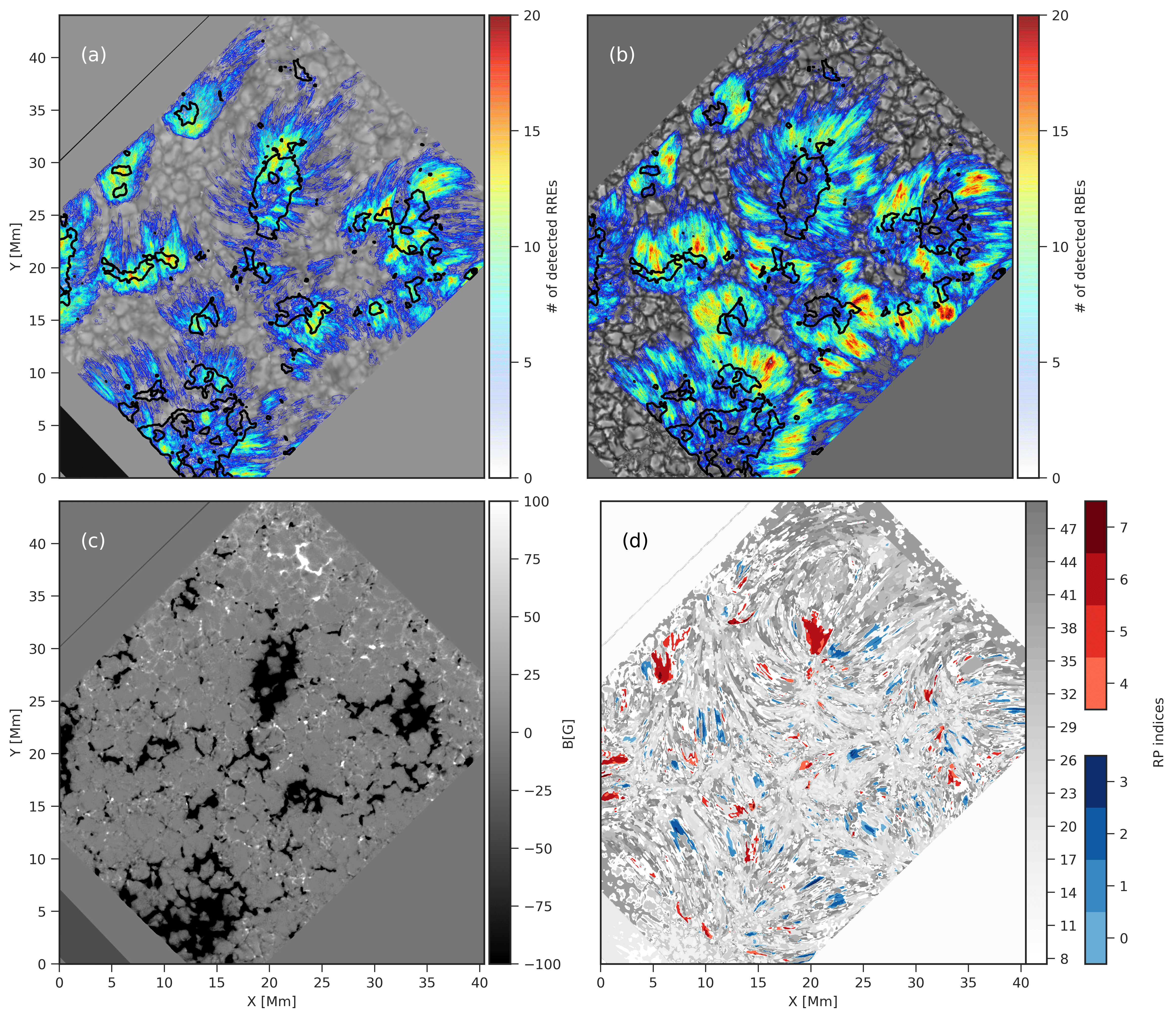}
    \end{adjustbox}
    \caption{Overview of the detected on-disk spicules in dataset~1. (a) \halpha{} red-wing image at $+$96~\kms{} ($+$2.1~\AA\ from the line center) with an overplotted density distribution of the detected downflowing RREs/RREs. (b) CHROMIS continuum image at 400~nm with an overplotted density distribution of the detected RBEs. Both density distributions are shown in a common rainbow color scale. (c) LOS magnetogram dervied from the Milne-Eddington inversions of the \ion{Fe}{i}~630.2~nm line clipped at $\pm$100~G. (d) RP index map corresponding to a scan at given time step derived from the application of the $k$-means clustering algorithm. RBE and RRE/downflowing RRE RPs are indicated in shades of blue (RPs~0--3) and red (RPs~4--7) respectively, with darker shades indicating higher values of the associated Doppler offsets, line core widths and absorption measure. RPs 8--49 are indicated in shades of gray.  }
    \label{figure_ch-3:Spicule_detection}
\end{figure}

Paper~II \citep{2021A&A...647A.147B} of this thesis, provides a comprehensive method of detecting on-disk spicules in \halpha{} that not only exploits the full spectral information, but also accounts for the variation in the Doppler offsets associated with spicules by including multiple RPs (i.e. RPs~0--3 for RBEs and 4--7 for RREs) in the detection. Moreover, this is followed by an application of morphological image processing operations that help in connecting spicular events and their associated 'halos' (structures with weaker Doppler offsets surrounding the centrally stronger offset regions shown in Figure 5 in Paper~II) spread over multiple spatial and temporal dimensions. The method of detection has been described extensively in Paper~II, and also in \cref{section:appendix_morph_process} to this thesis. For the sake of completeness, however, I briefly describe the criterion for selecting the RBE and RRE \halpha{} RPs below.

The Doppler offsets corresponding to each of the \halpha{} RPs were first determined by computing the differential profiles (difference between the average \halpha{} profile and the RPs), followed by determining the center-of-gravity (COG) of the positive part of all such differential profiles. The position of COG served as a representation of the Doppler offsets corresponding to each RP. The maximum value of the differential profiles was used as a measure of the absorption feature of the RPs that has a higher value for strong excursion asymmetries. Finally, the line-core width of each of the \halpha{} RPs was determined by following the method of \cite{2009A&A...503..577C}. Since RBEs and RREs generally show strong excursion in the wings of the \halpha{} spectral line, it leads to an enhancement in all of the above three parameters. As a result, only those RPs which had the values of the above parameters above a certain threshold (see Paper~II), were selected for further analysis involving morphological operations as described in \cref{section:appendix_morph_process}. Consequently, RPs~0--3 were assigned to RBEs and 4--7 were assigned to RREs in increasing order of their spectral strength. Roughly 15,000 RREs and 20,000 RBEs were detected in dataset~1 and an overview of their location along with the density of their occurrences in the FOV are shown in panels~(a) and (b) of \cref{figure_ch-3:Spicule_detection}. Panel~(c) shows the LOS magnetogram derived from the ME inversion of the \ion{Fe}{i}~630.2~nm spectral line and it serves as a context to indicate the overall photospheric magnetic field topology of the enhanced network target and most of the observed spicular activity were found to lie in the close vicinity of the strong field regions.  Panel~(d) shows the so-called RP index map derived from the $k$-means clustering (see \cref{appendix:k-means}) at a given time step where each pixel on the FOV is assigned to a particular cluster that constitutes an RP.
%To enable better visualization the RBE and RRE RPs are indicated in shades of blue and red with darker shades indicating higher values of the associated Doppler offsets, line core widths and absorption measure. The rest of the RPs are indicated in shades of gray. 

\section{Driving mechanisms}
\label{section:driving_mechanism}
A generation mechanism that could successfully explain \textit{all} the observed properties of spicules still remains elusive. Although type-I spicules show strong evidence for leakage of photospheric oscillations both in numerical simulations and observations \citep{2004Natur.430..536D,2006ApJ...647L..73H,2007ApJ...655..624D}, a convincing mechanism responsible for driving the more energetic and faster type-IIs is highly contested. Several high-quality ground and space-based observations, such as the ones described earlier in this chapter, provided effective constraints to the numerical simulations regarding the most common drivers of type-II spicules. Apart from the shock driven scenario stated above, spicules are also thought to be driven by several other mechanisms such as \alfvenic{} waves \citep{2007Sci...318.1574D,2014Sci...346D.315D,2015ApJ...812...71C,2017ApJ...848...38I}, magnetic reconnection \citep{2011Sci...331...55D,2011ApJ...731L..18M,2016ApJ...828L...9S,2019Sci...366..890S,2020ApJ...893L..45S}, and release of amplified magnetic tension through ambipolar diffusion \citep{2017Sci...356.1269M}, to name a few. However, it is still unclear as to which among the above is the most common driver. 

One of the most challenging aspects from the perspective of numerical simulations is to model the observed ubiquity of type-IIs. In this regard, the studies of ion-neutral interaction effects by \cite{2017Sci...356.1269M} provided a quantum leap in terms of our understanding of the formation of type-II spicules. According to the above set of authors, the interaction between strong vertical magnetic field regions, that exists typically at the edges of plages or internetworks, and the relatively weak (more horizontal) small-scaled fields present in the granules leads to a build up of tension in the magnetic field. Now, ambipolar diffusion causes these regions of high magnetic tension to diffuse into the higher layers of the solar atmosphere, followed by a violent release in the magnetic tension that essentially drives plasma at high speeds (50-100~\kms). A comparison of the synthetic observables derived from this numerical simulation with high-resolution SST and IRIS observations revealed a decent match with each other, suggesting that the jets or flows so observed in the model are representatives of type-II spicules. The above mechanism appears to be a promising driver of type-II spicules but further studies, focusing on the relationship between the small-scaled granular fields and production of RBEs/RREs (for on-disk observations), are necessary to unambiguously establish this idea \citep{2019ARA&A..57..189C}.  

Flux cancellations between small scale emerging flux and ambient (dominant) magnetic fields of opposite polarities leading to enhanced spicular activity, as suggested by \cite{2019Sci...366..890S}, also provides an interesting alternate explanation regarding the generation mechanism of type-II spicules. However, since their observations were limited to only far wing images in \halpha{} ($\pm$0.08~nm with respect to the line center), it is rather non-trivial to say whether flux cancellations led to the generation of those spicules or caused the already existing spicules (in images closer to the line core of \halpha{}) to appear in the far wing positions via an increase in their Doppler shifts. Moreover, observations closer to the line core of \halpha{} reveal an increase in the number density of spicules \citep{1972ARA&A..10...73B,2016ApJ...824...65P,2019A&A...631L...5B,2021A&A...647A.147B}. Therefore, a correlation between the flux cancellations and the appearance of weakly Doppler shifted spicules would further strengthen the argument of \cite{2019Sci...366..890S}. Nonetheless, their study offers an exciting prospect on reconnection driven mechanism that can constrain future numerical simulations. Formation of spicules due to vorticity driven shock waves, as hypothesized by \cite{2017ApJ...848...38I}, is also another possible driver since torsional motions are commonly observed in spicules \cite{Bart_3_motions,2014Sci...346D.315D}. However, observing small scaled photospheric vorticities with current instrumentation capabilities is rather demanding, and relating them to spicules (that are observed in the chromosphere and beyond) is a very challenging task that needs a careful accounting of the delay necessary for the propagation of these (\alfvenic{}) waves from the photosphere to the chromosphere. Promising results in this direction have recently been obtained by \cite{2019NatCo..10.3504L,2021A&A...645A...3Y} for example, where there are evidences of energy transport from the photosphere to chromosphere via ubiquitous \alfvenic{} waves. In this thesis however, I do not discuss or attempt to speculate the major driver of type-II spicules (RBEs/RREs) in our datasets, rather the focus is on the synthetic observables derived from the numerical simulation of \cite{2017Sci...356.1269M} via radiative transfer, to compare the properties of simulated spicules with actual observations. This is discussed further in Paper~III and briefly in \cref{subsection:TR_redshifts}.

\section{Mass flows associated with spicules}
\label{sectiom:mass_flows}

The idea that spicules may be responsible for driving the solar wind and may be an important aspect that could explain coronal heating has been prevalent since the studies of \cite{1977A&A....55..305P}, \cite{1982ApJ...255..743A,1984ApJ...287..412A}. As discussed earlier in this chapter, with the discovery of type-II spicules with Hinode and follow-up investigations by IRIS, their role in heating and mass loading the upper atmospheres of the Sun have sparked massive interest in the solar physics community.

\cite{2009ApJ...701L...1D} suggested that the footpoints of the million-degree coronal loops were correlated with strong ubiquitous plasma upflows (blueshifts of the order of 50-80~\kms) in the form of chromospheric jets, regardless of the magnetic activity. Their analysis showed that even if 1--5\% of the chromospheric jets (which are likely type-II spicules as per their velocities) are heated to coronal temperatures, it would be sufficient for them to provide hot plasma necessary to fill the corona. \cite{2011Sci...331...55D} further supported this proposition by linking the RBEs found in close vicinity of active regions, observed with the Hinode's SOT \halpha{}, with transient brightenings in the SDO/AIA channels. Statistical estimates by \cite{Vasco_2016} suggested that RBEs and RREs seen in quiet Sun \halpha{} observations are also associated with brightenings in AIA 30.4 and 17.1~nm channels. Recently, \cite{2019Sci...366..890S} also found evidences of brightenings in AIA 17.1~nm channel in tandem with RBEs observed in the chromosphere with \halpha{}, suggesting that many of these ubiquitous events could be heated to at least coronal temperatures. With imaging and spectroscopic IRIS observations, \cite{2014Sci...346A.315T} reported the presence of high-speed (100--300~\kms) intermittent network jets that are heated to at least 10$^{5}$~K. Many of these network jets were confirmed to be related to the heated signatures of on-disk RBEs and RREs by \cite{Luc_2015}, while some may be associated with small-scaled jetlets \citep{2014ApJ...787..118R,2018ApJ...868L..27P}. However, the high speed (apparent motions in excess of 100~\kms) associated with these network jets made it difficult to reconcile with the Doppler shifts associated with spicules that is generally found to be only between 30--70~\kms. Using the state-of-the-art 2.5D MHD spicule simulation of \cite{2017ApJ...847...36M,2017Sci...356.1269M} and IRIS observations, \cite{2017ApJ...849L...7D} explained that the discrepancy between actual mass motions (given by Doppler shifts) and the high-speed apparent motions could be attributed to rapidly propagating heating fronts that have much higher speeds than actual mass flows.

\subsection{Rapid downflows in the solar chromosphere}
\label{subsection:spicules_and_downflows}

\begin{figure}[!p]
    \centering
    \begin{adjustbox}{minipage=\linewidth,scale=1.}
    \includegraphics[width=\textwidth]{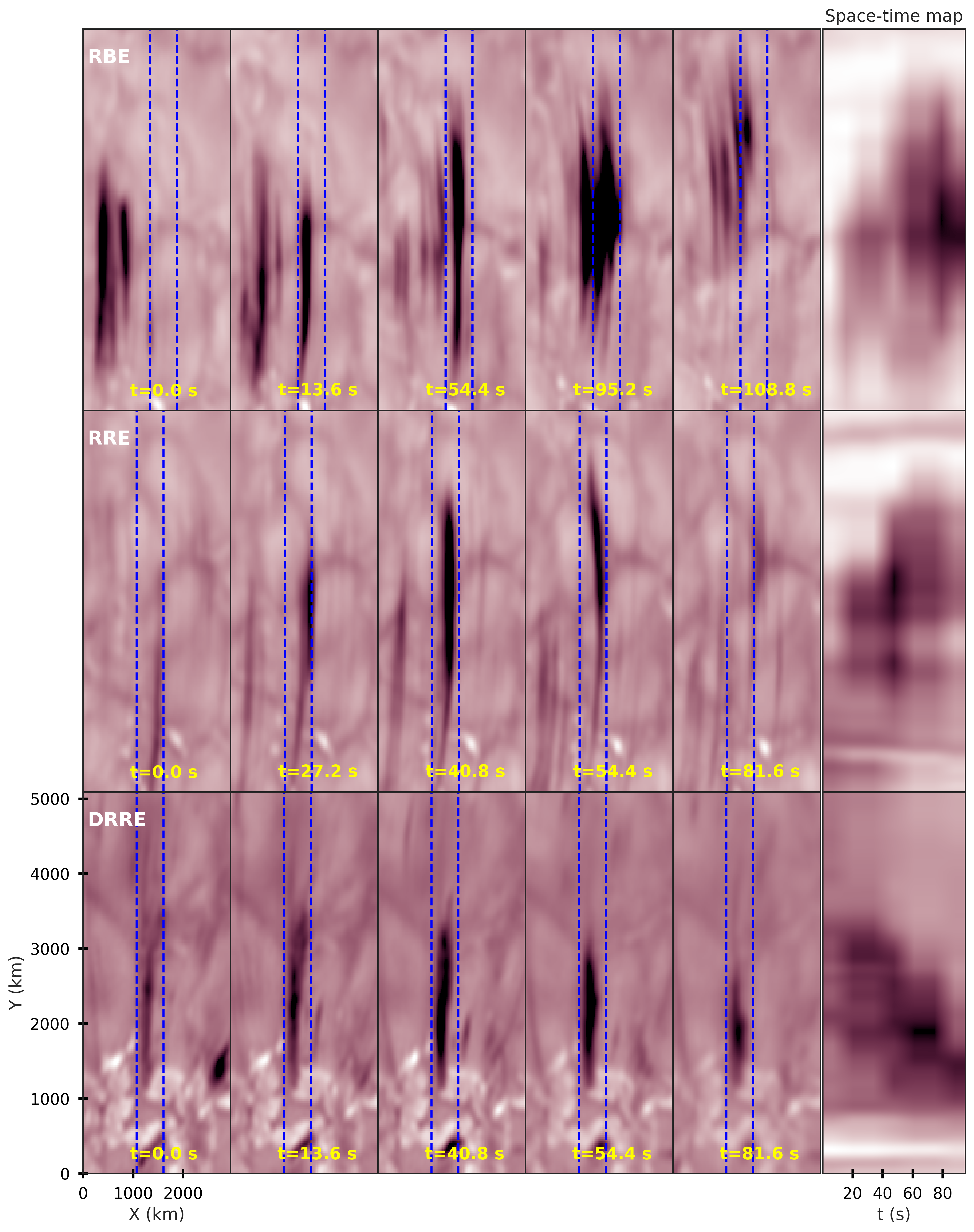}
    \end{adjustbox}
    \caption{Examples of on-disk counterparts of type-II spicules from dataset~1 (that have also been shown in Paper~II). The top and middle rows show an RBE and an RRE observed at \halpha{} blue and red wings at $-$40 and $+$40~\kms (at $\pm$0.87~\AA), respectively. The bottom row shows the newly reported downflowing rapid redshifted excursion (downflowing RRE) also in the red wing ($+$40~\kms or $+$0.87~\AA) of the \halpha{} spectral line. For each event, the temporal evolution is shown along the respective rows. The last column in each row shows the corresponding space-time maps (which highlight the apparent motion of the three features) obtained by averaging the region bounded between the dashed blue lines for each time step. Clearly, the downflowing RRE shows an apparent inward motion that is opposite to the RBE and RRE. }
    \label{figure_ch-3:DRRE_sample}
\end{figure}

Upflows in the solar atmosphere, both large and small-scaled, have been a part of numerous studies in the past \citep[see][for a recent review on this topic]{2021SoPh..296...47T} because of their potential role in driving the solar wind and heating the solar corona. However, they do not form a major theme of this thesis, where the focus is primarily on the study of downflows associated with spicules. The motivation stems from the observations presented in datasets~1 and 2 which have been described in Paper~II and Paper~III. Generally, animations of high-resolution blue and red wing \halpha{} images (at Doppler offsets $\geq$ 30~\kms{} or $\pm$~0.65~\AA\ from the line center) are replete with RBEs and RREs that have a radially outward motion with respect to strong field magnetic network regions. However, observations presented in this thesis \citep[refer to the animations associated with Figure~1 in][and in Paper~III]{2021A&A...647A.147B} suggest that corresponding red-wing \halpha{} images show absorption features that are rather elongated and, contrary to RREs, appear to move towards the network region. \cref{figure_ch-3:DRRE_sample} provides a clear visualization of the apparently downward moving red wing features in comparison to the traditionally\footnote{The word traditional has been used in this study to indicate typical outward moving RBEs and RREs.} observed RBEs and RREs. 

Paper~II offers a systematic insight into these downflowing features that appear morphologically similar to RREs and RBEs. Upon further investigation, the spectral characteristics of these downflowing features were found to be remarkably similar to traditionally observed RREs, with strong absorption in the red wings of \halpha{}. Moreover, their $\lambda t$ diagrams were also like traditional RREs which show sudden excursions in the redward side of the \halpha{} line core without any preceding blueward excursions. They were therefore termed as downflowing RREs. Consequently, the distinction between RREs and downflowing RREs was not possible while identifying the RPs via the $k$-means clustering process. However, a statistical analysis of the apparent motion of the detected spicular events (for RBEs and RREs/downflowing RREs) revealed that an overwhelmingly large number of redshifted events followed an apparent motion (with speeds ranging between 30--50~\kms{}) that were directed towards the strong network regions, contrary to the typical RBEs and RREs (as shown in the example in \cref{figure_ch-3:DRRE_sample}). Though it is difficult to provide an exact quantification of the percentage of downflowing RREs in comparison to RBEs and RREs, qualitatively their population seems to be comparable with the traditional excursion events as is evident from the apparent motion analysis described in Paper~II. In addition, the downflowing RREs were also found to be associated with multiple substructures owing to the wide range of Doppler offsets and \halpha{} line core widths, that are also commonly observed among RBEs and RREs. Being close to the disk-center and assuming that an overwhelming majority of spicules have a non-zero (but rarely beyond 50~$^{\circ}$) inclination with respect to the local surface normal \citep[see,][]{Tiago_2012}, the apparent speed of the downflows would be comparable with the Doppler offsets derived from spectroscopic measurements. Such a comparison for the downflowing RREs revealed that they represented actual mass flows, unlike the rapid heating fronts associated with TR network jets described in \cite{2014Sci...346A.315T}. It was therefore speculated that these ubiquitous downflows could well be the chromospheric counterparts of the TR downflows which formed the major subject of investigation in Paper~III.

\subsection{On the persistent redshifts in the TR}
\label{subsection:TR_redshifts}
A large number of observational studies dedicated to TR have suggested that the emission lines formed at temperatures (roughly) between 0.05~MK to \textasciitilde0.25~MK show persistent downflows (redshifts in the emission spectra) of the order of 10--15~\kms \citep{1976ApJ...205L.177D,1981ApJ...251L.115G,1982SoPh...77...77D,1997SoPh..175..349B,1999ApJ...522.1148P,1999ApJ...516..490P,2011A&A...534A..90D,2012ApJ...749...60M}. The cause of such downflows has been attributed to several aspects both from the perspective of numerical simulations and observations. \cite{1993ApJ...402..741H}, for example, proposed that nanoflares in the solar corona lead to downward propagating acoustic waves which might be responsible for the observed redshifts. \cite{1977A&A....55..305P} and \cite{1982ApJ...255..743A}, however, attributed these downflows to the returning components of the previously heated spicular material that underwent cooling -- a scenario that was also supported to a large extent by the observational studies of \cite{2012ApJ...749...60M}. Furthermore, in their numerical simulations \cite{2010ApJ...718.1070H}, naturally reproduced downflowing signatures in the synthetic observations of the TR as a consequence of bi-directional flows formed at the interface between the TR and corona, due to episodic heating of the chromospheric material to coronal temperatures. Despite such a large number of propositions, a definitive consensus explaining the cause of the downflows, has not emerged so far. 

Surprisingly, observations of the spectral lines formed in the solar chromosphere do not show a similar trend and lack evidence of average (and persistent) downflows. A possible explanation can be attributed to the enormously larger plasma density in the solar chromosphere in comparison to the TR which might cause a breakdown of these flows as soon as they reach chromospheric heights. Moreover, coordinated high-resolution observations of both the TR and chromosphere have not been available until recently. The rapid spicular downflows reported in \cite{2021A&A...647A.147B} (Paper~II) were speculated to be (the much sought after) chromospheric signatures of the TR downflows. 

In Paper~III, I revisit the topic on the downflows in the TR from a perspective of observations and numerical simulation of spicules.
Using two sets of coordinated SST, IRIS and SDO observations, it was possible to show that the downflowing RREs are ubiquitous in nature and many of them show clear signatures across a broad range of temperature, ranging from the cooler chromosphere to the hotter corona. Spatio-temporal analysis indicates that the apparent motion of these features are directed towards the strong magnetic field regions and are in tandem with each other across multiple wavelength channels, including the far red wings of the \halpha{} spectral line. Furthermore, the excursion asymmetries seen in the IRIS \Mgk{} and \Si{}~139.4~nm $\lambda t$ diagrams are in concert with the excursions in \halpha{} and \cak{}, indicating a clear multithermal nature. Additionally, the TR Doppler offsets associated with these spicular downflows imply that these flows could at least in part be responsible for the persistent redshifts observed in the TR, as speculated by \cite{1977A&A....55..305P}, \cite{1982ApJ...255..743A} and \cite{2012ApJ...749...60M}.

A theoretical support to the above observations was provided by performing radiative transfer computations on the 2.5D MHD simulation of spicules computed by \cite{2017Sci...356.1269M} (described in \cref{section:simulation_description}). A comparison of the synthetic chromospheric (\cak{} and \Mgk{}) and TR (\Si{}) spectral lines corresponding to type-II spicules in the simulation reveal a distinct match with the observed spectral profiles. Moreover, as speculated in Paper~I and \cref{section:IRIS_spicule_spectra}, we also find that the synthetic \cak{} and \Mgk{} spectral profiles corresponding to the RRE/downflowing RREs are enhanced in comparison to their upflowing counterparts suggesting the possibility of heating. Upon further investigation, I find that the increase in the synthetic intensities is correlated with a marked rise in temperature during the downflowing phase of the spicules owing primarily to the ambipolar diffusion mechanism. Paper~III is referred here for more details. Furthermore, it is also found that the cause of such ubiquitous downflows in the numerical simulation is attributed to either a typical downflow that is preceded by an upflowing type-II spicule or downflows along a loop-like structure. The examples presented in Paper~II and Paper~III lack signatures of preceding co-spatial upflows (reflected as blueward excursions in the $\lambda t$ diagrams of the different spectral lines) corresponding to the downflowing RREs, suggesting that many of the observed downflows are likely caused along loops as predicted by the numerical model. The lack of preceding blueward excursions convincingly separates these downflowing RREs from being associated with typical type-I spicules where we find characteristic saw-tooth pattern in their $\lambda t$ diagrams (see \cref{figure_ch-3:DF_sawtooth_SST}). 

The results from the analysis presented in this thesis, especially in Paper~III, provides strong evidence of TR and coronal signatures associated with the recently reported ubiquitous chromospheric downflowing RREs (in Paper~II). Furthermore, with the help of coordinated ground and space-based observations, it was possible to provide a direct connection between many spicules and the redshifts observed in the TR -- a scenario that was already speculated by early studies of \cite{1977A&A....55..305P} and later by \cite{2012ApJ...749...60M}, strongly indicating the presence of a mass-cycling process between the chromosphere and the corona. 

    \chapter{Summary of the papers}
\label{chap:paper_summary}

In \cref{chap:spicules}, I have attempted to provide an introduction to spicules and their associated mass flows with an intention to provide a suitable context under which the research carried out as a part of this thesis fits the whole scenario. This chapter provides a brief summary of the papers that make up this thesis. It is spread over three papers (Paper~I, Paper~II, and Paper~III) which focus on spicules, their characterization, associated mass flows and their impact on solar atmospheric heating. The SST CRISP and CHROMIS data have been used in all the three papers. In addition, Paper~I included the co-aligned IRIS data and Paper~III included co-aligned SDO AIA and IRIS data. The summary of the papers are discussed below. 

\section{Characterization and formation of on-disk spicules in \cak{} and \Mgk{} spectral lines} 
\label{section:summary_pap1}
Paper~I, \cite{2019A&A...631L...5B}, deals with characterizing spicules seen on-disk with coordinated high-resolution observations from CHROMIS, CRISP, and IRIS instruments. In particular, the analysis presented in this paper provided the first characterization of RBE and RRE spectra in \cak{} 393.4~nm. Traditionally, RBEs and RREs have only been observed and analyzed in \halpha{} far wing images \citep{Luc_2009,Sekse_2012,2015ApJ...802...26K,Vasco_2016}. This is because the opacity associated with RBEs and RREs causes them to have a sharp contrast with respect to the bright photospheric background, making them appear as distinct dark threads. The advent of CHROMIS \cak{} observations enabled the possibility of observing spicules at an unprecedented high spatial resolution that is better than 70~km on the Sun. However, unlike \halpha{}, the images in the far blue/red wings of \cak{} do not have a sharp contrast with respect to the background. This is because of the lack of an opacity gap in \cak{} due to the shallow slope of its wings in comparison to \halpha{} which causes the latter to become almost transparent to the chromosphere at any wavelength out of the line center \citep{2004ApJ...603L.129S,2012A&A...540A..86R}. As a result, identifying RBEs and RREs becomes a challenging task. We therefore used the $k$-means clustering algorithm on combined \halpha{} and \cak{} observations that led to a systematic clustering of over millions of spectral profiles (over the whole time series lasting 97~min) which were partitioned into 50 clusters that were described as representative profiles (RPs). Thereafter, the RPs corresponding to the \halpha{} RBE and RRE-like spectral profiles served as references to obtain the corresponding RPs for \cak{} spectral line. The IRIS \Mgk{} spectra were not included in the $k$-means clustering because of the limited spatial coverage of the rasters in comparison to CRISP and CHROMIS. Instead, the RPs in \Mgk{} were obtained by spatially and temporally averaging the \Mgk{} profiles that coincided with a particular cluster (from $k$-means).

Analysis of their spectral lines led to the conclusion that optically thick line formation in spicules is dominated by opacity shifts with strong gradients in the LOS velocity. More specifically, the shift of the \ion{Ca}{ii}~K$_{3}$ line core was found to be in tandem with the asymmetry of the \halpha{} line, strongly suggesting that the position of the Doppler shifted K$_{3}$ best represents the velocity of spicules. The K$_{2}$ emission features were found to be either suppressed by the Doppler shift of the K$_{3}$ core or enhanced due to increased contribution from the deeper layers owing to upper layer opacity removal. This was termed as an opacity window effect. The \Mgk{} spectra also showed a similar behavior. To test our understanding, we performed several line formation experiments where we computed the synthetic \cak{} and \Mgk{} profiles with a modified FAL-C atmosphere \citep[][with ad hoc modifications in the LOS velocity]{1993ApJ...406..319F}, that revealed a good match with the observations. 

\section{Spicules and downflows in the solar chromosphere} 
\label{section:summary_pap2}

Paper~II, \cite{2021A&A...647A.147B}, deals with the observation and characterization of abundant rapid spicular downflows in the solar chromosphere, particularly in \halpha{}. High-speed downflows have been observed in the upper atmospheres of the Sun, especially the TR, for many decades ever since the launch of NASA's \textit{Skylab} mission in the seventies \citep{1976ApJ...205L.177D,1981ApJ...251L.115G,1999ApJ...522.1148P,2011A&A...534A..90D}. However, their signatures in the deeper layers of the solar atmosphere have remained elusive. This could be attributed to the enormously high difference of the plasma density in the deeper layers of the solar atmosphere (such as the chromosphere) in comparison to TR, that might cause a breakdown in the downflows as soon as they reach chromospheric heights. In this paper, we targeted an enhanced network region (dataset~1 in \cref{chap:Instruments_simulations} and same as the one used in Paper~I), and report abundant occurrences of rapid spicular downflows in the solar chromosphere. 

We use the results directly from the $k$-means clustering analysis from Paper~I and classify \halpha{} and \cak{} spectral profiles that resembled on-disk spicules. We based our analysis on \halpha{} RPs since on-disk spicules show characteristic absorption asymmetries in the far wings of \halpha{}. Thereafter, to select the spectral profiles of interest, we computed the Doppler offsets corresponding to the different \halpha{} RPs and estimated the absorption in their wings by obtaining the difference between the RPs and average (over the entire FOV and time) \halpha{} profiles. Additionally, we also computed the line core widths of the \halpha{} RPs since RBEs and RREs are known to have enhanced \halpha{} line widths, in addition to strong Doppler offsets \citep{2016ApJ...824...65P}. Finally, only those RPs which satisfied a certain thresholding criterion of all the quantities extracted above, were selected for further processing. Once the spectral profiles were selected, we developed an automated detection method, based on morphological image processing techniques, to detect the downflowing spicular features along with traditionally observed RBEs and RREs. 

Such an analysis, led to a systematic detection of a new category of RREs (termed as downflowing RREs) that were associated with chromospheric field aligned \textit{downflows}. This was in direct contradiction to the traditionally observed RBEs and RREs that were basically a manifestation of the same phenomenon, and depending on the Doppler shift and transverse motion with respect to the LOS of the observer, they could either appear in the blue or red wing of \halpha{} images. As a result, the apparent motion associated with downflowing RREs are directed towards (inwards) the strong magnetic field regions which is in contrast with their traditional counterparts. However, statistical analysis performed on over 35,000 on-disk spicules revealed that the newly reported downflowing RREs have similar length, area, and lifetimes as those of the traditional RBEs and RREs, except for their opposite apparent motion. Analysis of their respective $\lambda t$ diagrams indicate that the downflowing RREs (like traditional RREs) have no preceding blue ward excursions that is typical to type-I spicules. Moreover, the LOS velocities associated with these downflowing features were well correlated with their apparent velocity, indicating that downflowing RREs were associated with actual mass flows. Finally, they were speculated to be the representatives of the chromospheric counterparts of the TR downflows. 

\section{Evidence of multithermal nature of spicular downflows.
} 
\label{section:summary_pap3}

Paper~III, \cite{2021arXiv210802153B}, directly follows the results of Paper~II and attempts to find TR and coronal evidence of the newly reported chromospheric downflowing RREs. The motivation behind this work stems from the studies of \cite{1977A&A....55..305P}, \cite{1982ApJ...255..743A}, and \cite{1984ApJ...287..412A}, which related the persistent redshifts in the TR to the returning spicular plasma that were previously heated and injected into the solar corona. However, lack of adequate observations and failure to reproduce these signatures in more advanced theoretical models, such as the one by \cite{1987ApJ...319..465M} raised serious skepticism regarding the contribution of spicules towards the observed downflows in the TR. It was not until the advent of high resolution space-based observations by Hinode and SDO, that the interest in the role of spicules in mass and energy transfer to the upper atmospheres was revived. With the help of space-based (E)UV observations, several studies like \cite{2007Sci...318.1574D}, \cite{2009ApJ...701L...1D}, \cite{2009ApJ...707..524M}, \cite{2011Sci...331...55D}, and \cite{2012ApJ...749...60M}, found strong evidences of the role of spicules in heating and mass-loading of upper solar atmospheres, i.e. the TR and the corona. \cite{2012ApJ...749...60M}, in particular, found evidences of spicular downflows in the cooler coronal AIA (13.1~nm) channel that were speculated to be the returning (previously heated) spicular plasma.

In this paper, we used two different sets of coordinated SST, IRIS and SDO data (referred to as dataset~1 and 2 in \cref{chap:Instruments_simulations}), to show that downflowing RREs are ubiquitous in the solar chromosphere and many of them are multithermal in nature with distinct emission signatures in the hotter TR and lower coronal passbands. The high resolution ground-based observations enabled us to unambiguously relate the TR downflows to spicules seen in the chromosphere, that was not possible in any of the earlier studies like \cite{2012ApJ...749...60M}. Furthermore, the Doppler offsets associated with these downflows in the TR (\ion{Si}{iv}) spectral line were found to be close to the average observed red shifts in the TR, which suggested that spicular downflows could (in part) explain the reason behind the persistent red shifts. 

To provide a strong theoretical support, we used the 2.5D MHD spicule simulation from \cite{2017ApJ...847...36M,2017Sci...356.1269M} (described in \cref{chap:Instruments_simulations}) and computed synthetic \cak{}, \Mgk{} and \Si{} spectra that successfully reproduced the observations, thereby reinforcing the conclusions from the preliminary studies on \cak{} and \Mgk{} line formation in spicules conducted in Paper~I. Moreover, it was also found that the returning phases of type-II spicules were heated in comparison to the upflowing stages that was further reflected in the form of enhanced emission in the peaks of the (observed and synthetic) chromospheric \cak{} and \Mgk{} spectral lines. Ambipolar diffusion was found to be a crucial ingredient that caused this heating during the downflowing stages, which was in-turn responsible for increasing the temperature by as much as 3000~K. This suggested a direct impact in heating of the solar chromosphere in association with downflowing RREs. Finally, we also proposed two mechanisms -- an upflow followed by a downflow along the spicule body and flows along a loop leading to one footpoint being downflowing, from the viewpoint of numerical simulation that could potentially explain the ubiquitous nature of these downflows. 

\section{Concluding remarks and future prospects}
\label{section:conclusions}

The enigmatic spicules have long been subjected to various observations from ground-based telescopes. With the advent of seeing-free, space-based, and high-resolution observations, their role of mass balance and heating the outer atmospheres of the Sun has sparked massive interest in the solar physics community over the past decade or so. These observations, in coordination with ground-based facilities, have provided significant constraints that lead to ever improving numerical models which could eventually explain their observed properties. However, finding a major driving mechanism that could possibly lead to generation of spicules and explain their subsequent behavior is still considered a very challenging task. This is primarily because complex physical processes, for example ion-neutral interaction effects and non-equilibrium ionization, along with the requirement of 3D geometry, appear to be at play among spicules (and in general in the chromosphere) that makes it challenging to include them with our current computing capabilities \citep{2020LRSP...17....3L}. Moreover, this is further compounded by the fact that the solar atmosphere is strongly coupled. Observing (multithermal) spicules with fine spatial (sub)structures that evolve simultaneously in very short time periods across multiple layers of the solar atmosphere can be a daunting task, even with our current instrumentation capabilities. 

This thesis does not delve into the detailed generation mechanism of spicules, but rather focuses on their on-disk manifestation including characterization in the different spectral lines observed in the chromosphere and TR, their dynamics, and associated mass flows in the chromosphere, TR and corona. With the help of the $k$-means clustering technique, it was possible to provide the first systematic characterization of type-II spicules simultaneously in \cak{}, \halpha{}, \Mgk{} and \Si{} spectra. A detailed, analysis of over 20 million spectral profiles indicated that the Doppler shift (between 20--50~\kms) of the \ion{Ca}{ii}~K$_{3}$ line core serves as a representative of the LOS velocity associated with on-disk spicules, which is found to be in tandem with earlier measurements. Additionally, the K$_{2}$ peaks in the opposite side of the Doppler shift were partially enhanced due to Doppler-driven opacity effects. An immediate application of such clustering in the future could be in the direction of performing single or multi-line spectral inversions of on-disk spicules with \cak{}, \Mgk{} or \ion{Ca}{ii}~854.2~nm RPs using the STockholm Inversion Code \citep[STIC,][for example]{2019A&A...623A..74D}. Since RPs are a mean over all the profiles in a cluster, inverting them would lead to a representative model atmosphere that could serve as an initial guess model for performing more detailed spectral inversions of individual spicules as they evolve. Such an approach (especially with multi-lines) could also provide essential constraints in determining the height stratification of temperature, mass-density and LOS velocity associated with an ensemble of spicules that have so far been limited to individual events \citep{2021ApJ...908..168K}.

The detection of on-disk spicules carried out with $k$-means clustering and morphological image processing leads to a novel approach that enabled the detection of nearly 35,000 on-disk spicules (roughly 20,000 RBEs and 15,000 RREs/downflowing RREs) in our \textasciitilde97~min long dataset. This is far more than all other studies \citep[such as,][]{Luc_2009,Sekse_2012,Tiago_2012,2015ApJ...802...26K,Vasco_2016} targeting spicules, and we attribute this to the detection based on the complete spectral information of the \halpha{} line profile, rather than analyzing the images at single wavelength positions far in the red or blue wings of the \halpha{} line profiles. Despite such a compendious approach, the algorithm was not able to distinguish between traditional RREs and their downflowing counterparts since they have the same spectral characteristics. This is a limitation of the $k$-means clustering technique that accounts only for the spectral lines. However, a statistical analysis based on their apparent motion showed that a substantial number of the strongest redshifted excursions (out of 15,000 events) were downflowing in nature, and had lifetimes and morphology that strongly resembled traditional type-II spicules. The transverse motion associated with the downflowing RREs (observed as zig-zag motions in the plane-of-sky) hints at the possibility of observing the corresponding blue wing counterparts, that would be termed as \textit{downflowing RBEs}. Future high-resolution observations could shed more light in this direction whereby the \alfvenic{} nature associated with them could be investigated in detail.

The downflowing RREs are thought to be the returning components of the previously generated RBEs. Evidence of TR and coronal response to the downflowing RREs provides a very compelling case that attempts to unambiguously relate these returning flows in the chromosphere with the redshifts observed in the TR (and the lower corona), which was not observed in any of the earlier studies. Despite such findings, it is rather challenging to estimate the spatio-temporal filling factor of these events based on our observations from the two datasets used in this thesis. The scenario painted by \cite{2012ApJ...749...60M} points at a longer-lasting, and slowly cooling downflowing component in the TR, in addition to the rapidly upflowing component that is possibly generated by chromospheric heating. The downflowing RREs reported in this thesis are short lived in comparison to the persistent downflows of \cite{2012ApJ...749...60M}. Therefore, it is definitely possible that spicules might contribute to the ensemble downflows observed in the TR but it is not easy to reconcile what fraction of the long-lasting TR redshifts are actually caused by the downflowing spicules. Moreover, the observed regional differences in the strength of the TR downflows \citep[see][]{2007ApJ...654..650M,2020ApJ...903...68P} makes it difficult to understand how spicules would show such differences based on the magnetic activity of the regions. It is possible that coronal loops might play a role in modulating the redshifts in TR in tandem with spicules, but this is pure speculation which needs further studies to reach a definitive consensus. In the future, coordinated ground and space-based observations can focus on a variety of enhanced network (or plage), and quiet Sun targets, corresponding to different heliocentric angles, and a statistical study of the observed downflows can be performed which would significantly aid in addressing some of the issues raised above. Apart from IRIS and SST, observations from the Spectral Imaging of the Coronal Environment \citep[SPICE,][]{2020A&A...642A..14S} instrument onboard the newly launched Solar Orbiter mission \citep{2020A&A...642A...1M}, in coordination with future missions such as the recently approved JAXA's Solar-C EUV High-Throughput Spectroscopic Telescope \citep[EUVST,][]{2019SPIE11118E..07S} and NASA's MUlti-slit Solar Explorer \citep[MUSE,][currently in phase A study]{2020ApJ...888....3D,2021arXiv210615584D}, along with the 4-m Daniel K. Inouye Solar Telescope \citep[DKIST,][]{2020SoPh..295..172R} can reveal new insights in the understanding of the observed TR redshifts and their association with spicules. I am looking forward to such developments.
    \appendix
    \appendixpage 
    \chapter{Detection of on-disk spicules using $k$-means clustering and morphological operations}
\label{appendix:k-means}

\section{$k$-means clustering}
\label{section:appendix_kmeans}

The $k$-means clustering technique is a popular unsupervised learning technique that is used to cluster (or group) an unlabeled data set. Unsupervised learning algorithms basically extract patterns in a dataset that has no predefined tags. In other words, $k$-means partitions a certain number of observations in a dataset into $k$ clusters, where a particular observation is represented by a cluster with the closest mean (also called cluster center). It is an iterative algorithm that is primarily based on minimizing the Euclidean distances between the observations and the cluster centers. The detailed steps are outlined below:

\begin{enumerate}

    \item We begin by choosing $k$ initial cluster centers among which we intend to partition our data.
    
    \item The Euclidean distances between each observed point (say) $\mathbf{\emph{x}}$ and the cluster centers \emph{$\mu$} are computed via $\sum_{i=1}^{n}\sum_{j=1}^{k} ||\mathbf{\emph{x}^{(i)}}-\mathbf{\emph{$\mu$}^{(j)}}||$, where $n$ represents the total number of observed data points in the dataset.
    
    \item The corresponding observed point $\mathbf{\emph{x}}$ is then assigned to a particular cluster with the closest mean by minimizing the residual sum of squares within each cluster. 
    
    \item New cluster centers are computed by obtaining the mean over all the data points (including the new ones from step 3) in each cluster.
    
    \item Steps 2--4 above are repeated in an iterative fashion until a convergence is reached and none of the observed data points change its cluster in two successive iterations.
\end{enumerate}

For a successful and fast convergence of the above steps, it is crucial to carefully choose a certain initial number of cluster centers as a starting point. Randomized selection of $k$ points from the whole dataset may lead to the algorithm being stuck at local minima. Therefore, the \verb|kmeans++| \citep{arthur2007k} initialization method is used where the goal is to choose the initial points that are located as far as possible from each other. This helps the algorithm to converge to a global minimum in relatively fewer iterations. 

From the perspective of the observations presented in this thesis, the cross-aligned \cak{} and \halpha{} datasets (dataset~1) are 4-dimensional, having two spatial, one spectral, and one temporal dimension. The clustering was performed after combining the cross-aligned dataset along the spectral dimension, for each spatial location at a given time. This ensured that a single cluster would effectively represent both \cak{} and \halpha{} spectra, and also allow identification of features in \cak{} based on their spectral signatures in \halpha{}. Both the \halpha{} and \cak{} spectra were normalized appropriately (see Appendix B of Paper~I) before applying $k$-means so as to avoid large pixel-to-pixel intensity fluctuations in the respective line wings. 

The choice of the number of clusters that can optimally divide the dataset into different groups is an important parameter that determines the value of $k$ (in $k$-means). For this purpose, we applied this algorithm to a subset of our observation (ten scans chosen from the entire dataset) with $k$ varying from 15 to 100. Consequently, the mean of standard deviation within the clusters ($\hat{\sigma}_{k}$) was computed for each value of $k$ (ranging between 50--100). Based on the variation of $\hat{\sigma}_{k}$, as a function of $k$, we choose to partition our dataset into 50 clusters each with their own representative mean. We refer to Figure~B.1 in the appendix of Paper~I for a graphical representation of the variation of $\hat{\sigma}_{k}$.

Ten scans (each with 2 spatial and one spectral dimension with more than 2$\times$10$^{7}$ pixels) spread over different times in the data sequence were chosen for training the $k$-means model with 50 clusters. The choice of the scans was primarily based on good seeing conditions (quantitatively described as root mean square contrast variation\footnote{The ratio between standard deviation and the mean over the FOV for each time step.}), and visually assessing the RBE/RRE activity in the \halpha{} images. This allowed the $k$-means model to assign all the pixels in the training dataset to a particular cluster ranging from 0 to 49. \cref{figure_ch-3:RP_H_alpha} and \cref{figure_ch-3:RP_Ca_K} shows the \halpha{} and \cak{} along with their representative profiles (RPs), that are the average over all spectral profiles in a cluster. Later, this model was applied to all the images in the sequence that led to an automated and unsupervised clustering of the spectra corresponding to all pixels in the FOV. Consequently, each pixel on the FOV at any time step is assigned to a particular cluster that is denoted by a representative profile (RP) index. Panel~(d) of \cref{figure_ch-3:Spicule_detection} shows an RP index map at a given time step (scan) in the data generated by the method described above.

\section{Morphological processing}
\label{section:appendix_morph_process}

\begin{figure}[!h]
    \centering
    \begin{adjustbox}{minipage=\linewidth,scale=1.}
    \includegraphics[width=\textwidth]{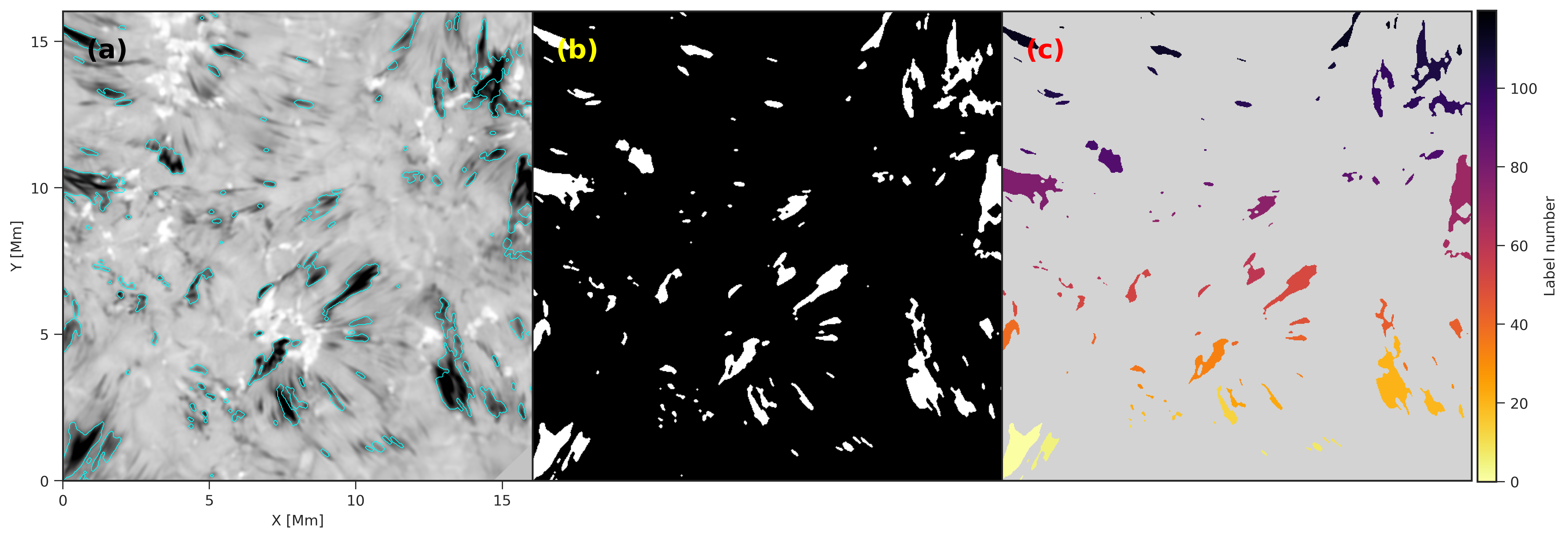}
    \end{adjustbox}
    \caption{An overview of the automated on-disk spicule detection method. Panel~(a) shows a cut out of the full FOV from dataset~1 in the blue wing ($-$38~\kms) of \halpha{} at a given time step. Cyan contours show the detected RBEs from the RPs of interest; panel~(b) shows the binary mask of the detected RBEs in panel~(a), and panel~(c) indicates the different labels after performing connected component labeling. Note that one pixel detections are removed, and one pixel holes are filled in panel~(c). The colorbar associated with panel~(c) indicates the label numbers attributed to the different RBEs. }
    \label{figure_append:detection_method}
\end{figure}

Morphological image processing refers to a collection of non-linear operations associated with shape or overall morphology of features in an image. Morph operations rely on the position of the pixels with reference to each other rather than absolute values of the pixels. Therefore, it is suited to process binary images. In this section, we discuss the morphological processes that enabled the detection of on-disk spicules in an automated fashion from dataset~1 after assigning RPs~0--3 to RBEs and RPs~4--7 to RREs/downflowing RREs based on the arguments presented in \cref{section:spicule_detection}. The method described below is for the detection of RBEs from the dataset, but the exact procedure was also used to detect RREs/downflowing RREs. The details of the steps undertaken are as follows:

\begin{enumerate}
    \item First a 3D binary mask (in space and time domains) is created by assigning a value of 1 (bright) to all pixels in the FOV belonging to RPs~0--3, while the remaining pixels are set to 0 (dark). 
    
    \item A morph open, followed by a close operation is applied to each of the binary masks on a per scan (time) basis with a 3$\times$3 diamond-shaped structuring element.
    
    \item Finally, a 3D connected component labeling \citep{labeling_1996} is employed that enables a unique identification of components based on a given heuristic. This step essentially 'labels' pixels in 3D space that are connected and have similar spectra as set by $k$-means, leading to a spicule detection. We use an 8-neighborhood connectivity in 3D to not bias for direction.
\end{enumerate}

The steps described above essentially allows a way to track all properties of each individual spicules such as size, Doppler offset, and lifetime. In turn, it permitted the study of frequency and distribution of the aforementioned quantities per RP, that is shown in Paper~II. Panels~(a)--(c) of \cref{figure_append:detection_method} provides an overview of the detection algorithm described above. A structuring element plays a central role in morphological image processing and it is analogous to convolution kernels in image filtering. The fundamental characteristics of a structural element are its shape and size. The shape determines a way of distinguishing some objects in an image from others based on their spatial orientation. A diamond shaped matrix can efficiently capture both the length and breadth of a spicule, while the relatively small size (3$\times$3) of this matrix preserves finer details of the image features that are three times the pixel scale of CHROMIS (0\farcs037). To avoid erroneous detections that are at the limit of the structuring element a lower threshold of the detected features was set at 4 CHROMIS pixels roughly corresponding to 100~km. 

The morphological opening is an erosion followed by a dilation operation that is useful to remove small (in this case 1-pixel) connectivity between different morphological structures in the 2D space as would be the commonly observed case of spicules moving in very close proximity to each other. Basically, it filters the noisy detections at spatial scales that are determined by the structuring element. An opening operation can create holes (dark features) between brighter structures which can be filled by performing a morph closing operation (opposite of opening) by using the same structural element as before. The sequence of morphological operations employed preserves dimensions of the spicules, as all dilates and erodes are performed with the same size of structuring element. Furthermore, the above set of operations were performed in 2D space so that the connection of spicules along the temporal domain remains unaffected, which would enable us to label them effectively. The 3D labeling step enabled the detection of roughly 15,000 RREs/downflowing RREs and 20,000 RBEs whose location and the density distribution are shown in panels~(a) and (b) of \cref{figure_ch-3:Spicule_detection}.
%    \include{sections/appendixB}
% \end{refsegment}      

    % \paper              % "Chapter" is renamed "Paper"
    % \paperpage          % Similar to \part*{Papers}, but appears in TOC
    % \numberofpapers{3}  % Specify size of thumb indices

    % \include{sections/paperI}
    % \include{sections/paperII}
    % \include{sections/paperIII}
    % \printbibliography{bibliographies/chapter1.bib}
    % \bibliographystyle{aa} % style aa.bst
    % \bibliography{bibliographies/chapter1.bib} 
    % \printbibliography
    \setquotestyle{english}
    \printbibliography[title={Bibliography}]
    % \include{sections/paperI}
    % \include{sections/paperII}
    % \include{sections/paperIII}

    % \appendix           % "Chapter" is renamed "Appendix"
    % \appendixpage       % Similar to \part*{Appendices}, but appears in TOC

    % \include{sections/appendixA}
    % \include{sections/appendixB}

\end{document}